\documentclass[a4paper,11pt]{article}
\pdfoutput=1
\usepackage{jheppub}

\usepackage{cleveref}
\usepackage{booktabs}
\usepackage{upgreek}
\usepackage{todonotes}
\usepackage{multirow}
\usepackage{amssymb}
\usepackage[T1]{fontenc}
\usepackage{dsfont}
\usepackage[squaren]{SIunits}
\usepackage{placeins}
\usepackage{xspace}
\usepackage{comment}

\newcommand{\sIK}{\ensuremath{s_{IK}}}
\newcommand{\sij}{\ensuremath{s_{ij}}}
\newcommand{\sjk}{\ensuremath{s_{jk}}}
\newcommand{\sik}{\ensuremath{s_{ik}}}
\newcommand{\yij}{\ensuremath{y_{ij}}}
\newcommand{\yjk}{\ensuremath{y_{jk}}}
\newcommand{\yik}{\ensuremath{y_{ik}}}
\newcommand{\sAB}{\ensuremath{s_{AB}}}
\newcommand{\yAB}{\ensuremath{y_{AB}}}
\newcommand{\sab}{\ensuremath{s_{ab}}}

\newcommand{\saj}{\ensuremath{s_{aj}}}
\newcommand{\yaj}{\ensuremath{y_{aj}}}
\newcommand{\sjb}{\ensuremath{s_{jb}}}
\newcommand{\yjb}{\ensuremath{y_{jb}}}
\newcommand{\sAK}{\ensuremath{s_{AK}}}
\newcommand{\yAK}{\ensuremath{y_{AK}}}
\newcommand{\yak}{\ensuremath{y_{ak}}}
\newcommand{\sak}{\ensuremath{s_{ak}}}
\newcommand{\expo}[1]{\ensuremath{\,\text{e}^{#1}}}
\newcommand{\Kallen}{\ensuremath{f^\mathrm{FF}_{\text{K\"all\'en}}}}
\newcommand{\KallenRF}{\ensuremath{f^\mathrm{RF}_{\text{K\"all\'en}}}}
\newcommand{\D}{\ensuremath{\text{d}}}
\newcommand{\alphastrong}{\ensuremath{\alpha_{\text{s}}}}
\newcommand{\gstrong}{\ensuremath{g_\text{s}}}

\numberwithin{equation}{section}
\allowdisplaybreaks

\preprint{CoEPP-MN-20-2, MCNET-20-09}

\title{Sector Showers for Hadron Collisions}

\author{Helen Brooks,}
\author{Christian T Preuss,}
\author{Peter Skands}

\emailAdd{helen.brooks@monash.edu}
\emailAdd{christian.preuss@monash.edu}
\emailAdd{peter.skands@monash.edu}

\affiliation{
	School of Physics and Astronomy, Monash University,\\
	Wellington Road, Clayton, VIC-3800, Australia}

\abstract{
	In conventional parton showers (including ones based on
        dipoles/antennae), a given $(\mathrm{Born}+m)$-parton configuration can typically be reached via  ${\mathcal O}(m!)$ different \textquotedblleft shower histories\textquotedblright. In the context of matrix-element-correction and merging procedures, accounting for these histories mandates fairly complex and resource-intensive algorithms.
	A so far little-explored alternative in the shower context is to divide the branching phase spaces into distinct \textquotedblleft sectors\textquotedblright, each of which only receives contributions from a single branching kernel. This has a number of  consequences including making the shower operator bijective; i.e., each parton configuration now has a \emph{single} unique \textquotedblleft inverse\textquotedblright. As a first step towards developing a full-fledged matrix-element-correction and merging procedure based on such showers, we here extend the sector approach for antenna showers to hadron-hadron collisions, including mass and helicity dependence.  
}

\begin{document}
\maketitle
\flushbottom

\newcommand{\Vincia}{V\protect\scalebox{0.8}{INCIA}\xspace}
\newcommand{\Pythia}{P\protect\scalebox{0.8}{YTHIA}\xspace}
\newcommand{\Rivet}{R\protect\scalebox{0.8}{IVET}\xspace}
\newcommand{\Sherpa}{S\protect\scalebox{0.8}{HERPA}\xspace}
\newcommand{\Ariadne}{A\protect\scalebox{0.8}{RIADNE}\xspace}
\newcommand{\mg}{M\protect\scalebox{0.8}{AD}G\protect\scalebox{0.8}{RAPH}5\_aM\protect\scalebox{0.8}{C}@N\protect\scalebox{0.8}{LO}}
\newcommand\varpm{\mathbin{\vcenter{\hbox{%
  \oalign{\hfil$\scriptstyle+$\hfil\cr
          \noalign{\kern-.3ex}
          $\scriptscriptstyle({-})$\cr}%
}}}}
\newcommand\varmp{\mathbin{\vcenter{\hbox{%
  \oalign{$\scriptstyle({+})$\cr
          \noalign{\kern-.3ex}
          \hfil$\scriptscriptstyle-$\hfil\cr}%
}}}}

\newcommand{\overbar}[1]{\mkern 5mu\overline{\mkern-5mu #1 }}
\newcommand{\Atrial}{\overbar{\vphantom{``}\mathcal{A}}}

\section{Introduction}
\label{sec:intro}
General-purpose Monte Carlo event generators such as \textsc{Herwig}~\cite{Bellm:2015jjp}, \Sherpa~\cite{Gleisberg:2008ta}, and \Pythia \cite{Sjostrand:2014zea} have become
indispensable tools to study the complex event structures produced in highly energetic hadronic collisions. At the heart of these tools, parton showers resum all leading-logarithmic (LL)
terms in the perturbative expansion of QCD, thereby connecting the perturbative, high-energy hard scale with the intrinsically non-perturbative, low-energy soft scale, at which the strong
coupling becomes large enough to invalidate perturbation theory. 

Early parton showers were built upon the
Dokshitzer-Gribov-Lipatov-Altarelli-Parisi (DGLAP) splitting kernels
\cite{Gribov:1972ri,Altarelli:1977zs,Dokshitzer:1977sg}, which are
suitable for describing collinear (small-angle) radiation off a hard partonic leg, but which fail to describe the interference of soft wide-angle radiation.
To incorporate coherent (wide-angle) emissions in the DGLAP formalism, one can either order emissions decreasing in angle, cf.\ \cite{Marchesini:1983bm},
or veto  emissions which are not  ordered in angle, cf.\ \cite{Bengtsson:1986et}. 

Alternatively, coherence effects can be directly taken into account
(to leading power in $1/N_C^2$) by considering colour dipoles, that is
by considering particle emissions as stemming from colour-anticolour
pairs. The first shower model based on this concept was
\Ariadne~\cite{Gustafson:1987rq,Lonnblad:1992tz}, with
\Vincia\ following a similar approach many years
later~\cite{Giele:2007di,Giele:2011cb}. In the meantime, an
alternative type of dipole showers, based on the Catani-Seymour (CS)
approach~\cite{Catani:1996jh,Catani:1996vz}, also
appeared~\cite{Nagy:2005aa,Schumann:2007mg,Winter:2007ye,Platzer:2009jq,Hoche:2015sya}. We
will not give a complete comparison between all of these approaches
here. It will suffice to say that:
\begin{itemize}
\item For final-state showers, the original formulation of the Lund dipole model~\cite{Gustafson:1987rq} (as implemented in \Ariadne)
is equivalent to the \textquotedblleft global\textquotedblright\ antenna subtraction and antenna shower framework (as implemented in \Vincia)~\cite{GehrmannDeRidder:2011dm}.
\item For initial-state showers, \Vincia\ is based on a backwards-evolution paradigm~\cite{Sjostrand:1985xi,Ritzmann:2012ca}, whereas \Ariadne\ uses a forwards-evolution picture. 
\item In the Lund-dipole / antenna formalisms, the soft singularity associated with each parton pair is captured by a single \textquotedblleft antenna function\textquotedblright\ and an associated
kinematics map in which both of the parents in general share the (transverse and longitudinal) recoil. In the CS dipole formalism, the radiation from each colour-connected parton pair is instead
partitioned into two separate terms, in which each of the parents in turn act as the \textquotedblleft emitter\textquotedblright\ while the other, the \textquotedblleft recoiler\textquotedblright,
recoils purely longitudinally along the dipole axis. 
\item In all of the dipole/antenna formalisms, the collinear $g\mapsto gg$ singularity is partitioned into two terms, one for each colour-connected partner of the parent gluon.
This partitioning is done such that the  $g\mapsto gg$ DGLAP kernel is obtained by summing over the two neighbouring dipoles/antennae. The exact form of the partitioning factor differs between
different models/implementations. 
\end{itemize}
 
To improve the precision away from the collinear and soft limits,
showers are usually matched or merged to fixed-order
calculations. Examples of well-established approaches include
MC@NLO \cite{Frixione:2002ik} and POWHEG \cite{Nason:2004rx,Frixione:2007vw} in the case of the former, and CKKW(-L) \cite{Catani:2001cc,Lonnblad:2001iq,Lonnblad:2011xx}, MLM
\cite{Mangano:2001xp,Mangano:2006rw}, in the case of the latter.
In recent years there has been a proliferation of refinements to merging algorithms, such as UMEPS~\cite{Lonnblad:2012ng}, UNLOPS~\cite{Lonnblad:2012ix}, 
MEPS@NLO~\cite{Gehrmann:2012yg,Hoeche:2012yf}, FxFx~\cite{Frederix:2012ps}, MINLO~\cite{Hamilton:2012np}, and MINNLOPS~\cite{Monni:2019whf}.

Matching algorithms correct the hardest emission of the parton shower to the stated fixed-order accuracy, typically next-to-leading order (NLO). Merging algorithms instead combine inclusive event-samples, 
each of a given accuracy but of increasing multiplicity. Merging is typically favoured where it is desirable to describe multijet final states. 
To consistently combine multiple event samples with the parton shower, it is necessary to reweight each event with a Sudakov factor, which in turn requires obtaining a sequence of scales
that represents the ``parton shower history''. It is possible to obtain these scales through ``winner-takes-it-all'' clustering methods, as is done in the case of CKKW. However,
this does not correspond to a direct inversion of the parton shower and potentially results in missed areas of phase space. Ideally one would obtain all paths of possible clusterings back to
the Born topology (and selecting one with its relative probability) as in CKKW-L.  This, however, comes at the price of a significant computational
overhead, rendering such calculations intractable for many legs (see, e.g.,~\cite{LopezVillarejo:2011ap,Fischer:2017yja,Hoeche:2019rti}). 

Owing to the very formulation of both DGLAP and dipole showers, namely that every colour charge in the event can radiate another parton, successive radiation leads to a proliferation of terms
in the cascade from the initial scale down to the hadronisation scale. For a process with $n$ shower branchings, there are thus in general $2^nn!$ histories in a CS dipole shower. 
Conventional antenna (and Lund-dipole) showers reduce this number by $2^n$, as there is now only one term for each \emph{pair} of colour-connected partons, but the 
scaling\footnote{This can in principle be reduced to just $n$ terms
  after the $n$-th branching by imposing a strictly Markovian ordering
  condition~\cite{Giele:2011cb}. This is, however, likely to lead to
  undesired side effects~\cite{Hartgring:2013jma,Li:2016yez} and hence is not 
  considered here.} still goes like $n!$.
Multileg merging therefore to date remains impeded by the sheer number of possible histories. This is especially true for CS-style approaches while  antenna-based ones 
(see, e.g.,~\cite{Fischer:2017yja})
should be expected to exhibit a somewhat lower computational overhead, see \cref{tab:histories}. 

\begin{table}[t]
    \centering
    Number of Histories for $n$ Branchings 
    \begin{tabular}{c|rrrrrrr}
\toprule
& $n=1$ & $n=2$ & $n=3$ & $n=4$ & $n=5$ & $n=6$ & $n=7$\\
\cmidrule{1-8}
        CS Dipole & 2 & 8 & 48 & 384 & 3840 & 46080 & 645120\\
        Global Antenna & 1 & 2 & 6 & 24 & 120 & 720 & 5040 \\        \bottomrule
    \end{tabular}
    \caption{The number of possible shower/clustering histories (ordered + unordered) that can contribute to a given (colour-ordered) parton configuration, after $n$ branchings starting
    from a single colour-anticolour pair.}
    \label{tab:histories}
\end{table}

A promising alternative that could reduce the complexity even further is the \emph{sector antenna} formalism \cite{Kosower:1997zr,Kosower:2003bh,Larkoski:2009ah,Larkoski:2011fd,LopezVillarejo:2011ap}.
In the context of the early \Vincia\ final-state shower, it was shown in \cite{LopezVillarejo:2011ap} that an antenna-shower history can be made unique if instead of the conventional (henceforth 
called \emph{global}) shower algorithm, a \emph{sector} formulation of the antenna framework is used, in which only a single shower term is allowed to contribute to each specific $(n+1)$-parton
phase-space point; all other potential contributions are vetoed. 
In this framework, each antenna splitting kernel must incorporate both
the full soft and collinear limits of the respective phase-space
sectors, in contrast to global antenna functions (such as those used
in \textsc{Ariadne} and
in earlier versions of \textsc{Vincia}) which smoothly
partition the collinear $g\mapsto gg$ singularities between two
neighbouring antennae.  
Although there remains an ambiguity in how to treat cases with
multiple interfering Born processes\footnote{An example of two such interfering Born
processes is $H\to gg$ and $H\to b\bar{b}$,
which mix at the $H\to b \bar{b} g$ level, via the $gg\mapsto b\bar{b} g$
and $b\bar{b}\mapsto b\bar{b}g$ branchings, respectively. Note that we use the slightly different symbols
$\to$ and $\mapsto$ to distinguish between hard processes and shower
evolution steps.}, the potential number of these
remains small for all $n$.

In the sector framework, the sharing of the gluon-collinear limits is
viewed as a discrete partitioning of phase space into two
\emph{sectors}, which in the collinear (DGLAP) limit correspond to
$z<1/2$ and $z>1/2$ respectively; in this picture, the gluon with
lower energy fraction $z$ is \emph{always} regarded as the emitted one.  
To generalise this \emph{sector decomposition}, we use a notion of transverse momentum ($p_\perp^2$)  which tends to $z Q^2$ in the collinear limits (with $Q^2$ the virtuality of the branching particle),
so that it remains unique, well-defined, and exact outside the singular limits. It is then the gluon with the lowest $p_\perp^2$ which is regarded as the emitted one.
Inside each sector only one antenna, which captures the full soft singularity of the respective phase-space sector and the full and $z<\frac12$ collinear
singularities for quarks and gluons respectively,  is allowed to contribute. 
This brings the number of possibilities to go from a given final state
all the way back to the Born configuration down to one (modulo the
question of interfering Born processes already alluded to), thus
creating a unique shower history. Sector showers thus have great potential for speeding up matched and merged multijet calculations.

Helicity- and mass-dependent sector antenna functions for initial- and final-state radiation were already presented in
\cite{Larkoski:2009ah,Larkoski:2011fd}, although without a dedicated implementation in a shower algorithm.
We here present a new implementation of a sector shower based on the \Vincia\ antenna shower in the \textsc{Pythia} 8.3 framework\footnote{
We note that, at the technical level, the major undertaking of integrating the \Vincia\ shower model fully within \textsc{Pythia}~8.3 (where previously it was developed a stand-alone plugin to 
\textsc{Pythia}~8.2) was just recently completed. 
To simplify this work, only a minimal subset of \Vincia\ --- which
only includes the global antenna-shower model --- was incorporated
into \textsc{Pythia} version 8.301. The sector shower, along with
other features such as matrix-element corrections, will be included in
forthcoming updates in the near future. The results we present in this
paper are obtained with a development version that essentially
represents  \textsc{Pythia}~8.301 + sector shower.}, including
helicity dependence~\cite{Larkoski:2013yi,Fischer:2017htu}, mass
corrections~\cite{GehrmannDeRidder:2011dm}, initial-state
radiation~\cite{Ritzmann:2012ca,Fischer:2016Vincia}, resonance
decays~\cite{Brooks:2019xso}, and interleaved coherent QED branchings~\cite{Skands:2020lkd}.
We define a full set of helicity-dependent sector antenna functions for initial- and final-state radiation entirely based on crossings and sums of global final-final antenna functions and validate
the new shower against leading-order (sector-sh) matrix elements, the global \Vincia\ shower, \Pythia\ 8.3, and experimental data.

The paper is structured as follows. After reviewing the foundations of the \Vincia\ antenna shower in \cref{sec:VinciaBasics}, the sector-shower implementation is explained in detail in \cref{sec:SectorShowerAlgorithm}.
The implementation is validated in \cref{sec:Validation} and we
conclude and give an outlook on matching and merging applications
in \cref{sec:Conclusion}.

\section{The \Vincia Antenna Shower}
In this section, the main focus is on aspects that are common to both
sector and global showers, including phase-space factorisations,
kinematics maps (a.k.a.\ recoil strategies),
and our choices of evolution variables. 
In addition to providing the starting point for our discussion of sector showers in \cref{sec:SectorShowerAlgorithm}, the material in this section therefore also serves as an up-to-date summary of the current implementation of  \Vincia's global showers in \textsc{Pythia}~8.3. 

\label{sec:VinciaBasics}
The starting point for antenna showers is the factorisation of QCD amplitudes in the soft-gluon limit,
\begin{equation}
    \vert \mathcal M_{n+1}(1,\ldots,i,j,k,\ldots,n+1) \vert^2 \overset{g_j \text{ soft}}{\longrightarrow} a_{j/IK}(\sij,\sjk) \vert \mathcal{M}_n(1,\ldots, I,K,\ldots,n) \vert^2 \, , \label{eq:SoftGluonFactorisation}
\end{equation}
where we adopt the convention of \cite{Fischer:2016Vincia} and label
parent or \textquotedblleft pre-branching\textquotedblright\ partons (i.e., ones in the  $n$-parton
configuration) by capital letters and daughter or ``post-branching''
partons (i.e., ones in $(n+1)$-parton configuration) by lowercase
letters. Furthermore, legs that are in the initial state will be
denoted with letters from the beginning of the alphabet, $a$, $b$,
while final-state legs are denoted with letters starting from $i$, $j$, $k$. 
Dimensionful invariants, regardless of whether the partons are incoming or outgoing, are cast in terms of dot products
\begin{equation}
    s_{jk} \equiv 2p_j \cdot p_k~,
\end{equation}
while invariant masses for final-final (FF) and initial-initial (II)
antennae are denoted as 
\begin{equation}
    m_{jk}^2 = (p_j + p_k)^2 = m_j^2 + m_k^2 + s_{jk} \, ,
\end{equation}  
and momentum transfers for initial-final (IF) and resonance-final (RF) antennae as  
\begin{equation}
     q_{ai}^2 = (p_a - p_i)^2 = m_a^2 + m_i^2 - s_{ai} \, .
\end{equation}
Energy-momentum conservation then implies:
\begin{align}
    \text{FF} & : 
    & \sIK + m_I^2 +m_K^2 = \sij+\sik+\sjk +m_i^2 +m_j^2 +m_k^2 
    \label{eq:epConsFF}\\[1mm]
\text{RF \& IF} & : 
    & \sAK -m_A^2 -m_K^2 = \saj+\sak-\sjk -m_a^2-m_j^2-m_k^2
    \label{eq:epConsIF}
    \\[1mm]
    \text{II} & : 
    & \sAB + m_A^2 +m_B^2 = \sab-\saj-\sjb +m_a^2+m_j^2+m_b^2 
    \label{eq:epConsII}
\end{align}
and we define dimensionless invariants $y\in [0,1]$
\begin{align}
 & \yij = \frac{\sij}{s}  && \mu_j^2 = \frac{m_j^2}{s} &
 \label{eq:DimensionlessInvariants}
\end{align}
by scaling by the largest dynamical invariant\footnote{In \cite{Brooks:2019xso,Fischer:2016Vincia}
  the normalisation $\sAK + \sjk$ was used, which differs from
  \cref{eq:largestInvariant} only for the case of massive splittings. 
The alternative normalisation used here was chosen in order to have
better uniformity of conventions.},  $s$: 
\begin{equation}
    s = \begin{cases}
    \sIK & \text{FF}\\
    \saj + \sak & \text{RF and IF}\\
    \sab & \text{II}
    \end{cases} \, .
    \label{eq:largestInvariant}
\end{equation} 
It should be emphasised, that parton $j$ is always in the final state, as we do not consider emission into the initial state. 
In \cref{sec:Kinematics} on branching kinematics below,
we provide some translations between the explicitly Lorentz-invariant $y$ variables defined here and the $z$ or $x$ energy fractions that are used in other shower formalisms and in PDFs.
Note also that for the specific case of gluon emission, $m_j=0$ and the
masses of the pre-branching partons are the same as those of the post-branching ones, hence all the $m^2$ terms in \cref{eq:epConsFF,eq:epConsII} cancel. Those terms are only relevant for branching processes involving a change in the number of massive quarks.

Considering it a better approximation to use massless kinematics for 
incoming heavy-flavour legs than to assign them their nominal on-shell
values, the current treatment of initial-state mass effects in \textsc{Vincia}
is a compromise, with all initial-state legs forced to have massless
kinematics. This is similar to the choice made in  \textsc{Pythia}'s
default showers~\cite{Sjostrand:2004ef}. Nevertheless, to maintain
maximum generality and to avoid needlessly obscuring the crossing
relations between initial- and final-state antennae, we maintain a
general language where possible, which at least in principle allows for 
massive initial-state partons. Mass terms for initial-state partons are therefore included in the antenna functions, cf.\ \cref{sec:AppendixAntennaFcts}, and in our evolution and sector resolution variables, cf.\ \cref{sec:EvolutionVariables,sec:SectorResolution}. 
When interpreting these in the context of massless initial-state kinematics, we use the following conventions:
\begin{itemize}
    \item Motivated by \cite{Abelof:2011jv}, we include mass terms for both quarks in initial-state $q_A \mapsto g_a q_j$ branchings, which, from the forward-evolution point of view, look like the gluon splitting $g_a \mapsto \bar q_A q_j$.
We justify this choice by noting that this branching is a crossing $j \to a$ of the final-state gluon emission process $q_I \mapsto q_i g_j$, with gluon $g_j$ being crossed to be identified with $g_a$, and therefore no massive parton is crossed into the initial state.
\item 
Following the same logic, we include mass terms only for the final-state quark in quark conversion $g_A \mapsto q_a q_j$, which, from the forward-evolution point of view, looks like an initial-state quark is radiating a gluon. 
There, the massive quark $q_i$ in the $g_I \mapsto \bar q_i q_j$ splitting had to be crossed into the initial state to become the massive quark $q_a$.
\end{itemize}
Although some initial-state mass effects might be implemented without major inconsistencies in the near future, a full-fledged and consistent treatment of massive initial-state legs depends upon the availability of massive PDFs and is outside the scope of this work.
First steps towards a consistent inclusion of initial-state mass effects have been presented in \cite{Krauss:2017wmx}, where an implementation in the Catani-Seymour dipole shower in \Sherpa\ has been outlined. As antennae can effectively be thought of as the sum of two CS dipoles, a similar scheme might be implemented in the \Vincia\ antenna shower.

\subsection{Antenna Functions}
The antenna function $a_{j/IK}$ in \cref{eq:SoftGluonFactorisation} acts as the splitting kernel of the coherent $2\mapsto 3$ branching $IK \mapsto ijk$ and may formally be represented as
\begin{equation}
    a_{j/IK}(\sij,\sjk,m_i^2,m_j^2,m_k^2) = \frac{\vert \mathcal M_3(i,j,k) \vert^2}{\vert \mathcal M_2(I,K)\vert^2} \, .
\end{equation}
in the factorised form of the integral over the three-body matrix element
\begin{equation}
    \vert \mathcal M_3(i,j,k) \vert^2 \D \Phi_3 = \vert \mathcal M_2(I,K) \vert^2 \D \Phi_2\, a_{j/IK}(\sij,\sjk,m_i^2,m_j^2,m_k^2) \frac{\D \Phi_3}{\D\Phi_2}.
\end{equation}
By construction, it reproduces the full eikonal factor in the limit of gluon $g_j$ becoming soft,
\begin{equation}
    a_{j/IK}(\sij,\sjk,m_i^2,0,m_k^2) \overset{g_j \text{ soft}}{\longrightarrow} \gstrong^2 \mathcal C_{j/IK} \left[\frac{2\sik}{\sij\sjk} 
    - \frac{2m_i^2}{s_{ij}^2} - \frac{2m_k^2}{s_{jk}^2}\right] \, ,
\end{equation}
with a process-dependent colour factor $\mathcal C_{j/IK}$. When $g_j$ becomes (quasi-)collinear with a quark $i$, it reproduces the full (quasi-collinear) DGLAP splitting kernel $P(z)$ (or $P(z)/z$ for an initial-state parton),
\begin{equation}
     a_{j/IK}(\sij,\sjk,0,0,0) \overset{i\parallel j}{\longrightarrow} \gstrong^2 \mathcal{C}_{j/IK} \frac{P(z)}{s_{ij}} \, ,
\end{equation}
where $z$ is the energy fraction taken by quark $i$. 

For the branchings $g\mapsto gg$, however, the treatment of the collinear limits differs between the global and sector formalisms. For the global antenna functions, 
the DGLAP splitting kernel is partitioned onto two neighbouring (colour-adjacent) antennae,
whose collinear limits are related by $z \leftrightarrow 1-z$ so that one of them incorporates the $1/(1-z)$ part of the DGLAP kernel while the other incorporates the $1/z$ part. 
For each value of $z$ the full collinear singularity is only recovered after summing over these two terms.  
In the sector formalism, instead, the shared gluon-collinear singularity is fully incorporated into \emph{both} of the neighbouring antenna functions;
these are then supplemented by a phase-space veto such that the $z\le 1/2$ part of the collinear boundary is covered by only one of them and the $z> 1/2$ part by the other. 
We return to this point in \cref{sec:SectorAntennae}.

By contrast, due to the lack of a soft singularity, gluon splittings are handled somewhat differently in the antenna framework.
In gluon splittings, a clear distinction between the splitter and
spectator is possible; therefore these could in principle be treated
differently to gluon emissions.
However, since energy and momentum conservation for on-shell partons requires at least a $2\mapsto3$ phase-space factorisation and since we want to generate a common (interleaved) sequence
in which all types of evolution steps are evolved in a common
resolution measure~\cite{Sjostrand:2004ef}, we use the same
phase-space and kinematics maps and ordering variables as for gluon
emissions\footnote{We note that this choice differs from that of the
  previous version of \textsc{Vincia}~\cite{Fischer:2016Vincia},
  in which gluon emissions were evolved in $p_\perp$ while gluon splittings were evolved in virtuality.}.

To distinguish between global and sector antenna functions, and between which partons are initial- and which are final-state ones, we label them as follows:
\begin{equation}
    a^{\text{state},\text{type}}_{j/IK} (\sij,\sjk,m_i^2,m_j^2,m_k^2) \, ,
\end{equation}
with \textquotedblleft state\textquotedblright\ $\in$ [II, IF, FF, RF], \textquotedblleft type\textquotedblright $\in$ [glb,sct].

Note that since $2\mapsto 3$ antenna functions have dimension -2 in natural units,
we usually cast them in terms of a dimensionless function of the $y$ and $\mu$ variables defined in \cref{eq:DimensionlessInvariants}
multiplied by a single (constant) dimensionful quantity\footnote{Specifically we normalise by $\sIK^{-1}$, $\sAK^{-1}$ and $\sAB^{-1}$ for the FF, IF(RF) and II antenna functions respectively.}, cf.\ \cref{sec:AppendixAntennaFcts}.
Moreover, throughout we use colour- and coupling-stripped antenna functions $\bar a$,
\begin{equation}
    a_{j/IK}(\sij,\sjk,m_i^2,m_j^2,m_k^2) = 4\uppi \, \alphastrong\, \mathcal C_{j/IK} \, \bar a_{j/IK}(\yij,\yjk,\mu_i^2,\mu_j^2,\mu_k^2) \, .
\label{eq:antDimless}
\end{equation}
Below, we will often let the mass dependence be implicit to avoid clutter.
The colour factor $\mathcal C_{j/IK}$ is chosen in a convention in which both $C_\text{F}$ and $C_\text{A}$ tend to $N_\text{C}$ in the large-$N_\text{C}$ limit and where $T_\text{R}$ is unity. In this convention, the ambiguity related to the colour factor of a $qg$ antenna is explicitly a subleading-colour effect \cite{Giele:2007di}. 

The complete set of antenna functions used in \Vincia\ are collected in \cref{sec:AppendixAntennaFcts}, including both mass and helicity dependence.

\subsection{Phase-Space Factorisation}
The Lorentz-invariant $(n+1)$-particle phase space measure in four dimensions,
\begin{equation}
    \D \Phi_{n+1}(p_a,p_b;p_1,\ldots,p_{n+1}) = (2\uppi)^4 \delta^{(4)}\left(p_a+p_b- \sum\limits_{\ell=1}^{n+1} p_\ell \right) \prod\limits_{\ell = 1}^{n+1} \frac{\D^4 p_\ell}{(2\uppi)^3} \delta^{(4)}(p_\ell^2 - m_\ell^2) \, ,
\end{equation}
exactly factorises into the ($n-m+1)$-particle phase-space measure $\D \Phi_{n-m+1}$ and the $m$-particle branching measure $\widetilde{\D \Phi}_m$, 
\begin{equation}
    \D \Phi_{n+1}(p_a, p_b; p_1,\ldots,p_j,\ldots,p_{n+1}) = \D \Phi_{n-m+1}(p_A,p_B;p_1,\ldots,p_{n+1}) \widetilde{\D \Phi}_m \, .
\end{equation}

\paragraph{FF Branchings} The branching measure $\widetilde{\D \Phi}_m$ is proportional to the three-particle phase space measure $\D \Phi_3$ which can be factorised into the product of the two-particle pre-branching phase space and the FF antenna phase space measure,
\begin{equation}
    \D \Phi_3(p_A,p_B;p_i,p_j,p_k) = \D \Phi_2(p_A,p_B; p_I, p_K) \D \Phi_\text{ant}^\text{FF}(p_i,p_j,p_k) \, . \label{eq:PhaseSpaceFactorisationFF}
\end{equation}
In terms of the dimensionless invariants $\yij$, $\yjk$ and an angle $\phi$ between the branching plane and the parent dipole, the final-final antenna phase space can be written as~\cite{GehrmannDeRidder:2011dm}
\begin{equation}
    \D \Phi_\text{ant}^\text{FF} = \frac{1}{16\uppi^2} \Kallen \sIK \Theta(\Gamma_{ijk}) \D \yij \D \yjk \frac{\D\phi}{2\uppi} \, , \label{eq:PhaseSpaceFF}
\end{equation}
with the dimensionless three-body Gram determinant expressing  boundaries of the physical phase space
\begin{equation}
    \Gamma_{ijk} = \yij\yjk\yik - \yjk^2\mu_i^2 - \yik^2\mu_j^2 - \yij^2\mu_k^2 + 4 \mu_i^2\mu_j^2\mu_k^2 \, ,
    \label{eq:GramDet}
\end{equation}
again scaled by the appropriate normalisation according to \cref{eq:DimensionlessInvariants}. Note that for massless partons, the $\Theta(\Gamma)$ factor just reduces to the standard triangular phase space defined by $\yij \ge 0$, $\yjk \ge 0$, $\yij + \yjk \le 1$, while for massive ones it defines a smaller region inside this hull, see, e.g.~\cite{GehrmannDeRidder:2011dm}. 
The volume of the two-particle phase-space of the parent antenna is taken into account by the K\"all\'en factor,
\begin{equation}
    \Kallen = \frac{\sIK}{\sqrt{\lambda(m_{IK}^2,m_I^2,m_K^2)}} \, ,
\end{equation}
in terms of the K\"all\'en function
\begin{equation}
    \lambda(a,b,c) = a^2+b^2+c^2-2(ab+ac+bc) \, .
\end{equation}
Note that $\Kallen = 1$ if either or both of partons $I$ and  $K$ are massless, while when both are massive we have $\Kallen > 1$. This does not present a  problem for generating numerical overestimates in the shower; $\Kallen$ is just an overall constant and can be factored out of the branching integrals.

\paragraph{RF Branchings}
The phase space factorisation for resonance-final branchings is largely unchanged relative to the final-final case \cite{Brooks:2019xso}:
\begin{equation}
    \D \Phi_\text{ant}^\text{RF} = \frac{1}{16\uppi^2} \KallenRF \frac{\sAK + m^2_j+ m^2_k - m^2_K}{(1-\yjk)^3} \Theta(\Gamma_{ajk}) \D \yaj \D \yjk \frac{\D\phi}{2\uppi} \, , \label{eq:PhaseSpaceRF}
\end{equation}
where now the K\"all\'en factor is given by:
\begin{equation}
 \KallenRF = \frac{\sAK + m^2_j+ m^2_k - m^2_K}{\sqrt{\lambda(m_{A}^2,m_{AK}^2,m_K^2)}} \, .
\end{equation}

\paragraph{IF Branchings} Keeping initial-state partons explicitly massless but allowing for massive final-state ones, the factorisation \cref{eq:PhaseSpaceFactorisationFF} is replaced by the convolution~\cite{Abelof:2011jv,Daleo:2006xa,Ritzmann:2012ca}
\begin{multline}
    \int \frac{\D x_a}{x_a} \Theta(1-x_a) \frac{\D x_B}{x_B} \Theta(1-x_B) \D \Phi_3(p_a, p_B; p_j, p_k, p_R) =\\ \int \frac{\D x_A}{x_A} \Theta(1-x_A) \frac{\D x_B}{x_B} \Theta(1-x_B) \D \Phi_2(p_A, p_B; p_K, p_R) \D \Phi_\text{ant}^\text{IF}(p_a,p_j,p_k) \, ,
\end{multline}
where the IF antenna phase space measure can be written in terms of dimensionless invariants $\yaj$, $\yjk$, and the angle $\phi$ as~\cite{Verheyen2019}
\begin{equation}
    \D \Phi_\text{ant}^\text{IF} = \frac{1}{16\uppi^2} \frac{1}{1-\yjk}\sAK \Theta(\Gamma_{ajk}) \D \yaj \D \yjk \frac{\D\phi}{2\uppi} \, .
\end{equation}

\paragraph{II Branchings} With a massive emitted final-state parton, the branching measure is given by the convolution~\cite{Daleo:2006xa,Ritzmann:2012ca}
\begin{multline}
    \int \frac{\D x_a}{x_a} \Theta(1-x_a) \frac{\D x_b}{x_b} \Theta(1-x_b) \D \Phi_2(p_a, p_b; p_j, p_r) =\\ \int \frac{\D x_A}{x_A} \Theta(1-x_A) \frac{\D x_B}{x_B} \Theta(1-x_B) \D \Phi_1(p_A, p_B; p_R) \D \Phi_\text{ant}^\text{II}(p_a,p_j,p_b) \, ,
\end{multline}
where again, the antenna phase space can be written in terms of dimensionless invariants $\yaj$, $\yjb$, and the angle $\phi$ as~\cite{Verheyen2019}
\begin{equation}
    \D \Phi_\text{ant}^\text{II} = \frac{1}{16\uppi^2} \frac{1}{1-\yaj-\yjb} \sAB \Theta(\Gamma_{ajb}) \D \yaj \D \yjb \frac{\D\phi}{2\uppi} \, .
\end{equation}

\paragraph{Common Form}
In general the antenna phase space can be written more compactly as
\begin{equation}
    \D \Phi_\text{ant} = \frac{1}{16 \uppi^2} F_\Phi \Theta(\Gamma_{ijk}) \D \yij \D \yjk \frac{\D \phi}{2\uppi} \, ,
\end{equation}
where we have introduced the (possibly dynamical) phase space factor $F_\Phi$ defined by
\begin{equation}
    F_\Phi = \begin{cases}
        \Kallen \sIK & \text{FF} \\
        \KallenRF \frac{\sAK + m_j ^2 +m_k^2 -m_K^2}{(1-\yjk)^3} & \text{RF} \\
        \frac{\sAK}{1-\yjk} & \text{IF} \\
        \frac{\sAB}{1-\yaj-\yjb} & \text{II}
    \end{cases} \, . \label{eq:phaseSpaceFac}
\end{equation}

\subsection{Branching Kinematics}\label{sec:Kinematics}
In the antenna-shower formalism both of the parents $I$ and $K$ in an FF branching $IK \mapsto ijk$ act collectively as emitters, and the transverse recoil of parton $j$ is shared between them. 
In IF or II branchings, this picture is changed slightly by the requirement that initial-state partons must be aligned with the beam axis. 
In these cases, the kinematic mappings are closer related to those used in the CS-type dipole picture \cite{Schumann:2007mg}.
In RF branchings, the on-shell momentum of the initial decaying resonance is fixed; in the resonance rest frame, the transverse recoil is shared between the two daughters directly involved
in the branching, with all remaining decay products taking the longitudinal recoil required to preserve the decaying resonance's mass~\cite{Brooks:2019xso}.
In all cases, all partons remain on-shell and all four-momenta are conserved at every step in the antenna shower evolution. 

Although we derive all of our antenna functions for general masses of the involved partons, cf.\ \cref{sec:AppendixAntennaFcts},
we construct the kinematics for massless initial-state partons in order to not contravene the assumptions of collinear
PDF evolution. We do however let any emitted final-state quark, denoted by $j$, acquire a finite mass if required and refer the reader to  \cref{sec:VinciaBasics} for our conventions
on massive quarks in initial-state gluon splittings. Should it become feasible to incorporate initial-state masses in the future, the kinematics maps below could be altered for instance 
by means of the kinematic maps in~\cite{Hoche:2015sya}.

\paragraph{FF Branchings:} 
the post-branching momenta are constructed in the centre-of-mass frame of the parent antenna in two steps. In the first step, the momenta are defined in a way that is agnostic to the overall orientation of the parent partons. Choosing (arbitrarily) to align parton $i$ with the $z$ axis of this temporary coordinate system~\cite{GehrmannDeRidder:2011dm}, we have:
\begin{align}
    p_i^\mu &= \left(E_i,0,0,\vert \vec p_i \vert\right) \, , \\
    p_j^\mu &= \left(E_j,-\vert \vec p_j \vert \sin \theta_{ij},0,\vert \vec p_j \vert \cos \theta_{ij}\right) \, , \\
    p_k^\mu &= \left(E_k,\vert \vec p_k \vert\sin \theta_{ik},0,\vert \vec p_k \vert\cos \theta_{ik}\right) \, ,
\end{align}
with energies, $E$, (and energy fractions, $x_E$)
\begin{align}
    E_i = \frac{\sij+\sik+2m_i^2}{2m_{IK}} \, ,\qquad
    E_j = \frac{\sij+\sjk+2m_j^2}{2m_{IK}} \, ,\qquad
    E_k = \frac{\sik+\sjk+2m_k^2}{2m_{IK}} \, ,
\end{align}
\begin{align}
 \implies ~~ x_{Ei} \equiv \frac{2E_{i}}{m_{IK}} = 1 - y_{jk} - \frac{\mu_{k}^2}{1+\mu_{k}^2}(2-y_{jk})~,\qquad \text{same for~$x_{Ek}$ with $i\leftrightarrow k$}~,
\end{align}
and angles
\begin{align}
    \cos \theta_{ij} = 
    \frac{2E_iE_j - \sij}{2\vert \vec p_i \vert \vert \vec p_j \vert} \, ,
    \qquad
    \cos \theta_{ik}  =  \frac{2E_iE_k - \sik}{2\vert \vec p_i \vert \vert \vec p_k \vert} \, ,
\end{align}
and the on-shell conditions for $\vert \vec p_{i,j,k} \vert = \sqrt{E^2_{i,j,k}-m_{i,j,k}^2}$.

In the second step, the branching plane is rotated by an angle $\phi$, uniformly sampled in $[0,2\uppi]$, in the $x$-$y$-plane, and by the angle $\psi$ between the mother parton $I$ and the daughter parton $i$. It is this latter rotation, by $\psi$ which establishes the relative orientation of the post-branching partons with respect to the axis defined by the pre-branching ones. 
The choice of $\psi$ is not unique outside of the collinear limits\footnote{In the $K$-collinear limit, $\psi\to 0$ ensures that parton $i$ recoils purely longitudinally along the direction of parton $I$, and similarly $\psi \to \pi - \theta_{ik}$ in the $I$-collinear limit.} and a few different options for the choice of $\psi$ are implemented in \Vincia, see~\cite{Giele:2007di,GehrmannDeRidder:2011dm}. The ambiguity in the choice of $\psi$ can be expressed in terms of a single free parameter $r$, as
\begin{equation}
    \cos \psi = \frac{2E_I E_i -  2c_i m_i^2 - r \sij - (1-c_k) \sik}{2\vert \vec p_I \vert \vert \vec p_i \vert} \, ,
\end{equation}
where the (kinematics-dependent) constants $c_{i,k}$ are determined by requiring on-shellness of all particles in the $\{I,K\}\mapsto \{i,j,k\}$ mapping, cf.\ \cite{Kosower:1997zr,GehrmannDeRidder:2011dm},
\begin{align}
p_I & = c_i p_i + r p_j + (1-c_k)p_k \, , \\ 
p_K & = c_k p_k + (1-r)p_j + (1-c_i)p_i \, .
\end{align}
It is worth noting that in the limit $r\to 1$, these expressions imply that any transverse momentum carried by parton $j$ is fully absorbed 
by parton $i$, hence we require that any 
sensible choice of $r$ should tend to 
unity in the $I$-collinear limit. Similarly, $r$ should vanish in the $K$-collinear limit. 
The phase space boundaries are given by $0\leq \yij \leq 1$ and $0\leq \yjk \leq 1-\yij$ as well as the roots of the Gram determinant. 

For massless partons, our default choice for $r$ is:
\begin{equation}
r~ =~ \frac{\sjk}{\sij + \sjk }~\to~\begin{cases}
1 & I\text{-collinear limit, $s_{ij} \ll s_{jk}$} \\
0 & K\text{-collinear limit, $s_{jk} \ll s_{ij}$}
\end{cases} \, ,
\end{equation}
which therefore obeys the requirement given above. 
For massive partons, the form is more involved, see \cite{GehrmannDeRidder:2011dm}.

\paragraph{RF Branchings:} The post-branching kinematics are constructed according to the prescription in \cite{Brooks:2019xso}.
With the invariant mass of the resonance being kept fixed, the post-branching momenta are constructed in the resonance rest frame, such that
\begin{align}
p_A^\mu &= p_a^\mu = \left( m_A , 0, 0, 0 \right)\,, \\
p_k^\mu &= \left( E_k , 0, 0, \sqrt{E^2_j - m^2_k} \right)\,, \\
p_j^\mu &= \left(E_j , \sqrt{E^2_j - m^2_j} \sin \theta_{jk}, 0, \sqrt{E^2_j - m^2_j}\cos{\theta_{jk}} \right)\, , \\
p_{X'}^\mu &= \left(m_A - E_k - E_j , -\sqrt{E^2_j - m^2_j} \sin \theta_{jk}, 0, -\sqrt{E^2_j - m^2_k} -\sqrt{E^2_j - m^2_j}\cos{\theta_{jk}} \right)\, ,
\end{align}
where
\begin{equation}
E_j = \frac{\saj}{2m_a}, \quad E_k = \frac{\sak}{2m_a}, \quad \cos \theta_{jk} = \frac{2 E_b E_g -\sjk}{2 \sqrt{(E_k^2 -m_k^2)(E_j^2-m_j^2)}}\,, 
\end{equation}
and where $\{X' \}$ denote the remainder of the resonance decay system.  The frame orientation is chosen such that the $z$-axis is defined along $p_K$.
These momenta are rotated about the $y$-axis such that the set of recoilers are along -$z$, so that only $j$ and $k$ receive transverse recoil. 
The momenta are subsequently rotated by a flatly sampled azimuthal angle $\phi$ about the $z$-axis. Finally the original orientation of $K$ with respect to $z$ is recovered, 
before boosting back to the lab frame.

The conservation of the invariant mass of the system of recoilers, $p_X = \sum_{i \in \{X\}} p_i$ is automatically ensured by imposing \cref{eq:epConsIF}
and the positivity of the Gram determinant $\Gamma_{ajk}$ defined in \cref{eq:GramDet}. (Note, however, that this statement no longer applies if only a subset of $X$
are chosen to receive the longitudinal recoil.)

\paragraph{IF Branchings:} 
For massless initial-state partons, $A$ and $a$, and general final-state ones, $K$, $j$, and $k$, and requiring that the beam axis defined by the pre-branching incoming parton $A$ remains the same post-branching, the post-branching momenta in the $A$-$K$ rest frame are given by~\cite{Verheyen2019}
\begin{align}
    p_a^\mu &= \frac{1}{\yAK}p_A^\mu \, , \\
    p_j^\mu &= \frac{(\yak+\mu_j^2-\mu_k^2)+(\yak-\yaj)\mu^2_K-\yAK\yak}{\yAK}\,p_A^\mu + \yaj\, p_K^\mu + \sqrt{\Gamma_{ajk}}\,q_{\perp\mathrm{max}}^\mu \, , \\
    p_k^\mu &= \frac{(\yaj-\mu_j^2+\mu_k^2)+(\yaj-\yak)\mu_K^2-\yAK\yaj}{\yAK}\,p_A^\mu + \yak\, p_K^\mu - \sqrt{\Gamma_{ajk}}\,q_{\perp\mathrm{max}}^\mu \, ,
\end{align}
where $q_{\perp\mathrm{max}}^\mu$ is a space-like four-vector in the transverse direction, perpendicular to both $p_A$ and $p_K$, with $q_{\perp\mathrm{max}}^2 = -(\saj+\sak)$.
The angle about the branching plane is uniformly distributed in $[0,2\uppi]$.

In the massless case, this reduces to the form given in \cite{Fischer:2016Vincia}. The phase space boundaries are given by $0\leq \yaj \leq 1$ and $0\leq \yjk \leq \frac{1-x_A}{x_A}\yAK-\mu_j^2-\mu_k^2+\mu_K^2$ as well as the roots of the Gram determinant.

In the $a$-collinear limit, we see trivially that
\begin{equation}
	z_A \equiv \frac{x_A}{x_a} = \yAK = \frac{\sAK}{\saj + \sak},
\end{equation}
while in the $k$-(quasi)-collinear limit, we find
\begin{equation}
	z_K = 1-\yaj = \yak = \frac{\sak}{\saj + \sak}.
\end{equation}

We note that in addition to the ``local'' map discussed here, there is also a ``global'' map implemented in \Vincia, where instead of the final-state parton, it is the 
the initial-state parton which recoils transversely. Moreover, \Vincia\ offers the possibility to interpolate between these two maps by choosing one of the two maps according to some probability
imposed by comparison to matrix elements~\cite{Verheyen2019}. While this procedure ties in better with the recoiler-emitter-agnostic antenna formalism,
we leave a dedicated study regarding the effect of choosing either of the latter two alternatives in the sector shower to future studies and here only use the local map for IF branchings.

\paragraph{II Branchings:} In an initial-initial branching, both partons $a$ and $b$ are assumed to stay aligned with the beam momenta $p_A$ and $p_B$, respectively and only the emitted final-state parton $j$ acquires mass, the post-branching momenta can be constructed to be~\cite{Verheyen2019}:
\begin{align}
    p_a^\mu &= \sqrt{\frac{1}{\yAB}\frac{1-\yaj}{1-\yjb}} p_A^\mu \, , \\
    p_j^\mu &= \sqrt{\frac{\yjb^2}{\yAB}\frac{1-\yaj}{1-\yjb}}p_A^\mu + \sqrt{\frac{\yaj^2}{\yAB}\frac{1-\yjb}{1-\yaj}}p_B^\mu + \sqrt{\yaj\yjb-\mu_j^2}q_{\perp\max}^\mu \, , \\
    p_b^\mu &= \sqrt{\frac{1}{\yAB}\frac{1-\yjb}{1-\yaj}} p_B^\mu \, , \\
    p_r^\mu &= p_a^\mu +p_b^\mu - p_j^\mu,
\end{align}
again with $q_{\perp\mathrm{max}}^\mu$ a space-like four-vector perpendicular to $p_A$ and $p_B$ with $q_{\perp\mathrm{max}}^2 = -\sab$ and
the angle about the branching plane is uniformly distributed in $[0,2\uppi]$.
The transverse momentum obtained by the emitted parton $j$ is compensated for by the recoiler $r$, which denotes the rest of the system.
Again, in the massless case, this is identical to the recoil scheme used in \cite{Fischer:2016Vincia}. The phase space boundaries are $0\leq \yaj \leq \frac{1-x_Ax_B}{x_Ax_B}\yAB +\mu_j^2$ and $0\leq \yjk \leq \frac{1-x_Ax_B}{x_Ax_B}\yAB +\mu_j^2$ as well as the roots of the Gram determinant.

The condition that partons $a$ and $b$ stay aligned with the beam axis leads to the identity
\begin{equation}
	\frac{x_A x_B}{x_a x_a} = \frac{\sAB}{\sab} = \yAB \, . \label{eq:momFracsYAB}
\end{equation}
We round off by noting that \cref{eq:momFracsYAB} implies that the $z$ fractions on sides $A$ and $B$ are related by
\begin{equation}
z_A = z_B \frac{1-\yjb}{1-\yaj} \, .
\end{equation}

\subsection{No-Branching Probability}
Like other parton showers, the \Vincia\ antenna shower is built around the no-branching probability, i.e., the probability that no branching occurs between two scales $Q^2_{n} > Q^2_{n+1}$, given by~\cite{Ritzmann:2012ca,Fischer:2016Vincia}
\begin{equation}
    \Pi_n (Q^2_{n}, Q^2_{n+1}) = \exp\left(-\sum\limits_{i \in \{n \mapsto n+1\}} \mathcal{A}_i\left(Q^2_n, Q^2_{n+1}\right)\right) \, ,
    \label{eq:noBranching}
\end{equation}
where
\begin{equation}
	\mathcal{A}_i (Q^2_n, Q^2_{n+1}) = \int \limits_{Q^2_{n+1}}^{Q^2_{n}} 4\uppi \alphastrong(Q^2) \mathcal C \bar a_i(Q^2, \zeta) R_f \D \Phi_{\text{ant}}\, , \label{eq:sudakovKernel}
\end{equation}
with a PDF ratio for each initial-state leg,
\begin{equation}
    R_f = \begin{cases}\displaystyle
        1 & \text{FF \& RF} \\[2mm]
\displaystyle
\frac{f_a(x_a, Q^2)}{f_A(x_A,Q^2)} & \text{IF} \\[3mm]
\displaystyle
        \frac{f_a(x_a, Q^2)}{f_A(x_A,Q^2)} \frac{f_b(x_b, Q^2)}{f_B(x_B, Q^2)} & \text{II} 
            \end{cases} \, .
\end{equation}
When starting the evolution at a scale $Q_n^2$, the probability for a parton to branch at scale $Q^2_{n+1}$ is therefore given by
\begin{equation}
    P_\text{branch} (Q^2_{n},Q^2_{n+1}) = \frac{\D \Pi_n(Q_n^2,Q_{n+1}^2)}{\D \log(Q_{n+1}^2)} \, . \label{eq:BranchingProbability}
\end{equation}
Branchings are then generated by solving \cref{eq:BranchingProbability} by means of the veto algorithm \cite{Sjostrand:2006za}, by overestimating the antenna function by a larger trial function $\bar a_{\text{trial}}$, for which a trial integral with an ansatz like $\hat{R}_f = \mathrm{const}$ or $\hat{R}_f\propto (x_A x_B)/(x_a x_b)$
can be analytically solved to give simple expressions. 
By accepting each trial branching with a probability
\begin{equation}
    P_{\text{accept}} = \frac{\alphastrong}{\hat\alpha_{\text{s}}} \frac{\mathcal C}{\hat{\mathcal C}} \frac{R_f}{\hat R_{f}} \frac{\bar a}{\bar a_{\text{trial}}} \, , \label{eq:AcceptProb}
\end{equation}
the full integral is recovered post-facto. The additional ratios
\begin{equation}
   \frac{\alphastrong}{\hat\alpha_{\text{s}}}~,\quad \frac{\mathcal C}{\hat{\mathcal C}}~, \quad \text{and}\quad \frac{R_f}{\hat R_{f}} \, ,
\end{equation}
where hats denote trial quantities, take into account that nominally larger values than the physical ones may be used during the trial generation, cf.\ \cref{sec:TrialGenerators}. 

\subsection{Evolution Variables}
\label{sec:EvolutionVariables}
Both gluon emissions and gluon splittings are evolved in a scaled
notion of off-shellness based on the \Ariadne\ transverse
momentum, which we denote by $p_\perp^2$. 
An alternative (hypothetical) treatment permitting different definitions for the evolution measures of emissions and splittings 
(see, e.g.,~\cite{Fischer:2016Vincia}) would in
principle allow to use sector resolution variables, cf.\
\cref{sec:SectorResolution}, as evolution variables for the sector
shower. However in this case one would also have to address how this would impact
the interleaving of the shower branchings; we therefore leave a dedicated study
regarding this possibility to forthcoming work.  

\paragraph{FF Branchings} We evolve in
\begin{equation}
    p^2_{\perp,\text{FF}} = \frac{(m_{ij}^2-m_I^2)(m_{jk}^2-m_K^2)}{\sIK} = \begin{cases}\displaystyle
    \frac{\sij\sjk}{\sIK} & g\text{-emission} \\[4mm]
    \displaystyle\frac{m_{ij}^2(\sjk +m_j^2)}{\sIK} & g_I\text{-splitting}
    \end{cases} \, , \label{eq:FFevolutionVars}
\end{equation}
For gluon emissions, this evolution variable is identical to the \Ariadne\ $p_\perp^2$. For gluon splittings, the previous version of \Vincia~\cite{Fischer:2016Vincia} used the virtuality of the splitting gluon, $m_{ij}^2$, as the evolution variable, while we have here instead made the choice to evolve all branchings in a common measure of $p_\perp^2$ which, unlike $m^2_{ij}$, vanishes in the limit that the remaining $jk$ colour dipole becomes unresolved. 

\paragraph{RF Branchings}
Crossing partons $I$ and $i$ into the initial state to become partons $A$ and $a$, we define the evolution variable for RF branchings as:
\begin{equation}
    p^2_{\perp,\text{RF}} = \frac{(m_A^2-q_{aj}^2)(m_{jk}^2-m_K^2)}{\saj + \sak} = \begin{cases} 
        \displaystyle
        \frac{\saj\sjk}{\saj +\sak} & g\text{-emission} \\[4mm]
    \displaystyle
        \frac{m_{jk}^2(\saj-m_j^2)}{\saj+\sak} & g_K\text{-splitting}
    \end{cases} \, . \label{eq:RFevolutionVars}
\end{equation}

\paragraph{IF Branchings}
In principle, the evolution variable for IF branchings is chosen to be
the same as for RF ones, but with the additional cases of
initial-state splittings and gluon conversions that are not present for RF
branchings:  
\begin{equation}
    p^2_{\perp,\text{IF}} = \frac{(m_A^2 - q^2_{aj})(m_{jk}^2-m_K^2)}{\saj + \sak} = \begin{cases} 
        \displaystyle
        \frac{\saj\sjk}{\saj +\sak} & g\text{-emission} \\[4mm]
        \displaystyle
        \frac{m_{jk}^2(\saj-m_j^2)}{\saj+\sak} & g_K\text{-splitting}\\[4mm]
        \displaystyle
        \frac{(\saj-m_j^2)(\sjk+m_j^2)}{\saj+\sak} & q_A\text{-conversion} \\[4mm]
        \displaystyle
        \frac{\saj(\sjk+m_j^2)}{\saj+\sak} & g_A\text{-splitting} 
    \end{cases}\, , \label{eq:IFevolutionVars}
\end{equation}
where initial-state partons are treated as massless, $m_A = m_a = 0$, except for the case of a initial-state gluon splittings, where $m_a = m_j$, as alluded to above.

\paragraph{II Branchings} We take into account that both parents, $A$ and $B$ are in the initial state, so that our choice of the evolution variable reads:
\begin{equation}
     p^2_{\perp,\text{II}} = \frac{(m_A^2-q_{aj}^2)(m_B^2- q_{jb}^2)}{\sab} = \begin{cases}
     \displaystyle
        \frac{\saj \sjb}{\sab} & g\text{-emission} \\[4mm]
        \displaystyle
        \frac{(\saj-m_j^2)(\sjb-m_j^2)}{\sab} & q_A\text{-conversion}\\[4mm]
        \displaystyle
        \frac{\saj(\sjb-m_j^2)}{\sab} & g_A\text{-splitting}
     \end{cases}\, , \label{eq:IIevolutionVars}
\end{equation}
with the same convention for massive partons in the initial state as in the IF case; i.e.\ both parton $A$ and $B$ are treated as massless, $m_A = m_B = 0$, except for splittings of initial-state gluons $g_a$ or $g_b$, where $m_a = m_j$ or $m_b = m_j$, respectively.

\section{The Sector Shower Algorithm}
\label{sec:SectorShowerAlgorithm}
In this section, we discuss the differences between the  global antenna-shower implementation in  \textsc{Pythia}~8.3, and the sector-shower model we have developed, including all aspects that are relevant for a full implementation of the sector-shower algorithm. 

In \cref{sec:SectorAntennae} we derive the physical sector antenna functions from the  corresponding global ones. We then describe the overestimates used for trial generation and the components of the trial-generation algorithm in \cref{sec:TrialGenerators}.
We close by reviewing our choice for the resolution variable used to discriminate between phase-space sectors in \cref{sec:SectorResolution}.

\subsection{Sector Antenna Functions}
\label{sec:SectorAntennae}
In conventional (global) showers, each radiation kernel is allowed to radiate into any phase-point that is kinematically accessible to it, subject only to the following requirements:
(i) strong ordering (the evolution variable should be below the scale of the previous branching), (ii) perturbativity (it should be resolved with respect to the IR cutoff of the shower), 
and, potentially, (iii) that it should pass any angular-ordering or equivalent vetoes that are imposed in the given algorithm. 
Consequently, every phase-space point can receive contributions from multiple different parton-shower histories, which must be summed over to compute the 
\textquotedblleft shower weight\textquotedblright\ for the phase-space point in question.  

In the sector approach however, the phase space is divided into distinct sectors in each of which only a single antenna is allowed to contribute,
cf.\  \cref{fig:2to3SectorsFF,fig:2to3SectorsIF}.
In collinear limits corresponding to $g\mapsto gg$ or $g\mapsto q\bar{q}$ branchings, a single sector antenna function must therefore contain the same collinear singularity structure as the sum of two neighbouring global ones. 

For $g\mapsto q\bar{q}$ branchings, there is no overlap with any soft singularity to worry about as far as the antenna functions are concerned\footnote{There is a subtlety concerning how to choose the boundary between $g\mapsto q\bar{q}$ and $q \mapsto qg$ sectors however, which we return to below.}, and the sector antenna function can simply be taken to be twice the global one. 

For $g\mapsto gg$ branchings, global antenna functions only contain explicit poles in $1/(1-z)$, not in $1/z$. However, since the collinear limits of neighbouring antennae are related 
by $z\leftrightarrow 1-z$, the full DGLAP kernel (which is symmetric under $z\leftrightarrow 1-z$),
\begin{equation}
    P_{g\mapsto gg}(z) = 
    \frac{ (1 - z(1-z))^2}{z(1-z)}
    = \frac{z}{1-z} + \frac{1-z}{z} + z(1-z) \, ,
\end{equation}
 is recovered after summing over the two contributions\footnote{One consequence of this is that global antenna functions are only unique up to terms that are antisymmetric under $z\leftrightarrow 1-z$, with e.g.\ \Ariadne\ and GGG making different choices, while the global shower in \Vincia\ allows a user-defined choice; see~\cite{Giele:2011cb,Fischer:2017htu}.}, see e.g.~\cite{Fischer:2016Vincia}.
A sector antenna function on the other hand, must contain the full
pole structure on its own, at least within the phase-space region the
given sector is meant to cover. As we will discuss below, this
corresponds to $z>\frac12$ in the collinear limit since, in the sector
context it is always the softer of the two gluons that is perceived of
as the \textquotedblleft emitted\textquotedblright\ one. 

\begin{figure}[t]
    \centering
    \includegraphics[width=0.9\textwidth]{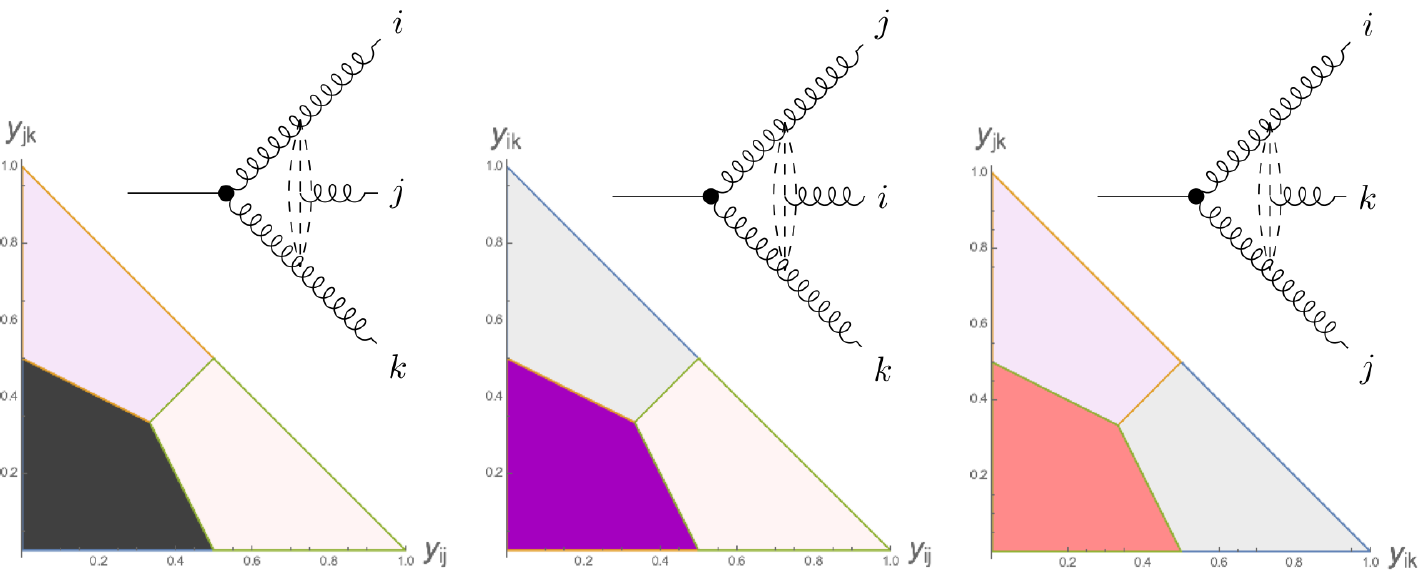}
    \caption{Illustration of the three different sectors of a $gg \mapsto ggg$ antenna, corresponding to the emission of gluon $g_j$ (black), gluon $g_i$ (blue), or gluon $g_k$ (red). The sectors are shown in the respective antenna phase spaces, with the axes labelled such that the associated soft singularity is always located in the bottom left corner. Note that $z_i = 1-y_{jk}$ in any $i$-collinear limit, and similarly for the $j$- and $k$-collinear limits.}
    \label{fig:gggSectors}
\end{figure}

To illustrate this situation, take for instance the decay process $H\to gg$ followed by an FF branching process $gg \mapsto ggg$.
There are three distinct shower histories contributing to the $H\to ggg$ final state (or six, in CS-style dipole approaches),
illustrated by the three diagrams shown above the phase-space triangles in \cref{fig:gggSectors}. 
In a global shower, each of the corresponding antenna functions (or, equivalently, CS dipole functions)
radiates over all of the available phase-space region $\yij+\yjk\leq 1$, and  for each value of $z$ the full collinear singularity involves an explicit sum over the antennae (or dipoles)
which share the given collinear pair. 

From the sector point of view, the phase space for the same $H\to ggg$ final state is regarded as composed of three distinct sectors, in which either $g_i$, $g_j$, or $g_k$, is considered as the emitted gluon, respectively.
These sectors are illustrated by the differently shaded regions in the phase-space triangles shown in \cref{fig:gggSectors},
using the $p^2_{\perp,\text{FF}}$ variable defined in \cref{eq:FFevolutionVars}
as the sector resolution criterion (discussed further in
\cref{sec:SectorResolution}).   
The phase-space sector which corresponds to, and is covered by, the antenna clustering shown above each triangle is shown with full shading, while the other two sectors are shown with partial shading.

Focusing on the left-hand plot, for instance, we see that the sector representing emission of gluon $j$ (shown with full black shading) contains the 
entire singularity associated with gluon $j$ becoming soft ($y_{ij} \to 0$ and $y_{jk} \to 0$) 
as well as the $z>\frac12$ parts of the  $ij$ and $jk$ collinear singularities ($y_{jk} = 1-z_i < \frac12$ and $y_{ij} = 1 - z_k < \frac12$ respectively).
In our approach, we nominally include the entire $(n+1)$-parton singularity structure in all three antenna functions, but only one of them is allowed to contribute in each sector. 

Similar arguments apply to the gluon-collinear singularities in antennae with quark-gluon or gluon-antiquark parents. In such antennae, the global antenna functions already contain the full quark-collinear singularities, but again the gluon-collinear side only contains the $1/(1-z)$ poles, which must be summed together with $1/z$ ones from the neighbouring antenna to reproduce the full DGLAP kernel, phase-space point by phase-space point. In the sector approach, the full DGLAP limit is incorporated into both of the neighbouring antennae, but only the one corresponding to emission of the softer of the two gluons is allowed to contribute in each phase-space point. In the collinear limit, this is again equivalent to imposing a boundary at $z=\frac12$ between the two histories.

We start from the global \Vincia\ FF 
antenna functions, cf.\ \cref{sec:AppendixAntennaFcts}; global RF, IF, and II antenna functions are derived from these by crossing symmetry. Among these crossings are, formally, ones representing crossings of the emitted parton, $j$, into the initial state. Antennae corresponding to such crossings (emission into the initial state) are not part of the current \Vincia\ shower framework; instead the collinear singularities associated with these crossings are added onto the initial-state legs of the II and IF antennae, such that already in the global case the initial-state legs contain the full gluon-collinear singularities~\cite{Ritzmann:2012ca,Fischer:2016Vincia}. 

\begin{figure}[t]
    \centering
    \begin{minipage}{0.8\textwidth}
    \includegraphics[width=0.45\textwidth]{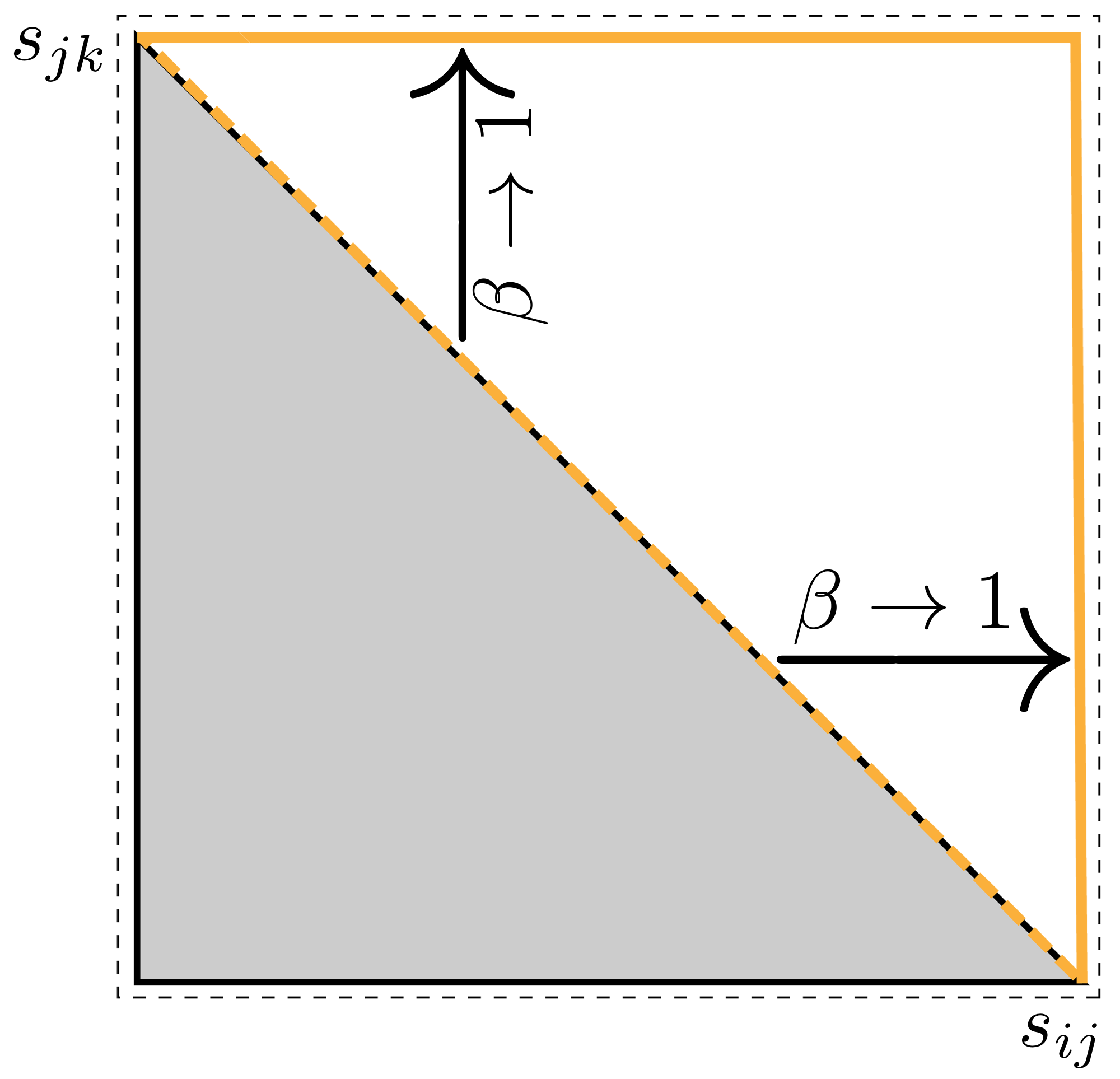}
    \hfill
    \includegraphics[width=0.45\textwidth]{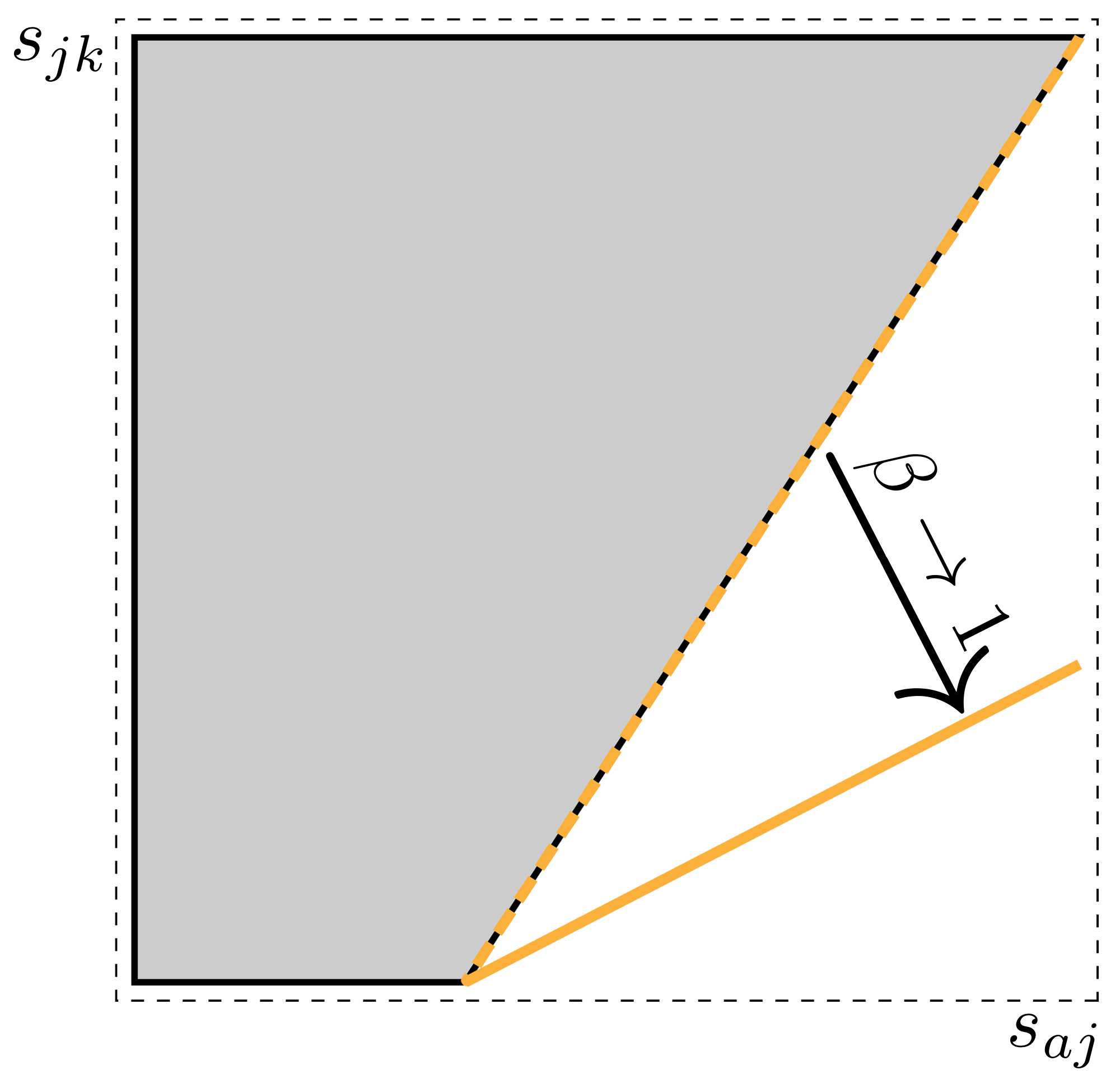}
    \end{minipage}
    \caption{Illustration of the effect of the sector damp parameter $\beta$. For $\beta \to 1$, the unphysical singularity, marked in orange, is pushed away from the hard phase-space boundary $\sij+\sjk=\sIK$ in the final-final phase-space (\textit{left}) and $\sjk=\saj-\sAK$ in the initial-final phase-space (\textit{right}).}
    \label{fig:SectorDamping}
\end{figure}

To derive sector antenna functions based on global ones, we distinguish between gluon emissions and gluon splittings. As mentioned above, for the latter, simply a factor of two has to be included if the splitting happened in the final state, to account for the fact that the gluon is part of two global antennae. For gluon emissions, we include the missing gluon-collinear parts by symmetrising over each final-state gluon-gluon pair. For instance, the eikonal component of the global antenna function for $g_Ig_K \mapsto g_ig_jg_k$, becomes
\begin{equation}
   \text{Global:~}\frac{2\sik}{\sij\sjk} \to \text{~Sector:~}
    \frac{2\sik}{\sij\sjk} 
    + \frac{2\sij}{\sjk(\sik + \beta\sjk)}  
    + \frac{2\sjk}{\sij(\sik + \beta\sij)} \, ,
\label{eq:antennaSectorisationExample}
\end{equation}
where the soft eikonal remains unchanged and the two additional terms correspond to the \textquotedblleft missing\textquotedblright\ collinear parts in the $ij$- and $jk$-collinear limits.
The auxiliary \textquotedblleft sector damp\textquotedblright\ parameter $\beta \in [0,1]$, we have introduced here, allows to push the singularities for $s_{ik}\to 0$, that would result from an exact symmetrisation, away from the phase-space boundaries, cf.~\cref{fig:SectorDamping}. Although this is not necessary for the construction of sector antenna functions, it ensures numerical stability in phase space configurations where the sector boundaries become close to the phase space boundary $\sik = 0$. Moreover, as will be alluded to in \cref{sec:TrialGenerators}, it permits to generate trials over all of phase space with an a-posteriori imposed sector veto. Hence, our default choice is $\beta = 1$, but it can be changeed with \texttt{Vincia:sectorDamp}.
We note that choosing $\beta \neq 0$ is not in disagreement with leading-logarithmic resummation as long as the sector resolution criterion reflects the singularity structure of the QCD matrix element, a point we return to in detail in \cref{sec:SectorResolution}. For the example \cref{eq:antennaSectorisationExample} considered here, this implies that phase space points for which $\sik \to 0$ always have to be assigned to the antenna where $k$ is the emitted parton. This situation is shown in \cref{fig:gggSectors} for a simple three-parton configuration, where the radiation phase space is covered by three antennae, each of which is constrained to a unique region. 

A further subtlety concerns the treatment of terms in the global antenna functions that are antisymmetric under $z \leftrightarrow 1-z$. As mentioned previously, the contributions of such terms to the collinear limits cancel in the sum over two neighbouring global antennae. 
Outside of those limits, however, nonsingular differences can remain. In \Vincia, this ambiguity is parameterised by a parameter called the octet-partitioning parameter\footnote{It can be set with \texttt{Vincia:octetPartitioning}.} $\alpha$ with default choice $\alpha = 0$, cf.\  \cref{sec:AppendixAntennaFcts}. To ensure positivity of the sector antenna functions over all of their respective  phase spaces, our sector antennae are based on global ones with $\alpha = 1$.

Below, we define our set of sector antenna functions derived from the
above principles. For FF branchings, there are only slight differences
with respect to the earlier sector-shower implementation
of~\cite{LopezVillarejo:2011ap} while the IF, II, and RF aspects are
new.  The construction principles are applied to both
helicity-averaged and, if the helicities of two collinear gluons
match, helicity-dependent antenna functions -- and helicity-dependence
is implicitly understood. 
We discuss validating the collinear limits of sector antenna functions in \cref{sec:CollinearLimits}. 

\paragraph{FF Branchings} For final-final gluon emission, we symmetrise over each gluon pair,
\begin{equation}
    \begin{split}
        \bar a_{g_j-\text{emit}}^{\text{FF,sct}}(\yij,\yjk) = \,&\bar a_{g_j-\text{emit}}^{\text{FF,gl}}(\yij,\yjk) \\ &+ \delta_{Ig}\delta_{h_Ih_j} \bar a_{g_i-\text{emit}}^{\text{FF,gl}}(\yij,1-\yjk-(1-\beta)\yij) \\ &+ \delta_{Kg}\delta_{h_Kh_j} \bar a_{g_k-\text{emit}}^{\text{FF,gl}}(1-\yij-(1-\beta)\yjk,\yjk),
    \end{split} \label{eq:DefSectorAntennaFF}
\end{equation}
with $\delta_{Ig}$ and $\delta_{Kg}$ being one if $I$ or $K$ is a gluon, respectively, and zero otherwise. Similarly, $\delta_{h_Ih_j}$ and $\delta_{h_Kh_j}$ are one if the helicity $h_I$ or $h_K$ of the parent gluon matches the one of the emitted gluon $h_i$.
For $\beta \to 1$, the symmetrised invariants reduce to $1-\yjk$ and $1-\yij$ for $i\leftrightarrow j$ and $j \leftrightarrow k$, respectively, ensuring that the additional, unphysical $i$-$k$-singularity is tamed.

For gluon splittings, we multiply by a factor of two,
\begin{equation}
    \bar a_{g-\text{split}}^{\text{FF,sct}}(\yij,\yjk) = 2 \bar a_{g-\text{split}}^{\text{FF,gl}}(\yij,\yjk)
\end{equation}
where the gluon may either be $g_I$ or $g_K$. 

\paragraph{RF Branchings} 
Although in principle the antenna shower formalism can handle resonances of any spin and colour representation, so far only the antenna functions for 
spin-$\frac{1}{2}$, resonances in the fundamental representation (e.g. top quarks) have been included\footnote{Although it should be straightforward to include these, we postpone such a study 
of such hypothetical coloured scalar (e.g. gluinos) or vector (e.g. squark) resonances to future work.}.
Thus in this case, the sector antenna functions for RF
branchings are essentially unchanged with respect to the IF ones, discussed below, with the only restriction that the resonance cannot ``backwards-evolve'' (i.e., $a = A$). 

\paragraph{IF Branchings} In initial-final antennae, only final-state
legs have to be sectorised, as there is no radiation into the initial
state. For gluon emissions, we therefore have 
\begin{equation}
    \begin{split}
        \bar a_{g_j-\text{emit}}^{\text{IF,sct}}(\yaj,\yjk) = \,&\bar a_{g_j-\text{emit}}^{\text{IF,gl}}(\yaj,\yjk) \\
            &+ \delta_{Kg} \delta_{h_K h_j} \bar a_{g_k-\text{emit}}^{\text{IF,gl}}(1-\yaj+\beta\yjk,\yjk),
    \end{split}
\end{equation}
where again the additional, unphysical $a$-$k$-singularity resulting from the $j\leftrightarrow k$ symmetrisation is tamed for $\beta \to 1$.

For gluon splittings we include a factor of two only if the gluon was
in the final state,
\begin{equation}
    \bar a_{g_K-\text{split}}^{\text{IF,sct}}(\yaj,\yjk) = 2 \bar a_{g_K-\text{split}}^{\text{IF,gl}}(\yaj,\yjk),
\end{equation}
and use global antenna functions for initial-state gluon splittings and quark conversions,
\begin{align}
    \bar a_{g_A-\text{split}}^{\text{IF,sct}}(\yaj,\yjk) = \bar a_{g_A-\text{split}}^{\text{IF,gl}}(\yaj,\yjk), \\
    \bar a_{g_A-\text{conv}}^{\text{IF,sct}}(\yaj,\yjk) = \bar a_{g_A-\text{conv}}^{\text{IF,gl}}(\yaj,\yjk).
\end{align}

\paragraph{II Branchings} As already discussed, by construction initial-state legs are already sectorised, and so we can readily use \Vincia's global antenna functions in the sector shower without modifications,
\begin{equation}
    \bar a^{\text{II,sct}}(\yaj,\yjb) = \bar a^{\text{II,gl}}(\yaj,\yjb).
\end{equation}

\subsection{Trial Generation}\label{sec:TrialGenerators}
Compared to the global case, sector antenna functions contain additional terms to incorporate the full gluon-collinear singularity in each sector. In the context of trial functions for gluon emissions, we note that the $ij$ collinear limit of the eikonal overestimate used in \Vincia's global showers $\propto 1/p_\perp^2$ is $1/(Q_{ij}^2(1-z_i))$.
This is sufficient to overestimate global antenna functions, since the $1/z_i$ part of the $I$-collinear singularity 
are contained in the neighbouring 
antenna, with the $ij$ collinear limits of the two related by $z_i\leftrightarrow 1-z_i$. 
For a sector shower, however, the eikonal $1/p_\perp^2$ does not overestimate the $1/z$ parts of the physical sector antenna functions. 
The sector shower therefore requires further trial generators for the collinear parts of the antennae, as we describe below.

We start by rewriting the antenna integral \cref{eq:sudakovKernel} in the no-branching probability, \cref{eq:noBranching}, in terms of dimensionless \textquotedblleft shower variables\textquotedblright, $x_\perp$ and $\zeta$,
\begin{equation}
    \mathcal A(x_{\perp1}, x_{\perp2}) = \frac{1}{16 \uppi^2} \int\limits_{x_{\perp2}}^{x_{\perp1}} 4\uppi \alphastrong\, \mathcal C\, \bar a(x_\perp, \zeta)\, R_{f}\, F_\Phi\, \vert J(x_\perp, \zeta)\vert \,\D x_\perp \D \zeta. \label{eq:TrialIntegralDef}
\end{equation}
with the (dimensionless) variable $x_\perp$ defined by
\begin{equation}
x_\perp = \frac{p_\perp^2}{s}~,
\end{equation}
where $p_\perp^2$ is the evolution variable as defined in \cref{eq:FFevolutionVars,eq:RFevolutionVars,eq:IFevolutionVars,eq:IIevolutionVars}, 
and a complementary phase-space variable $\zeta$ which must be chosen such that the mapping 
\begin{equation}
    (\yij,\yjk) \to (x_\perp,\zeta)
\end{equation}
is one-to-one. The factor $F_\Phi$ is the phase-space factor defined in \cref{eq:phaseSpaceFac} and $\vert J(x_\perp,\zeta)\vert$ denotes the Jacobian associated with the above variable transformation. As long as the Jacobian is properly accounted for, the choice of $\zeta$ does not affect physical observables. Below, this freedom in choosing $\zeta$ is exploited to produce relatively simple expressions for the trial integrals and to optimise the phase-space sampling for trial branchings. 

From a given starting scale, $x_{\perp1 }$, the next branching scale in the downwards evolution is found by solving
\begin{equation}
    R = \exp(-\mathcal A(x_{\perp1},x_{\perp2})) \label{eq:SudakovDistribution}
\end{equation}
for $x_{\perp2}$ with a uniformly distributed random number $R\in [0,1]$. 
In general, this is not feasible  analytically, hence we work instead with simple overestimates of the integrand in \cref{eq:TrialIntegralDef}
and use the Sudakov veto algorithm to ensure that the correct integral is recovered post-facto. That is, we replace physical antenna functions $\bar a$ by simpler trial functions $\bar a_\text{trial}$ that overestimate the physical antenna function everywhere in phase space and overestimate the PDF ratio $R_f$ by 
\begin{equation}
    \hat R_f = \left(\frac{x_A}{x_a}\frac{x_B}{x_b} \right)^\gamma \frac{f_a(x_A,p^2_{\perp \text{min}})f_b(x_B,p^2_{\perp \text{min}})}{f_A(x_A,p^2_{\perp\text{min}})f_B(x_B,p^2_{\perp\text{min}})} =: \left(\frac{x_A}{x_a}\frac{x_B}{x_b} \right)^\gamma \tilde R_f,
\end{equation}
with $x_{\perp\text{min}}$ the minimal scale for the current trial and $\gamma \in \{0, 1\}$. The additional momentum-fraction ratio takes into account that for many cases (in particular for gluon and sea-quark distributions), an assumption that the PDFs fall off as some power of $1/x$ for higher $x$ is a reasonable starting approximation. Where possible (except for valence-type flavours), our default choice is to use $\gamma=1$. 

We further assume $\alphastrong$ depends only on $p^2_\perp$ and define the one-loop running coupling
\begin{equation}
    \hat{\alpha}_\text{s}(x_\perp) = \frac{1}{b_0\log\left(\frac{x_\perp}{x_\Lambda} \right)}, \label{eq:TrialAlphaSoneLoop}
\end{equation}
where
\begin{equation}
    b_0 = \frac{33 - 2n_f}{12\uppi} \quad \text{and} \quad \frac{x_\perp}{x_\Lambda} = \frac{k_\text{R} p^2_\perp}{\Lambda_{\text{QCD}}^2} \label{eq:ConstantCouplingRep}
\end{equation}
with an arbitrary scaling factor $k_\text{R}$ to adjust the effective renormalisation scale.
We note that two-loop running of the physical coupling is also implemented, via a simple modification of the veto probability, as done in~\cite{Giele:2011cb}.

We overestimate all singular parts of the antenna functions individually and let different trials compete for the highest branching scale. A branching is then accepted with probability
\begin{equation}
    P^\text{sct}_{\text{accept}} = \frac{\alphastrong}{\hat\alpha_{\text{s}}} \frac{\mathcal C}{\hat{\mathcal C}} \frac{R_f}{\hat R_{f}} \frac{\bar a}{\sum\limits_\text{trials}\bar a_{\text{trial}}},
\end{equation}
Currently, we also utilise the veto algorithm to restrict branchings to the respective phase-space sectors, i.e., we generate trial branchings over all of phase space and accept only the one with minimal sector resolution variable of the post-branching configuration.

After accepting a branching, the complementary phase-space variable is generated by solving
\begin{equation}
    R_\zeta = \frac{I_\zeta(\zeta_\text{min}, \zeta)}{I_\zeta (\zeta_\text{min}, \zeta_\text{max})} \label{eq:ZetaTrial}
\end{equation}
for $\zeta$ with a second uniformly distributed random number $R_\zeta \in [0,1]$. Here, $I_\zeta$ denotes the integral over the $\zeta$-dependence in $\mathcal A$, which is carried out over a larger region than the physically allowed phase space with simpler $\zeta$ boundaries. If this generates a value outside the physical boundaries,
\begin{equation}
    \zeta < \zeta_{\text{min}}(x_{\perp}) \quad \vee \quad \zeta > \zeta_{\text{max}}(x_{\perp}),
\end{equation}
the trial is vetoed and a new branching is generated with starting scale $x_\perp$.

To solve \cref{eq:SudakovDistribution} numerically, we rewrite the trial integral \cref{eq:TrialIntegralDef} in terms of the function
\begin{equation}
   \Atrial(x_{\perp1}, x_{\perp2}, \chi(\zeta)) = \int\limits_{x_{\perp2}}^{x_{\perp1}} \int\limits_{\zeta_\text{min}(x_\perp)}^{\zeta_\text{max}(x_\perp)} \alphastrong(x_\perp) \chi(\zeta) \D \zeta \frac{\D x_\perp}{x_\perp} . \label{eq:EvolutionMasterInt}
\end{equation}
for $\chi(\zeta)$ a function of $\zeta$. 
For a one-loop running coupling, the solution to \cref{eq:SudakovDistribution} is then given by
\begin{equation}
    x_{\perp2} = x_\Lambda \left(\frac{x_{\perp1}}{x_\Lambda} \right)^{R^{\frac{4\uppi b_0}{\mathcal C \tilde R_f I(\zeta_\text{min}, \zeta_\text{max})}}},
\end{equation}
while for a constant trial coupling, it is given by
\begin{equation}
    x_{\perp2} = x_{\perp1} R^{\frac{4\uppi}{ \hat{\alpha}_\text{s}\mathcal C \tilde R_f I(\zeta_\text{min}, \zeta_\text{max})}},
\end{equation}
or alternatively
\begin{equation}
    x_{\perp 2} = \exp\left(-\sqrt{\log^2(x_{\perp 1}) - \frac{4\uppi}{\Kallen \alphastrong \mathcal C} \log(R)} \right).
\end{equation}
if the zeta-integral can be evaluated to
\begin{equation}
    I_\text{log}(x_\perp) = \log\left(\frac{1}{x_\perp}\right).
\end{equation}
Explicit expressions for \cref{eq:EvolutionMasterInt} are collected in \cref{sec:ExplicitTrialGens}.

\subsection{Choice of Sector Resolution Variables} \label{sec:SectorResolution}
\begin{figure}[t]
	\centering
	\begin{minipage}{0.8\textwidth}
		\includegraphics[width=0.4\textwidth]{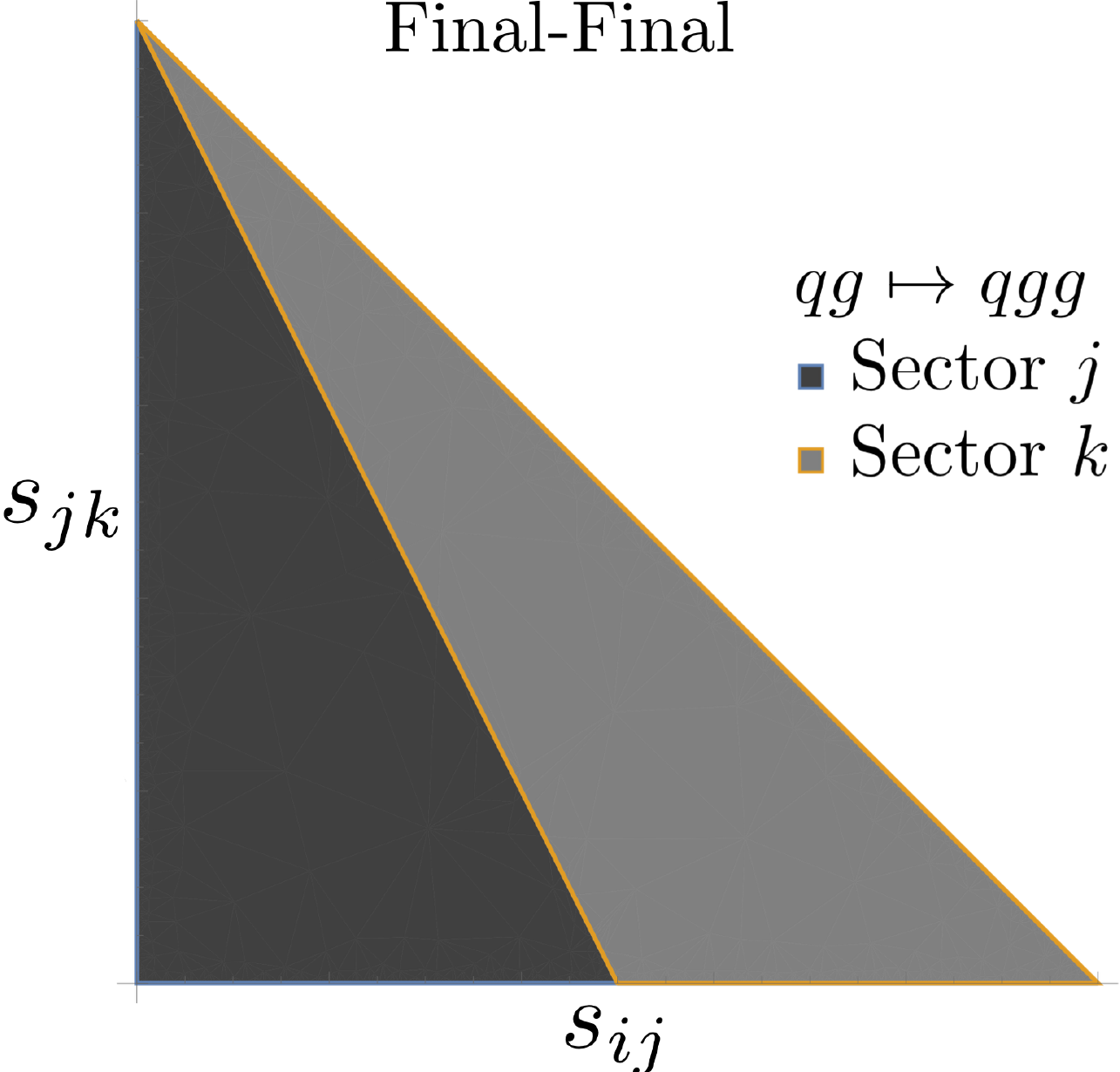}
		\hfill
		\includegraphics[width=0.4\textwidth]{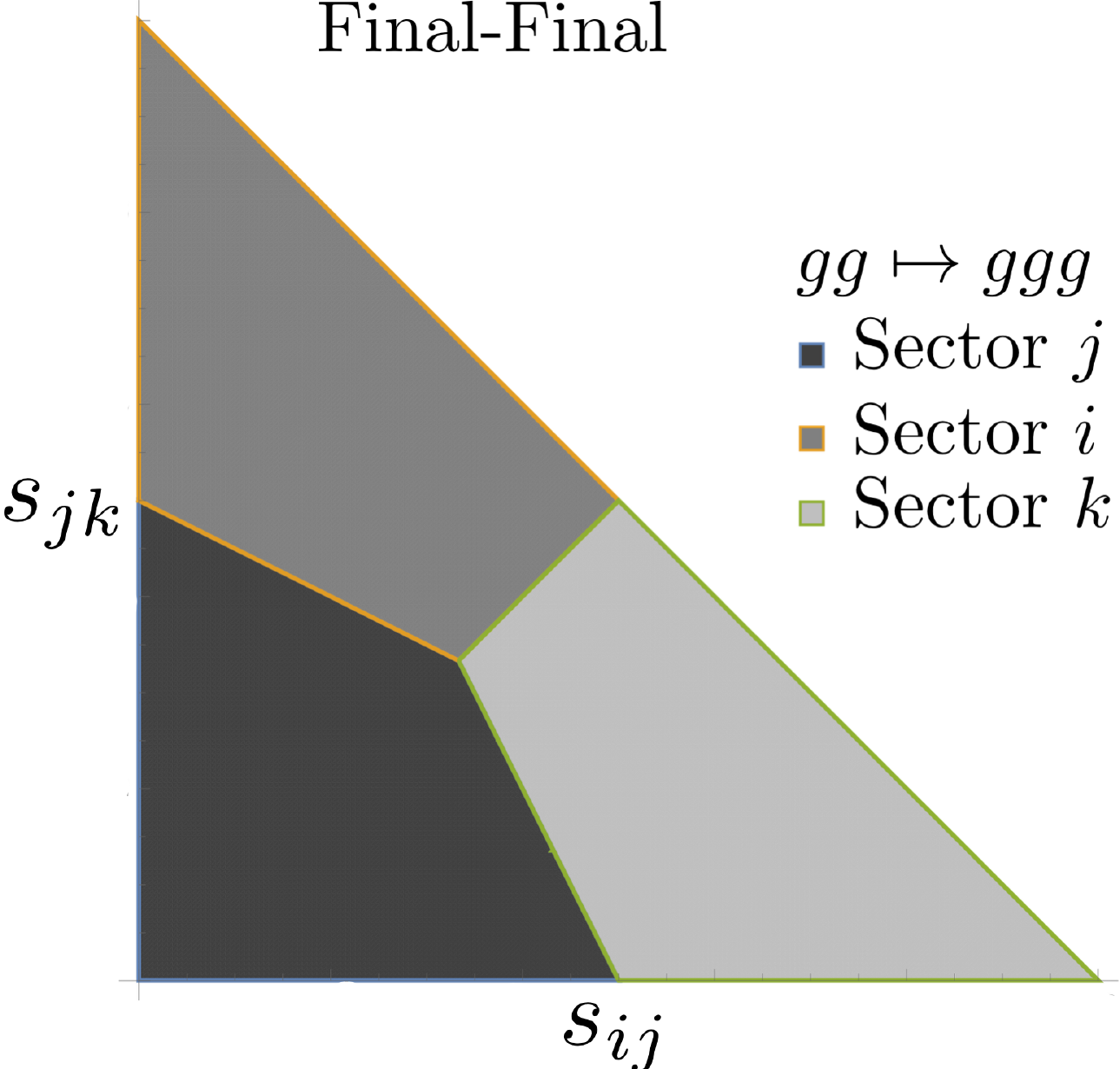}
	\end{minipage}
	\caption{Illustration of the sectors in the antenna phase spaces for a final-final $qg \mapsto qgg$ antenna (\textit{left}) and a final-final $gg \mapsto ggg$ antenna (\textit{right}), with the sector resolution according to \cref{eq:FFresolutionVars}.}
	\label{fig:2to3SectorsFF}
\end{figure}

\begin{figure}[ht]
	\centering
	\begin{minipage}{0.8\textwidth}
		\includegraphics[width=0.4\textwidth]{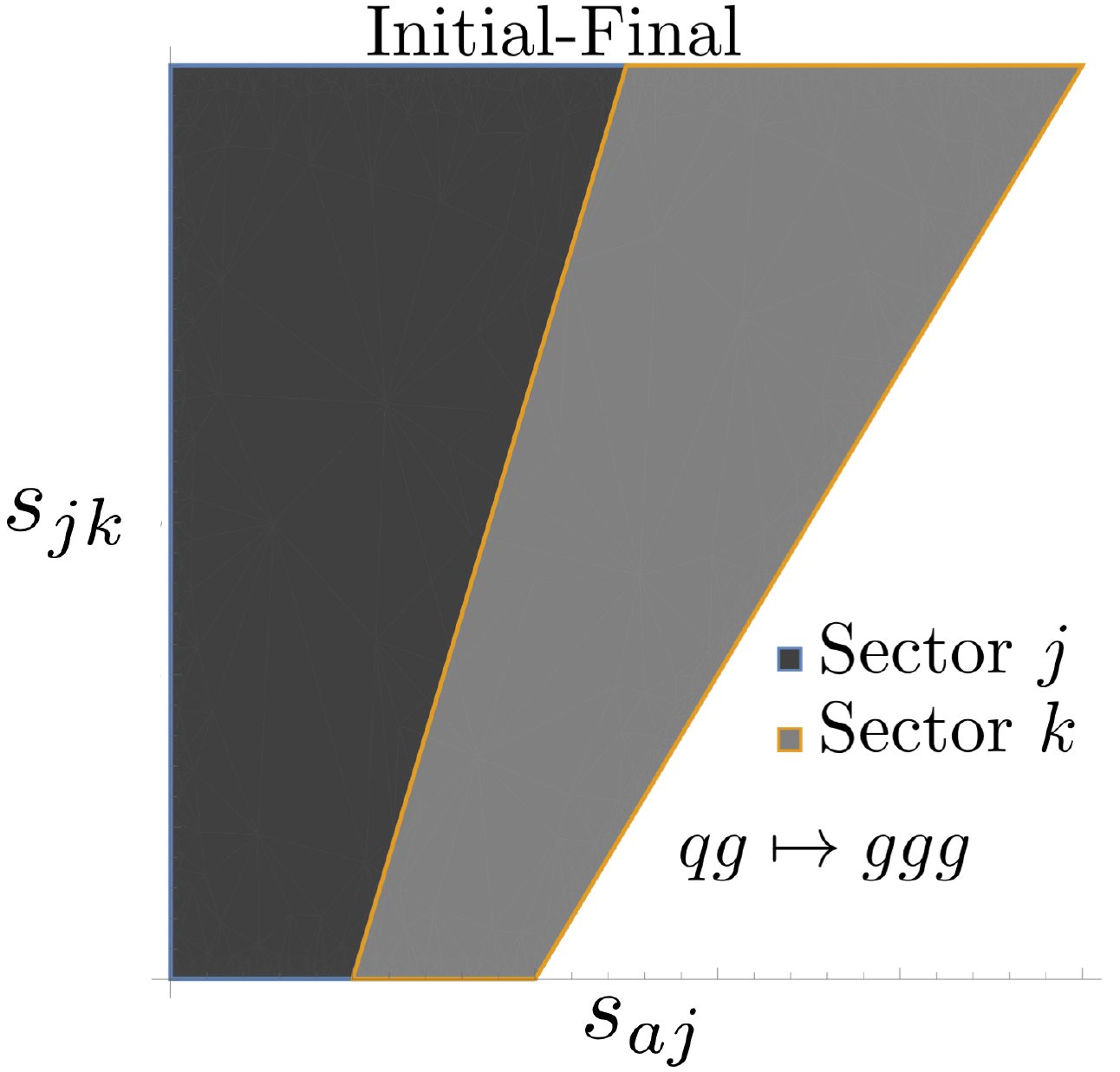}
		\hfill
		\includegraphics[width=0.4\textwidth]{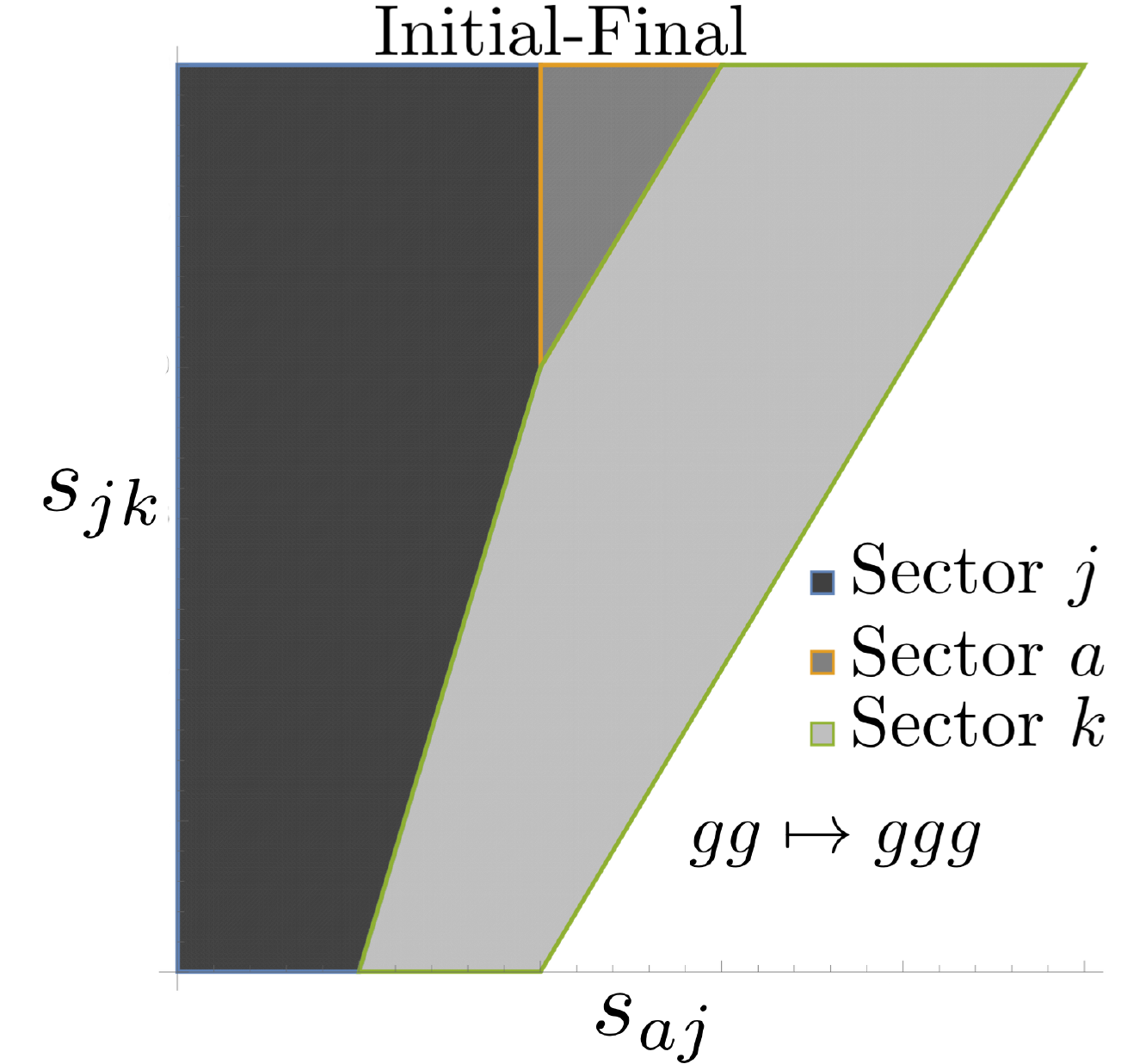}
	\end{minipage}
	\caption{Illustration of the sectors in the antenna phase spaces for an initial-final $q g \mapsto qgg$ antenna (\textit{left}) and an initial-final $gg \mapsto ggg$ antenna (\textit{right}),
 with the sector resolution according to \cref{eq:IFresolutionVars}. The sector accounting for $a$ as the emitted gluon, shown in the right pane, is not considered during the shower evolution, as emission into the initial state is not accounted for.}
\label{fig:2to3SectorsIF}
\end{figure}

As the antenna phase space is divided
into sectors corresponding to the radiation from different antennae, cf.\ \cref{fig:2to3SectorsFF,fig:2to3SectorsIF}, a criterion to decide which branching $IK \mapsto ijk$ to perform has to be chosen.
When thinking in terms of inverting the shower, this variable then determines which clustering $ijk \mapsto IK$ to perform.
The obvious choice would be to simply chose the ordering variables \cref{eq:FFevolutionVars,eq:RFevolutionVars,eq:IFevolutionVars,eq:IIevolutionVars}, as its proportionality to $\yij\yjk$ ensures that the most singular sector is picked if in gluon emissions either the soft or a collinear singularity is approached. 


However, this choice is not unique and for LL accuracy it must only ensure that the antenna function with the correct divergent terms is picked.
Different choices will then have different subleading logarithmic behaviour.

We follow the choice made in \cite{LopezVillarejo:2011ap} and take the dimensionful evolution variables $p_\perp$ to discriminate between sectors for gluon emissions and a $p_\perp$-weighted virtuality for gluon splittings, as summarised in \cref{eq:FFresolutionVars,eq:RFresolutionVars,eq:IFresolutionVars,eq:IIresolutionVars}. We refrain from choosing the dimensionless equivalents, as it was shown in \cite{LopezVillarejo:2011ap} to yield worse subleading-logarithmic behaviour for final-state gluon emissions.

\begin{figure}[t]
    \centering
    \includegraphics[width=0.9\textwidth]{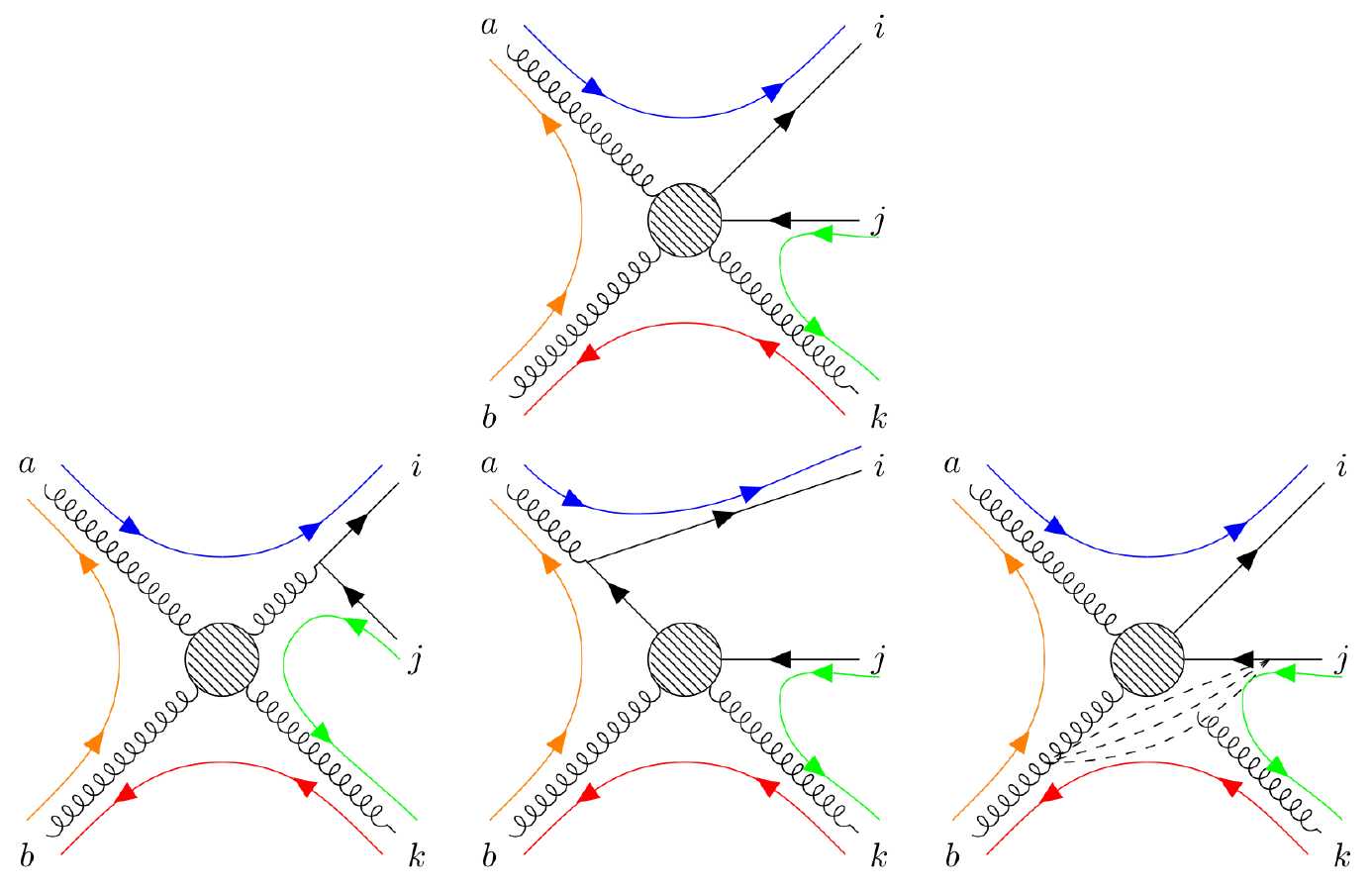}
    \caption{Illustration of all different histories (\textit{bottom row}) leading to the same colour-ordered process $gg \to q\bar q g$ (\textit{top row}).
    Going from left to right we have: a final-state gluon splitting, an initial state gluon splitting, and an initial-final gluon emission.}
    \label{fig:qbqgXSectorRes}
\end{figure}

The particular choice for quark pairs can be understood by considering the colour-ordered process $gg\to q\bar q g$, with a colour-ordering as shown
in \cref{fig:qbqgXSectorRes}. Three different histories contribute to this final state. Firstly, the emission of gluon $k$ could have happened before or after the creation of the $q$-$\bar q$ pair. Secondly, this splitting could have both happened in the initial or final state. In total, this leads to four different possible clusterings,
\begin{itemize}
    \item final-state gluon splitting: $ijk \mapsto IK$, $aij \mapsto AJ$
    \item initial-state gluon splitting: $aib \mapsto AB$
    \item gluon emission: $jkb \mapsto JB$
\end{itemize}
two for each $q_i$-$\bar q_j$ pair and one for the gluon $g_k$. However, while the $\bar q_j$-$g_k$-$g_b$ antenna contains the $q_j$-$g_k$-collinear singularity, the $q_i$-$\bar q_j$-$g_k$ antenna only contains the quark-antiquark-collinear singularity and is finite for $q_j$ and $g_k$ becoming collinear. Hence, by comparing only the transverse momentum $p_\perp$ for $\yjk \to 0$, one is prone to pick the wrong clustering, if the invariant on the $i$-$j$-side is smaller than the one on the $k$-$b$-side. Using an interpolation between the geometric mean of the transverse momentum $p_\perp$ and the virtuality of the quark pair as collected below in \cref{eq:FFresolutionVars,eq:RFresolutionVars,eq:IFresolutionVars,eq:IIresolutionVars} reflects this subtlety and guarantees that the entire $\bar q_j$-$g_k$-collinear limit is classified to belong to the $\bar q_j$-$g_k$-$g_b$ antenna.

\paragraph{FF Branchings} For final-final configurations, we choose the sector resolution variables:
\begin{equation}
    Q^2_{\text{res}_j \text{FF}} = \begin{cases}\displaystyle
    \frac{\sij\sjk}{\sIK} & \text{if } j \text{ is a gluon} \\[4mm]
    \displaystyle (\sij + 2 m_j^2) \sqrt{\yjk +\mu_j^2} & \text{if } (i,j) \text{ is a quark-antiquark pair}
    \end{cases}~, \label{eq:FFresolutionVars}
\end{equation}
with the respective choice for $i \leftrightarrow k$.

\paragraph{RF Branchings} The RF sector resolution variables are virtually the same as the FF ones,
\begin{equation}
    Q^2_{\text{res}_j \text{RF}} = \begin{cases} \displaystyle
        \frac{\saj\sjk}{\saj +\sak} & \text{if } j \text{ is a gluon} \\[4mm]
        \displaystyle(\sjk + 2m_j^2)\sqrt{\yaj-\mu_j^2} & \text{if } (j,k) \text{ is a quark-antiquark pair}
    \end{cases}~, \label{eq:RFresolutionVars}
\end{equation}
with the difference only in the sign of the mass correction to $\yaj$, due to the different momentum conservation, cf.\ \cref{eq:epConsIF}.

\paragraph{IF Branchings} For IF configurations, we choose the sector resolution variables to be the same as the RF ones, complemented by additional ones treating initial-state gluon splittings and conversions, not present in RF branchings:
\begin{equation}
    Q^2_{\text{res}_j \text{IF}} = \begin{cases} \displaystyle
        \frac{\saj\sjk}{\saj +\sak} & \text{if } j \text{ is a gluon} \\[4mm]
        \displaystyle(\saj-2m_j^2)\sqrt{\yjk+\mu_j^2} & \text{if } (a,j) \text{ is a quark-quark pair} \\[4mm]
        \displaystyle\saj\sqrt{\yjk+\mu_j^2} & \text{if } (a,j) \text{ is a gluon-(anti)quark pair} \\[4mm]
        \displaystyle(\sjk + 2m_j^2)\sqrt{\yaj-\mu_j^2} & \text{if } (j,k) \text{ is a quark-antiquark pair}
    \end{cases}~. \label{eq:IFresolutionVars}
\end{equation}

\paragraph{II Branchings} For initial-initial configurations, we choose the resolution variables 
\begin{equation}
     Q^2_{\text{res}_j \text{II}} = \begin{cases}\displaystyle
        \frac{\saj \sjb}{\sab} & \text{if } j \text{ is a gluon} \\[4mm]
        \displaystyle(\saj-2m_j^2)\sqrt{\yjb-\mu_j^2} & \text{if } (a,j) \text{ is a gluon-(anti)quark pair}\\[4mm]
        \displaystyle\saj\sqrt{\yjb-\mu_j^2} & \text{if } (a,j) \text{ is a quark-quark pair}
     \end{cases}~, \label{eq:IIresolutionVars}
\end{equation}
with the respective choice for $a \leftrightarrow b$.

\section{Validation and Preliminary Results}
\label{sec:Validation}

We validate the sector shower in two stages: first in \cref{sec:MERatios} by comparing its tree-level expansion to tree-level matrix elements (see \cite{Giele:2011cb,LopezVillarejo:2011ap,Fischer:2016Vincia} for equivalent plots for global showers), and second in \cref{sec:experimentalValidation} by comparing the full-fledged sector shower to experimental data. For the latter, comparisons are also included to the global \Vincia\ and default \Pythia\ 8.3  showers. 
Both \Rivet~\cite{Buckley:2010ar} and an internal \Vincia\ analysis package were used for performing analyses.

\subsection{Comparison to Tree-Level Matrix Elements}
\label{sec:MERatios}

To probe the quality of our sector antenna functions as well as our sector resolution criteria, we define the parton-shower-to-matrix-element ratio
\begin{equation}
	\begin{split}
	R^\text{sct}_n &=  \left( \prod\limits_{\substack{\text{\tiny{ordered}} \\ i}}^{n-n_\text{Born}}  \sum\limits_{j \in \{p\}_{n-i}} \Theta(Q^2_\text{min}(\{p\}_{n-i}) - Q^2_{\text{res}_j}) \gstrong^2 \mathcal{C}_{j} \bar{a}^\text{sct}_{j} \right) \frac{\left\vert \mathcal{M}_\text{Born} \right\vert^2}{\vert \mathcal{M}_n \vert^2} \\
	&= \left( \prod\limits_{\substack{\text{\tiny{ordered}} \\ i}} \gstrong^2 \mathcal{C}_{i} \bar{a}^\text{sct}_{i} \right) \frac{\left\vert \mathcal{M}_\text{Born} \right\vert^2}{\vert \mathcal{M}_n \vert^2}\, , \label{eq:mePSratio}
	\end{split}
\end{equation}
where $Q^2_\text{min}(\{p\}_{n-i}) $ is the minimal sector resolution over the state $\{p\}_{n-i}$ and the symbol $\prod\limits_\text{\tiny{ordered}}$ is meant to impose strong ordering, i.e., that
$p^2_{\perp,n} < p^2_{\perp, n-1} < p^2_{\perp, n-2} < \ldots < Q^2_\text{F}$ for each clustering sequence, where $Q^2_\text{F}$ denotes the factorisation scale.
By construction, \emph{only a single clustering} contributes at each $n$-parton phase space point, as indicated by the second line in \cref{eq:mePSratio}.

\begin{figure}[t]
    \centering
    \includegraphics[width=0.49\textwidth]{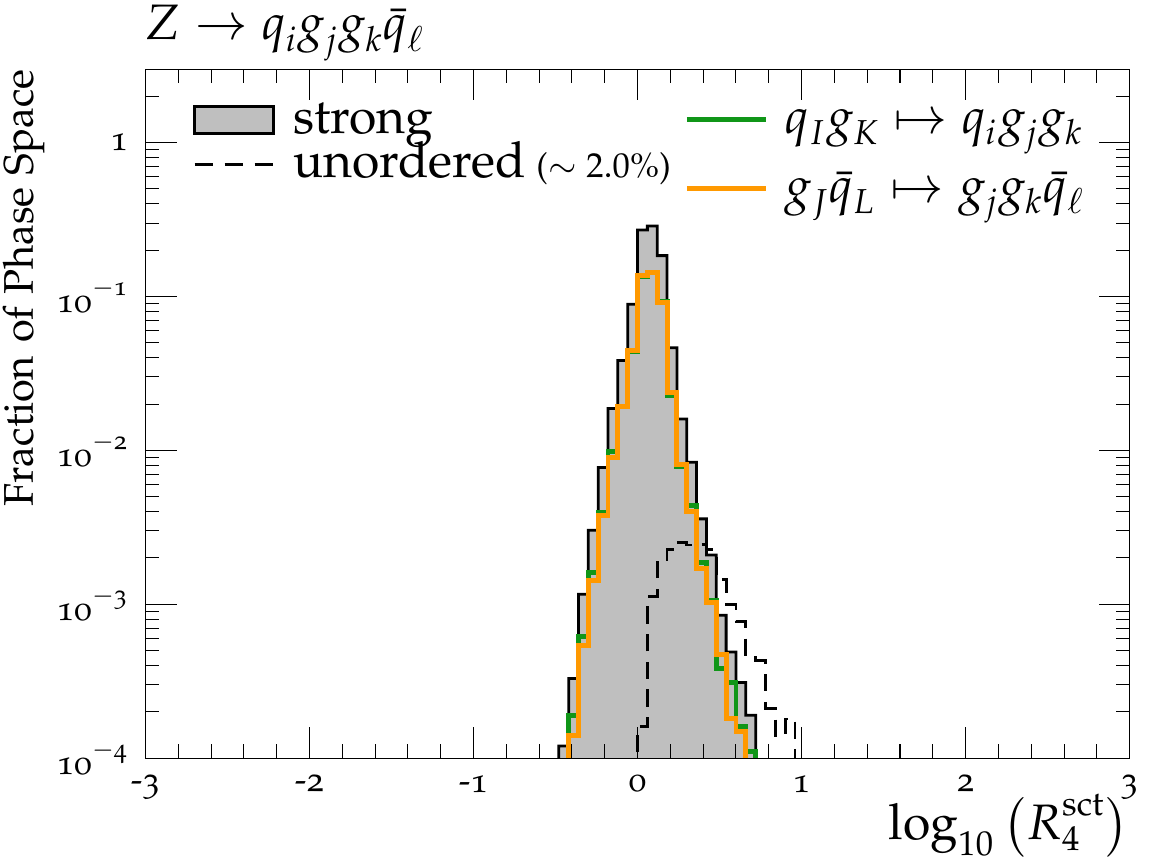}
    \includegraphics[width=0.49\textwidth]{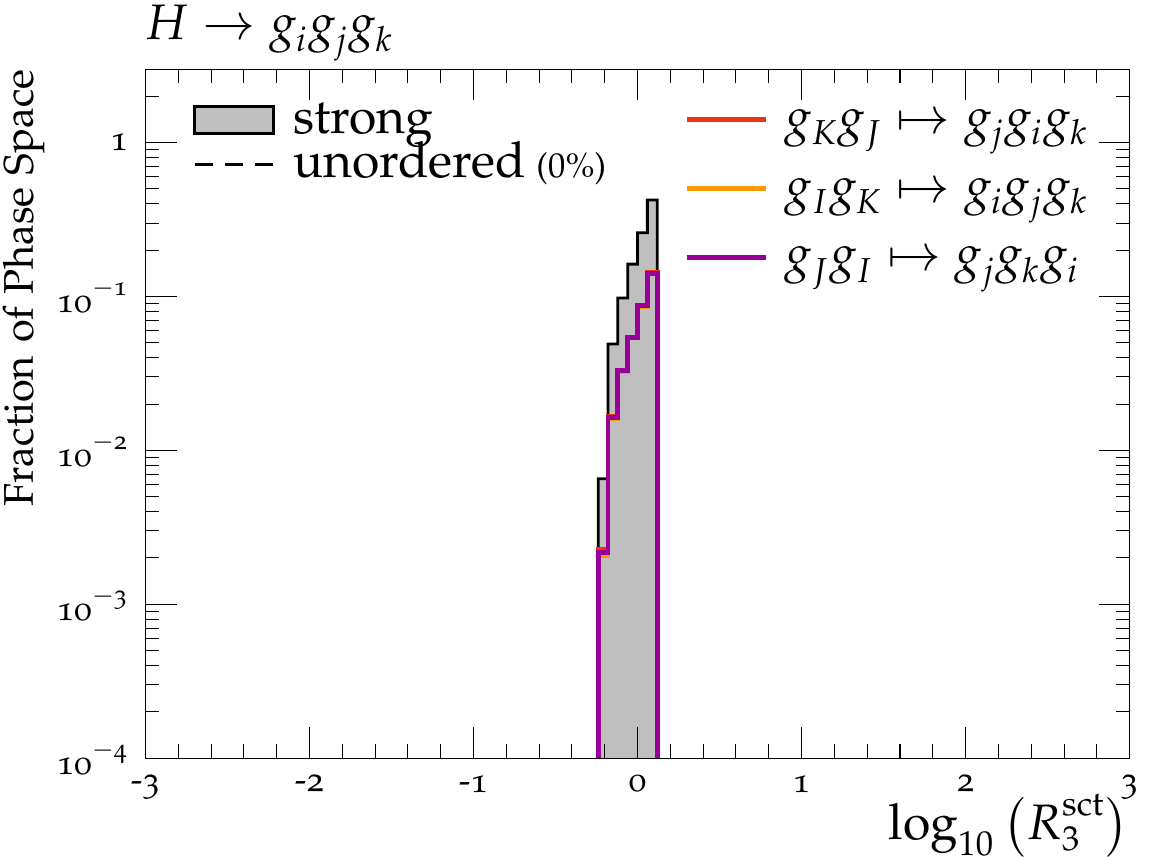}
    \caption{Ratios $R_n^\text{sct}$ of the sector-shower approximation to LO matrix elements for $Z \to q \bar q + 2g$ and $H \to 3g$ in a flat phase space scan.}
    \label{fig:MERatiosZH}
\end{figure}
\begin{figure}[t]
	\centering
	\includegraphics[width=0.49\textwidth]{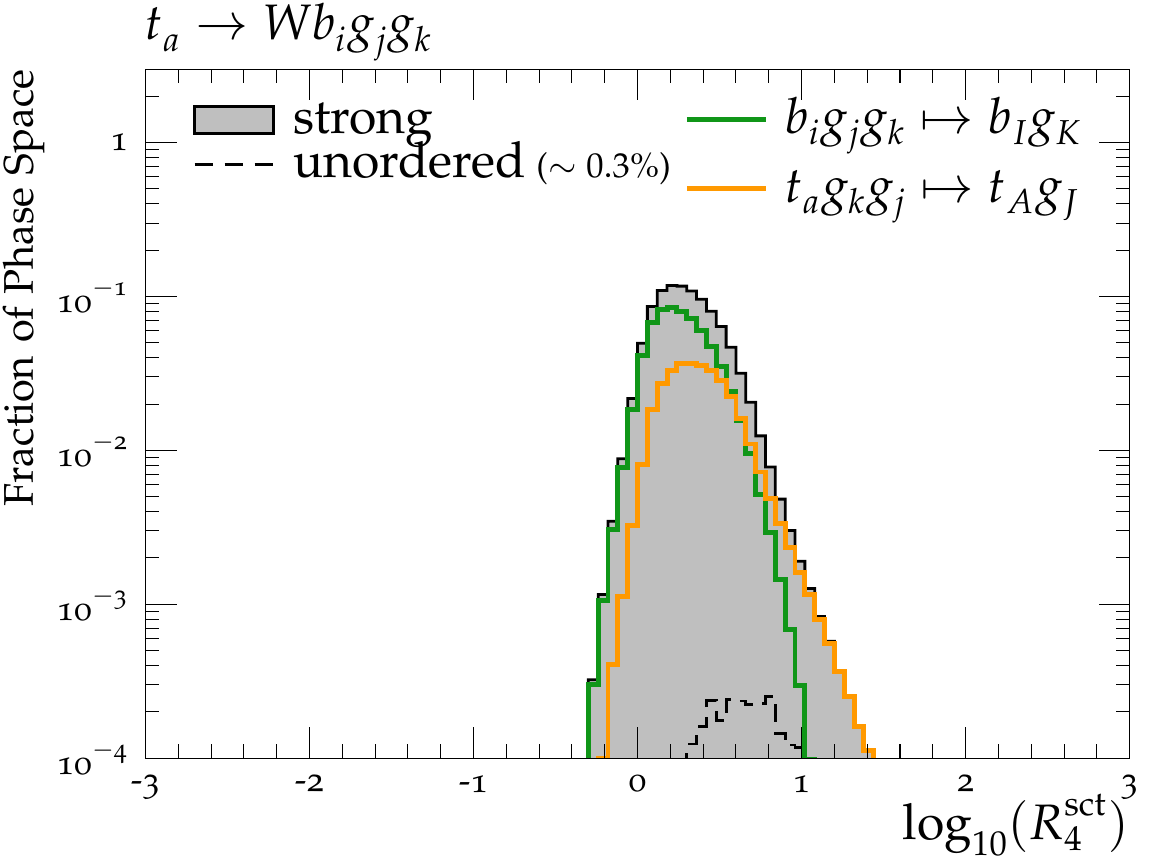}
	\includegraphics[width=0.49\textwidth]{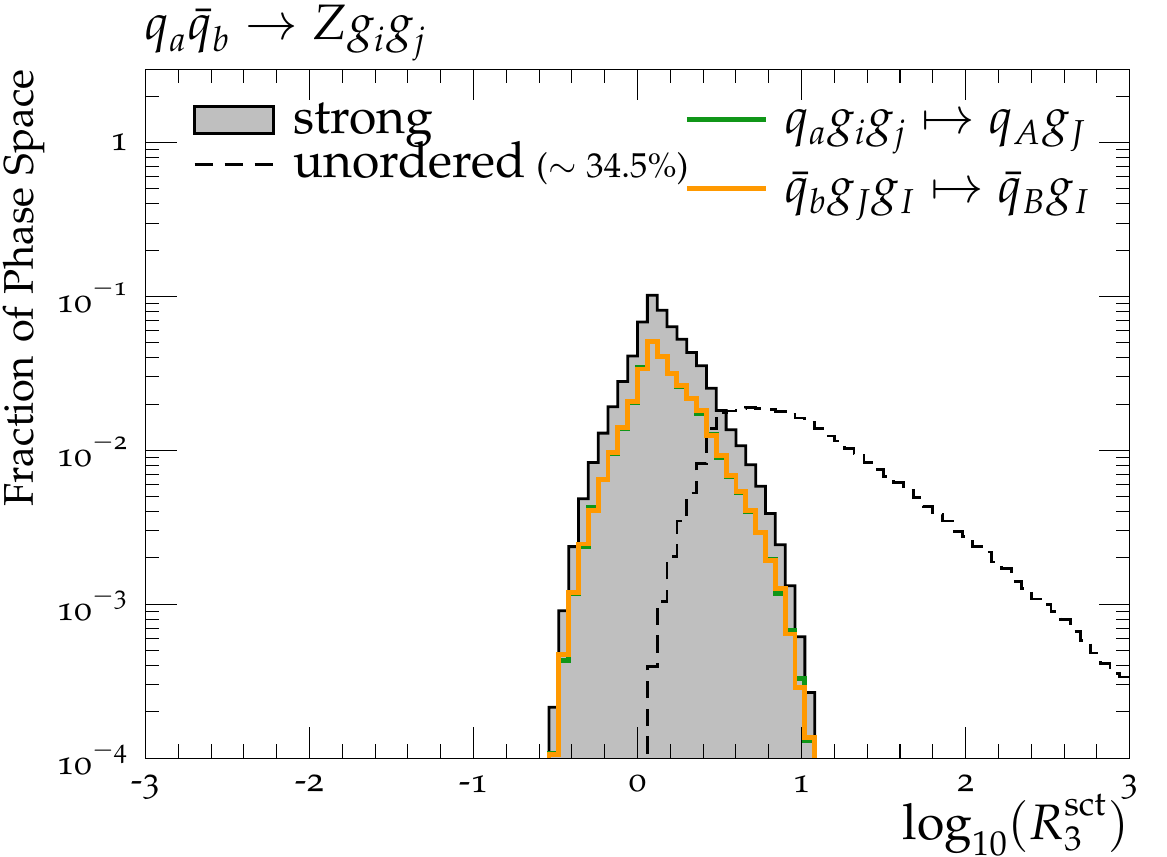}
	\caption{Ratios $R_n^\text{sct}$ of the sector-shower approximation to LO matrix elements for gluon emissions off $t \to W^+ b$ (\textit{left}) and $q\bar q \to Z$ (\textit{right}) in a flat phase space scan.}
	\label{fig:MERatiosResAndDY}
\end{figure}

For a fixed colour ordering, we compare the sector antenna shower approximation to leading-colour tree-level matrix elements from the \mg\ matrix element generator \cite{Alwall:2014hca}.
Using \textsc{Vincia}'s implementation of \textsc{Rambo} \cite{Kleiss:1985gy}, we generate large samples of the $n$-parton phase space in a flat phase-space scan. Initial-state momenta are fixed by a flat sampling of the momentum fractions for partonic scattering processes and to the pole mass for resonance decays.
We cluster the state back to the Born configuration according to the exact inverses of the corresponding $2 \mapsto 3$ kinematic maps, cf.\ \cref{sec:Kinematics}, and determine the sector-shower history according to the resolution variables \cref{eq:FFresolutionVars,eq:RFresolutionVars,eq:IFresolutionVars,eq:IIresolutionVars}.

\begin{figure}[ht]
	\centering
	\includegraphics[width=0.49\textwidth]{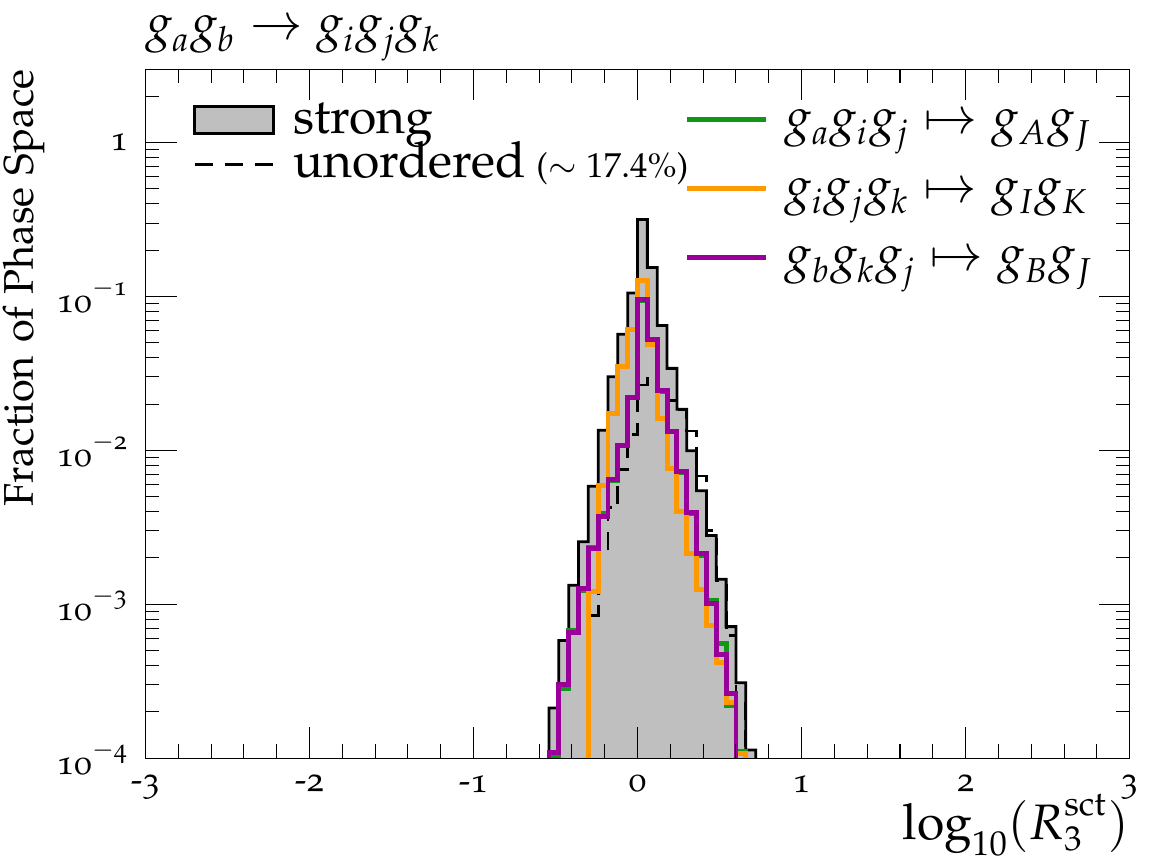}
	\includegraphics[width=0.49\textwidth]{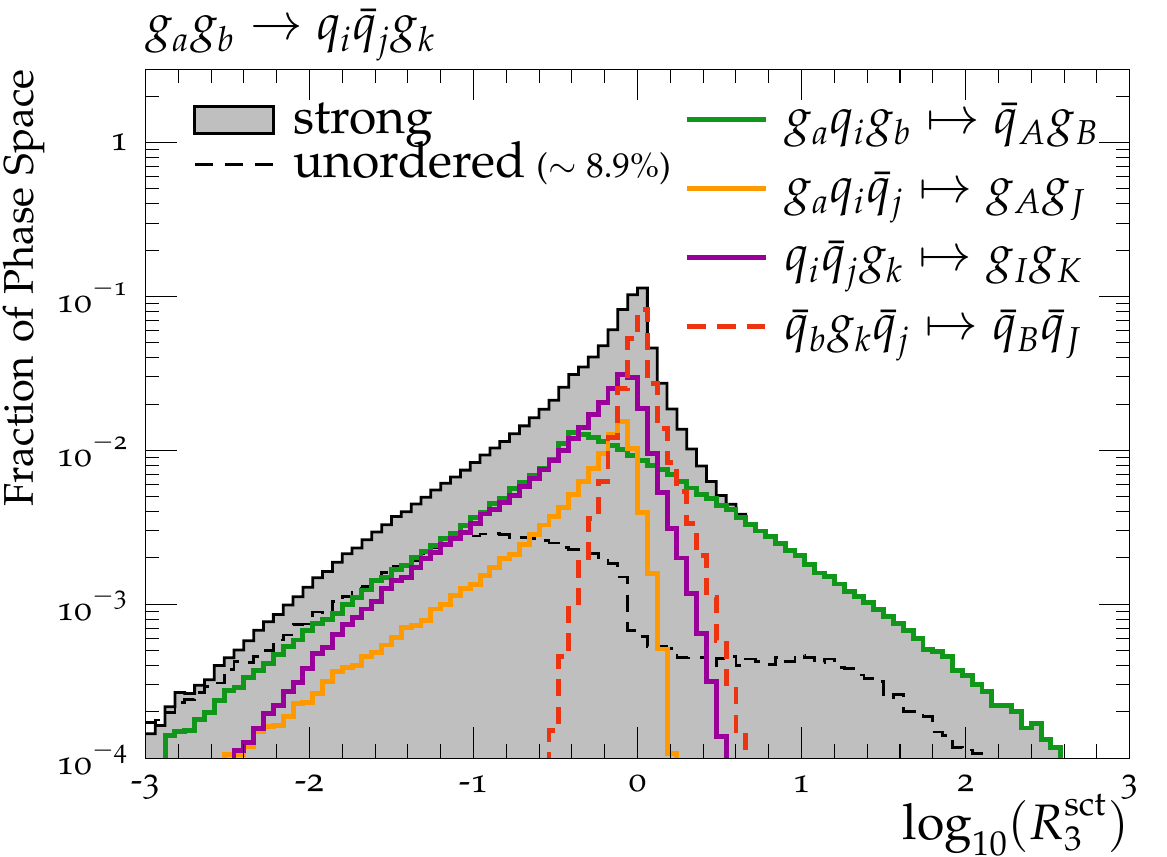}
	\caption{Ratios $R_3^\text{sct}$ of the sector-shower approximation to LO matrix elements for $gg$-fusion processes in a flat phase space scan.}
	\label{fig:MERatiosGG}
\end{figure}

\begin{figure}[ht]
    \centering
    \includegraphics[width=0.49\textwidth]{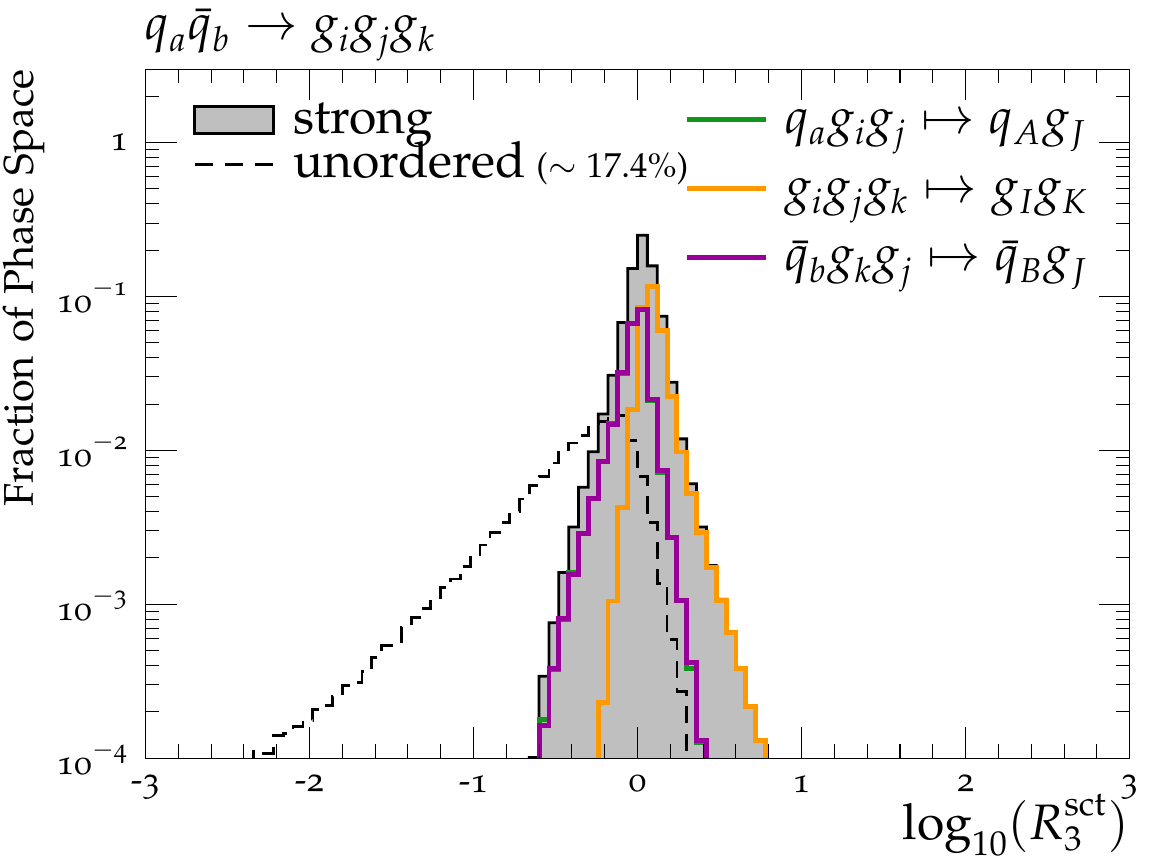}
    \includegraphics[width=0.49\textwidth]{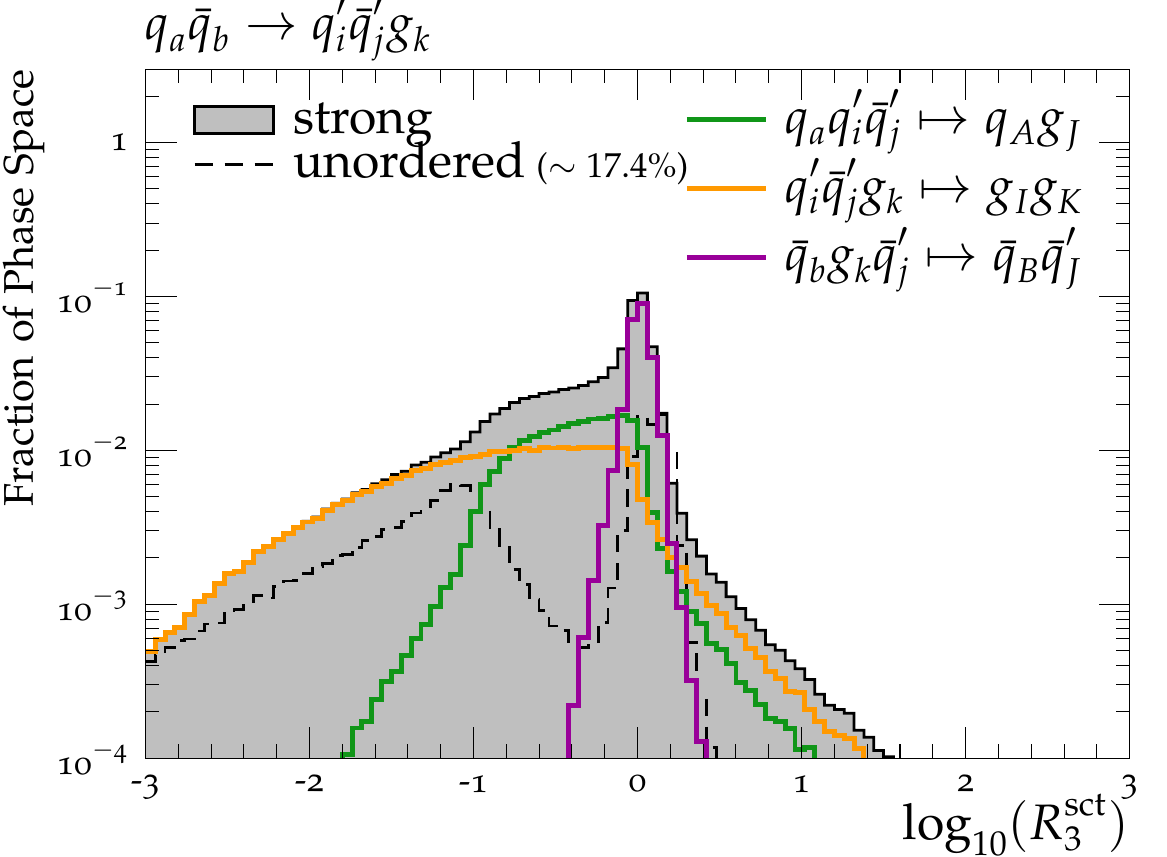}
    \caption{Ratios $R_n^\text{sct}$ of the sector-shower approximation to LO matrix elements for $q\bar q$-annihilation processes in a flat phase-space scan.}
    \label{fig:MERatiosQQb}
    \vspace{-1em}
\end{figure}

In \cref{fig:MERatiosZH}, we plot $R_n^\text{sct}$ for the resonance decay processes $Z \to q gg \bar q$ and $H \to ggg$,
while in \cref{fig:MERatiosResAndDY}, we show it for gluon emissions off the resonance decay $t\to W^+ b$ and the Drell-Yan process $q\bar q \to Z$.
In \cref{fig:MERatiosGG,fig:MERatiosQQb}, we consider $2\to 3$ gluon-gluon and quark-antiquark scattering processes respectively. 
We here denote light (i.e.\ effectively massless) quarks by $q$ and denote heavy flavours explicitly by $t$ and $b$; unequal flavours are denoted by dashes, $q'$.
For all cases, we show the shower approximations 
in both the ordered (solid) and unordered (dashed) regions of phase space. 
(The latter is shown for completeness; it would be vetoed in a strongly ordered shower.) In the ordered regions, 
we also include the breakdown per contributing sector (coloured lines).
In \cref{fig:MERatiosZhiMulti,fig:MERatiosHhiMulti,fig:MERatiosResHiMulti,fig:MERatiosDYhiMulti,fig:MERatiosGGhiMulti,fig:MERatiosQQbHiMulti} in \cref{sec:hiMultiMERatios}, we present $R_n^\text{sct}$ for higher multiplicities. 

The patterns we observe are the following.
For gluon emissions, we see quite narrow distributions centred on $\log(R_n) = 0$, indicating a generally very good agreement between the sector-shower approximation and the LO matrix elements. This is particularly true in the ordered regions of the phase space for each hard process. 
Branching processes involving gluon splittings (including initial-state splittings and quark conversions) are significantly less well approximated. This is due to a combination of two factors, as follows. Firstly, gluon-splitting kernels only contain single poles which hence are expected to dominate over a smaller fraction of phase space compared with the double poles of gluon-emission ones. Secondly, the numerator of the single poles of the $g\to q\bar{q}$ branchings has a significant dependence on the (linear) polarisation of the splitting gluon (which is currently not accounted for in our sector shower), while the polarisation dependence of gluon-emission processes is much milder~\cite{Webber:1986mc}. 
The overall quality of the approximations are consistent with those seen for global showers in \cite{Giele:2011cb,Fischer:2016Vincia} and for the final-final sector shower in \cite{LopezVillarejo:2011ap}. 

Furthermore, for several of the QCD scattering processes, we observe that even the distributions of \emph{unordered} histories are peaked close to $R_n^\text{sct} = 1$.
Many of these \textquotedblleft good\textquotedblright\ unordered paths are, however, only unordered in the first emission, i.e., the first emission appears at a higher scale than the factorisation scale.
Although this suggests to let the shower evolution start at a nominally larger scale than the factorisation scale, we refrain from doing so for now and leave a study of 
such \textquotedblleft power-shower\textquotedblright\ effects, cf.~\cite{Plehn:2005cq}, in sector showers to future work.

\subsection{Comparison to Experimental Data}
\label{sec:experimentalValidation}
We here present first results with the \Vincia\ sector shower, comparing it to experimental measurements and to the global \Vincia\ and \Pythia\ 8.3 showers. Although we included a large set of observables in our study, 
we here show only a minimal set which were deemed to be the most physically relevant and/or representative ones.

With the splitting kernels and kinematics maps fixed, the main quantity governing the perturbative shower evolution is the strong coupling. 
For both of the \Vincia\ showers, we evolve the coupling at two-loop order with an $\overline{\text{MS}}$ value of $\alphastrong (m_Z) = 0.118$. Since our shower model is coherent, we take the CMW scheme~\cite{Catani:1990rr} as the baseline scheme for defining the effective value of $\alpha_s$ for shower branchings, i.e., we use
\begin{equation}
    \alphastrong^\text{CMW} = \alphastrong^{\overline{\text{MS}}}
    \left(1 + \frac{\alphastrong^{\overline{\text{MS}}}}{2\pi} \left[
    C_A \left(\frac{67}{18} - \frac{\pi^2}{6}\right) - \frac{5 n_f}{9}
    \right]\right)~,
\end{equation}
which is technically imposed via an $n_f$-dependent rescaling of the $\Lambda_\mathrm{QCD}$ parameter. 
The CMW factor is derived for gluon emissions in the infinitely soft limit. To reflect remaining ambiguities in the scale definition and to 
allow the effective value of $\alpha_s$ used for physical branchings to deviate from this, we further allow renormalisation-scale prefactors $k_R$ 
to modify the evolution-scale $p_\perp$ argument of the running coupling, so that:
\begin{equation}
\alpha_s^\mathrm{Vincia}(p_\perp^2) 
  = \alpha_s^\mathrm{CMW}(k_R p_\perp^2) ~.
 \end{equation}
From preliminary tuning studies comparing \Vincia's global shower to LEP event shapes and Drell-Yan $p_\perp$ spectra, the default values for the additional scale prefactors are chosen as
\begin{align}
    k_\text{R,Emit}^\text{F} &= 0.66 \, , \quad 
    k_\text{R,Split}^\text{F} = 0.8 \, , \\
    k_\text{R,Emit}^\text{I} &= 0.66 \, , \quad
    k_\text{R,Split}^\text{I} = 0.5 \, , \quad 
    k_\text{R,Conv}^\text{I} = 0.5 \, .
\end{align}
We note that these choices roughly double the effect of the \text{CMW} scheme translation. We plan to return to an investigation of the seemingly large $1/k_R$ values needed to reproduce data in a future study including higher-order virtual effects (e.g., along the lines of~\cite{Hartgring:2013jma}) but for the moment note that the combined effect is similar to that of the rather large effective value of $\alpha_s(M_Z)=0.1365$ chosen in the baseline tune of \Pythia~\cite{Skands:2014pea}. 

The complete set of parameters changed relative to the default \Pythia\ tune are collected in \cref{sec:tune}.
\begin{figure}[ht]
    \centering
    \includegraphics[width=0.48\textwidth]{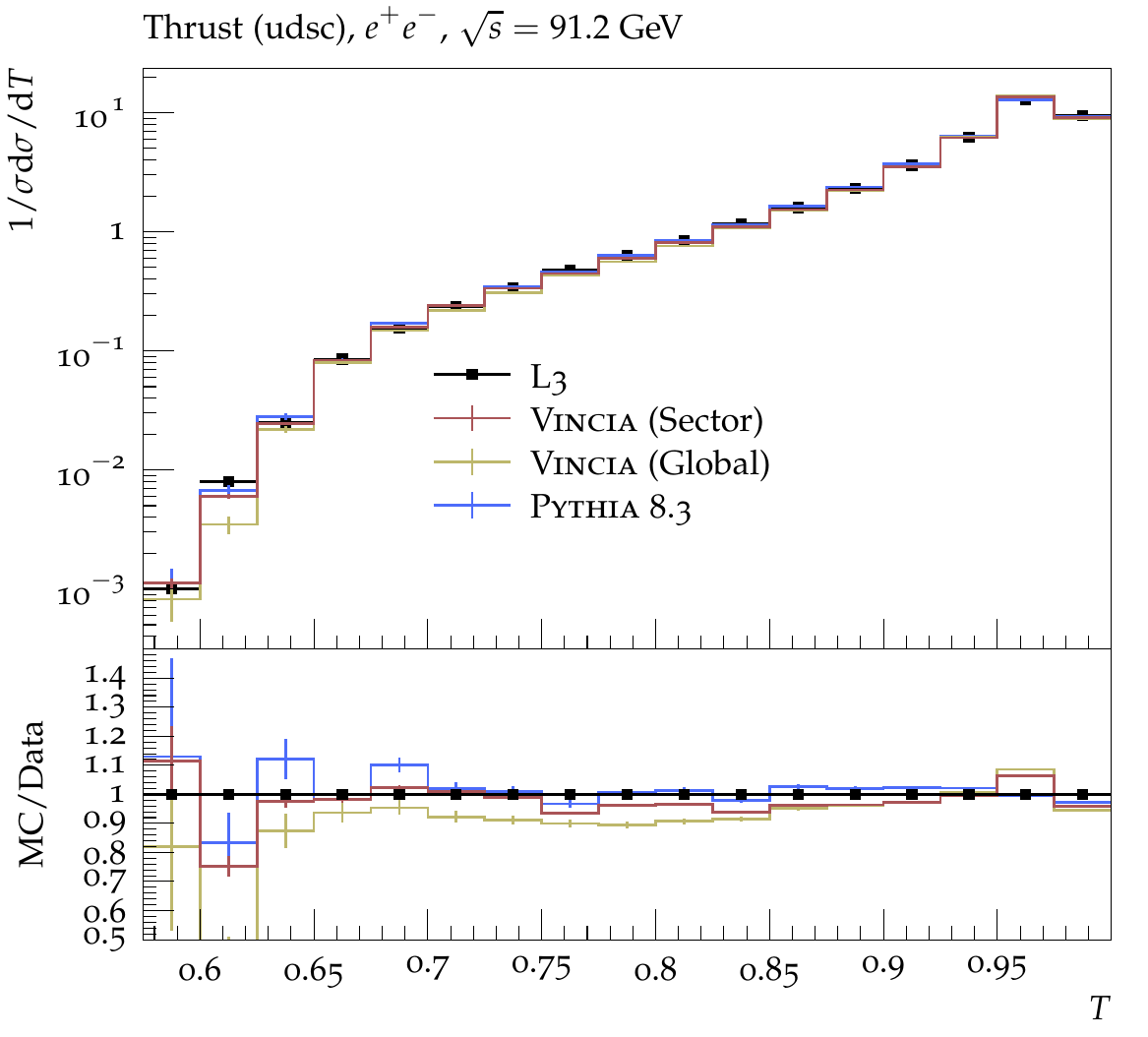}
    \includegraphics[width=0.48\textwidth]{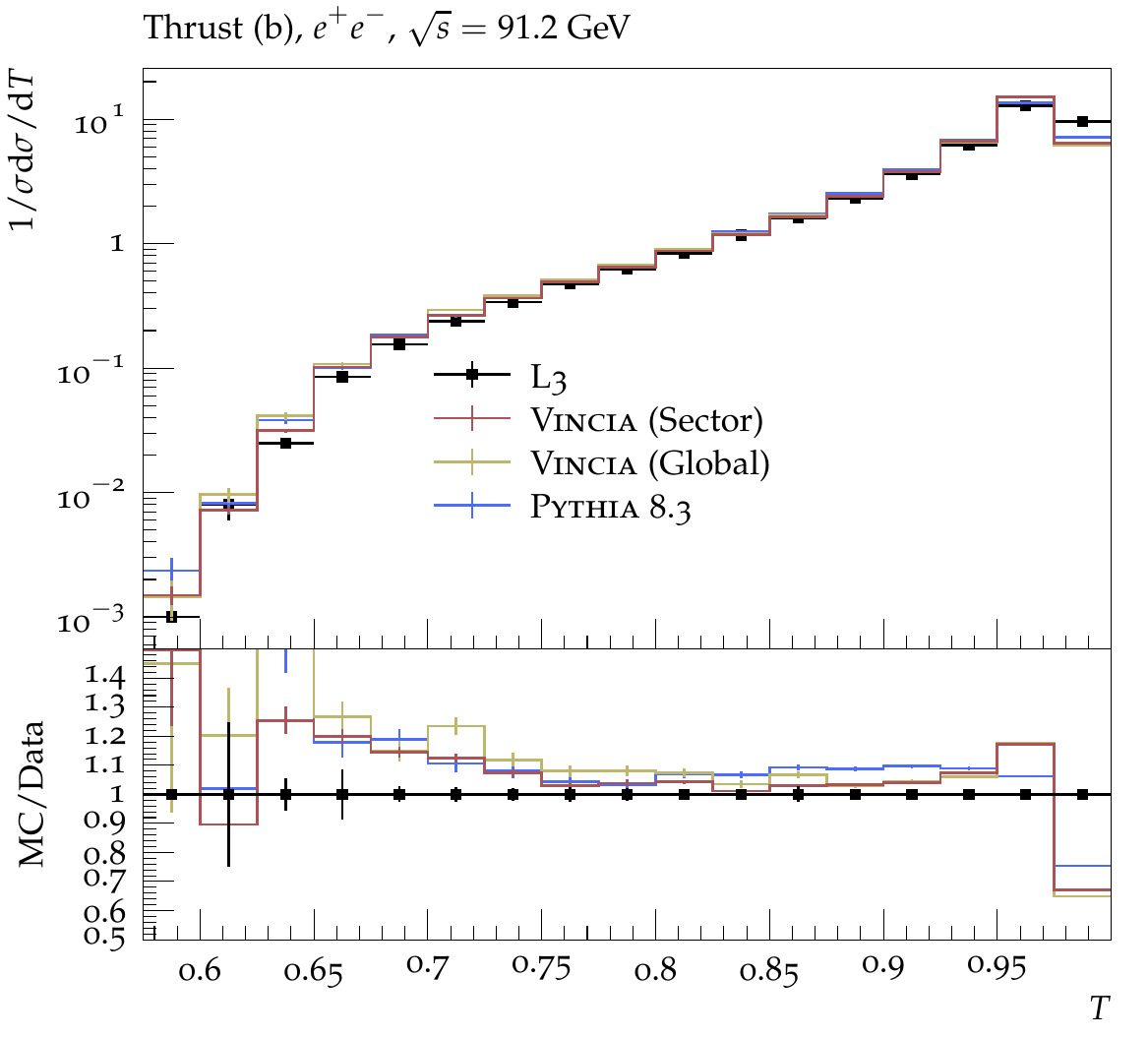}
    \includegraphics[width=0.48\textwidth]{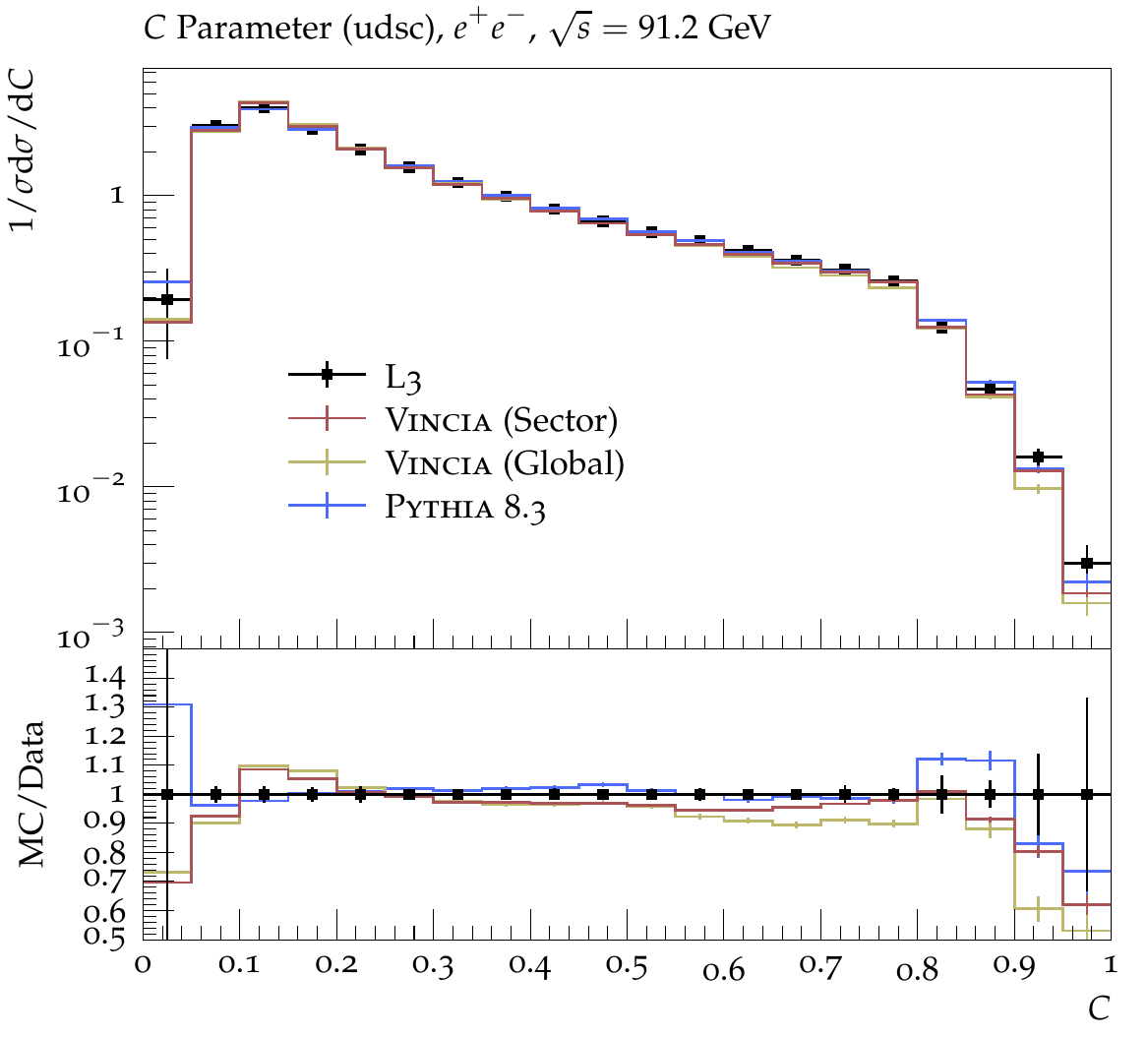}
    \includegraphics[width=0.48\textwidth]{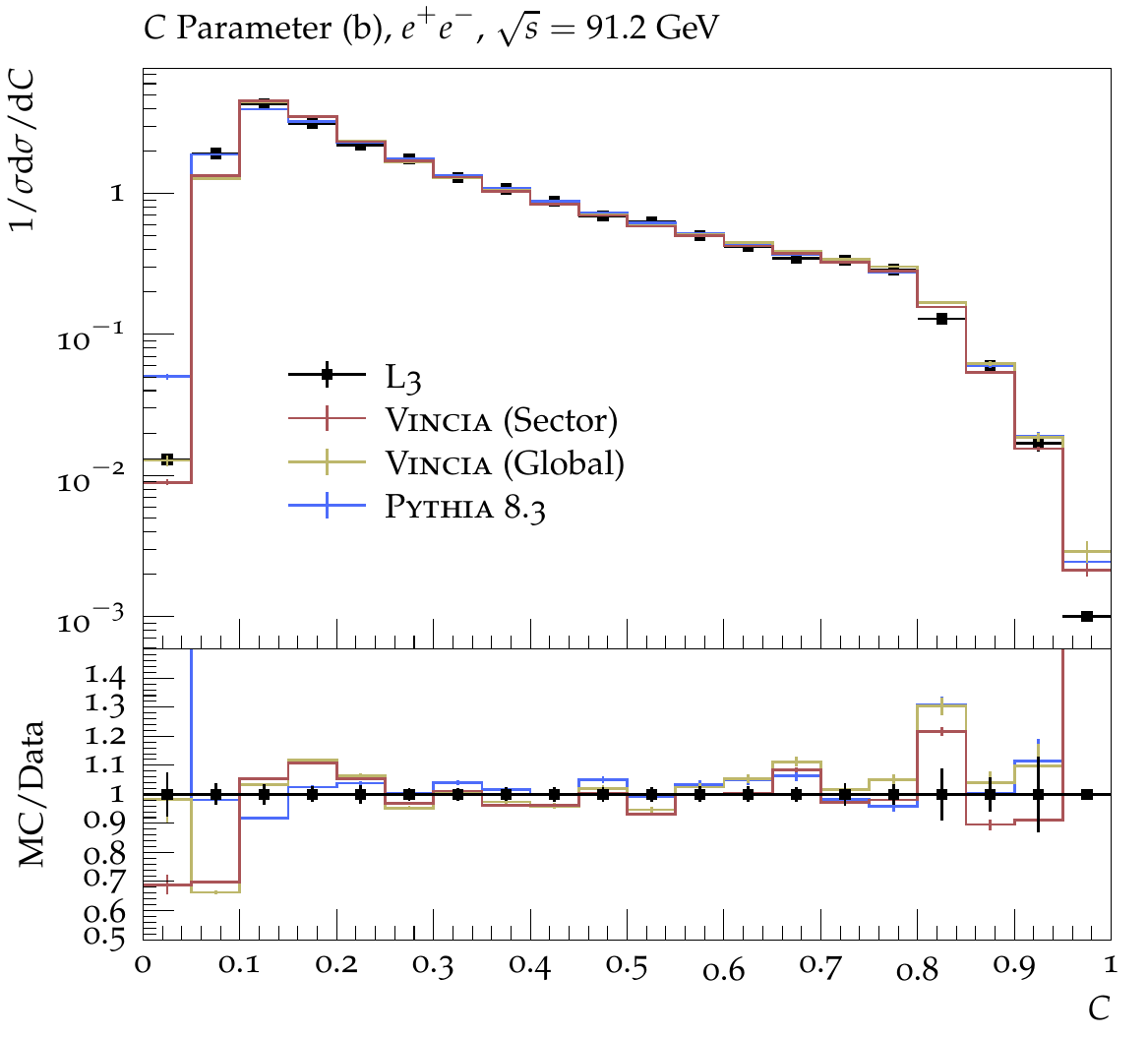}
    \includegraphics[width=0.48\textwidth]{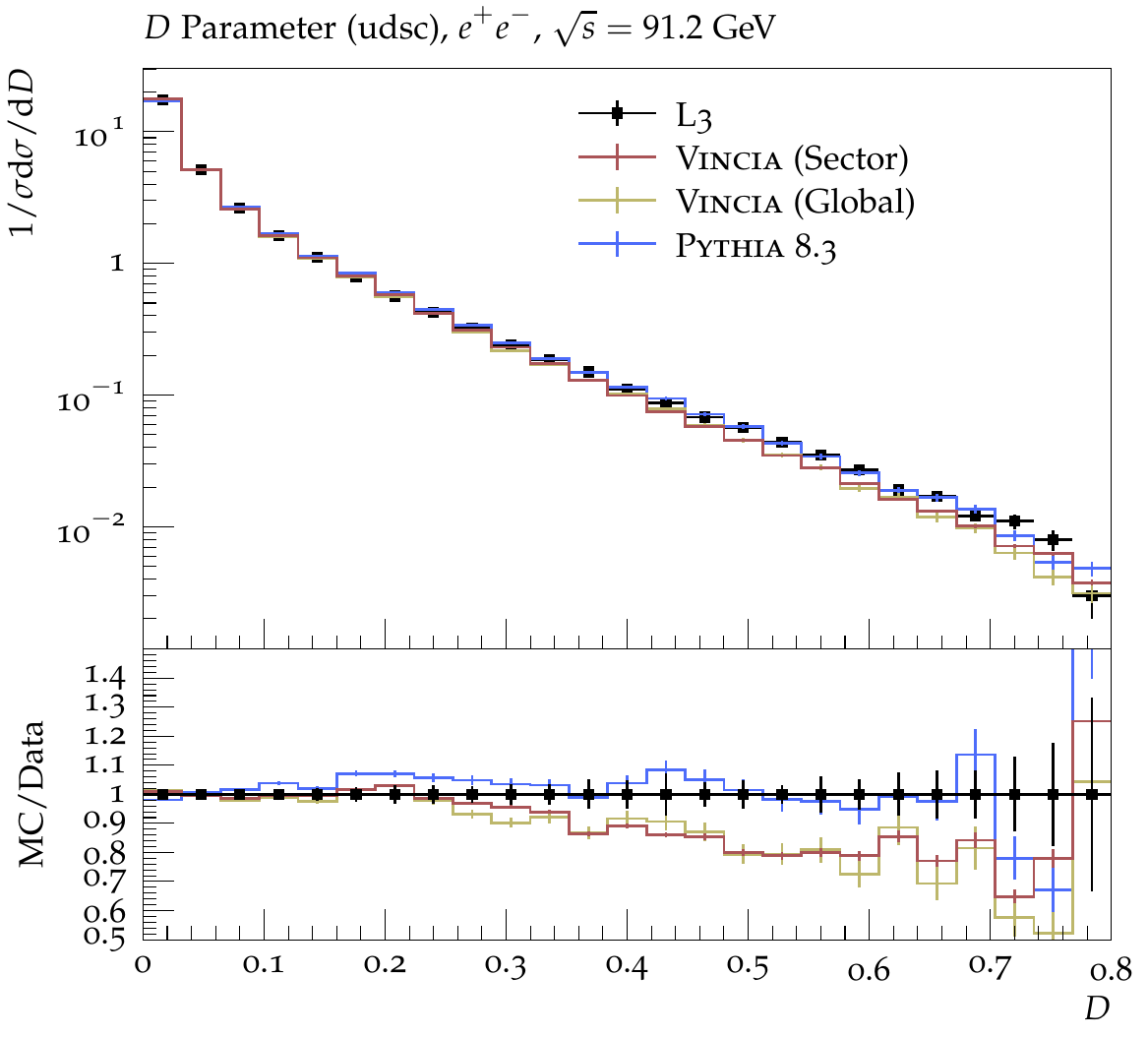}
    \includegraphics[width=0.48\textwidth]{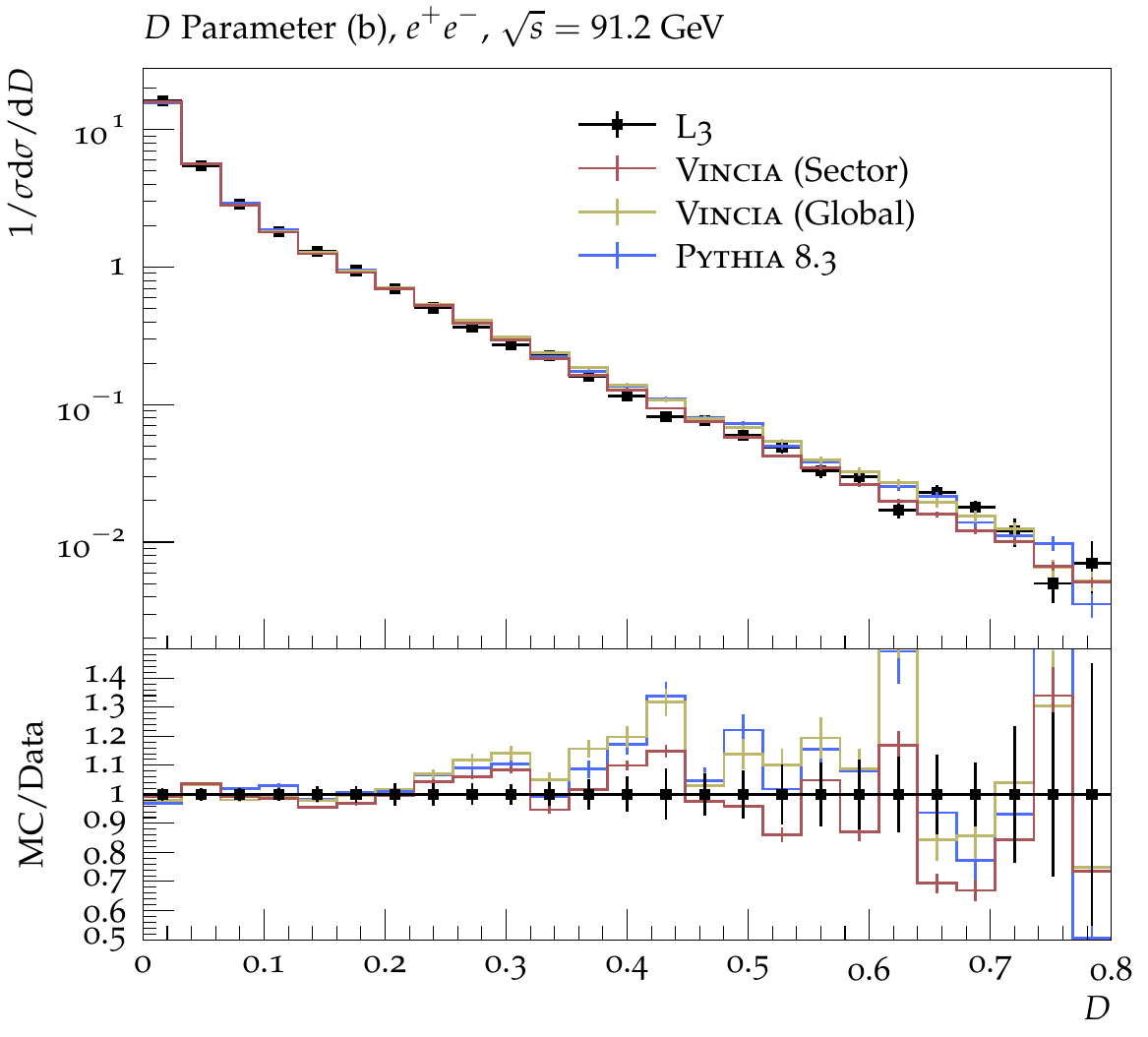}
    \caption{Normalised event shape distributions in $Z$ decays to light flavours (\textit{left}) and $b$ quarks only (\textit{right}) in $e^+e^-$ collisions at a centre-of-mass energy of $\sqrt{s} = 91.2~\giga e\volt$. 
    Comparison of the sector shower against the default (global) \Vincia\ shower, \textsc{Pythia} 8.3, and L3 data~\cite{Achard:2004sv}.}
    \label{fig:LEPEventShapes1}
\end{figure}

\paragraph{Electron-Positron Annihilation}
\begin{figure}[ht]
    \centering
    \includegraphics[width=0.49\textwidth]{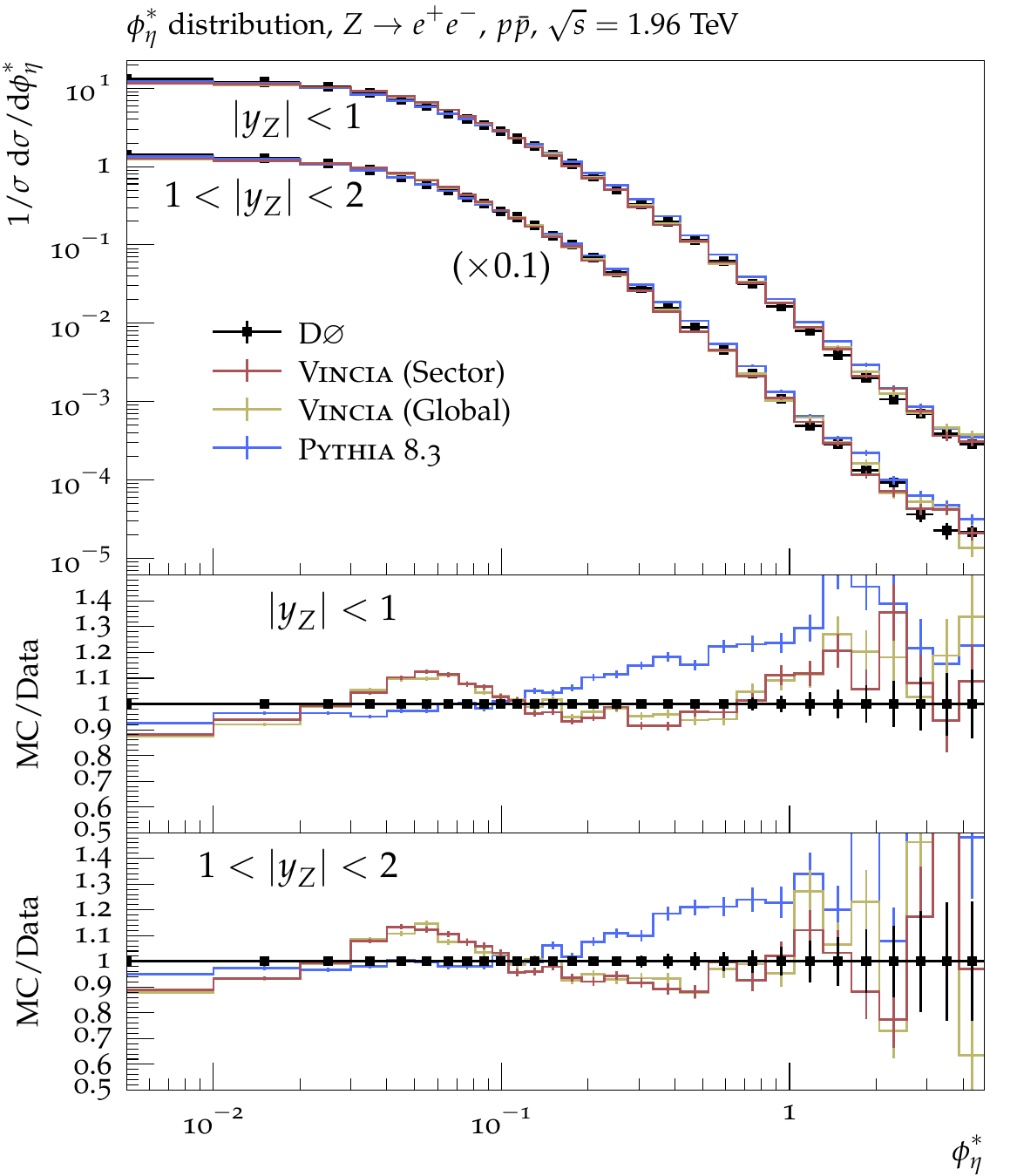}
    \includegraphics[width=0.49\textwidth]{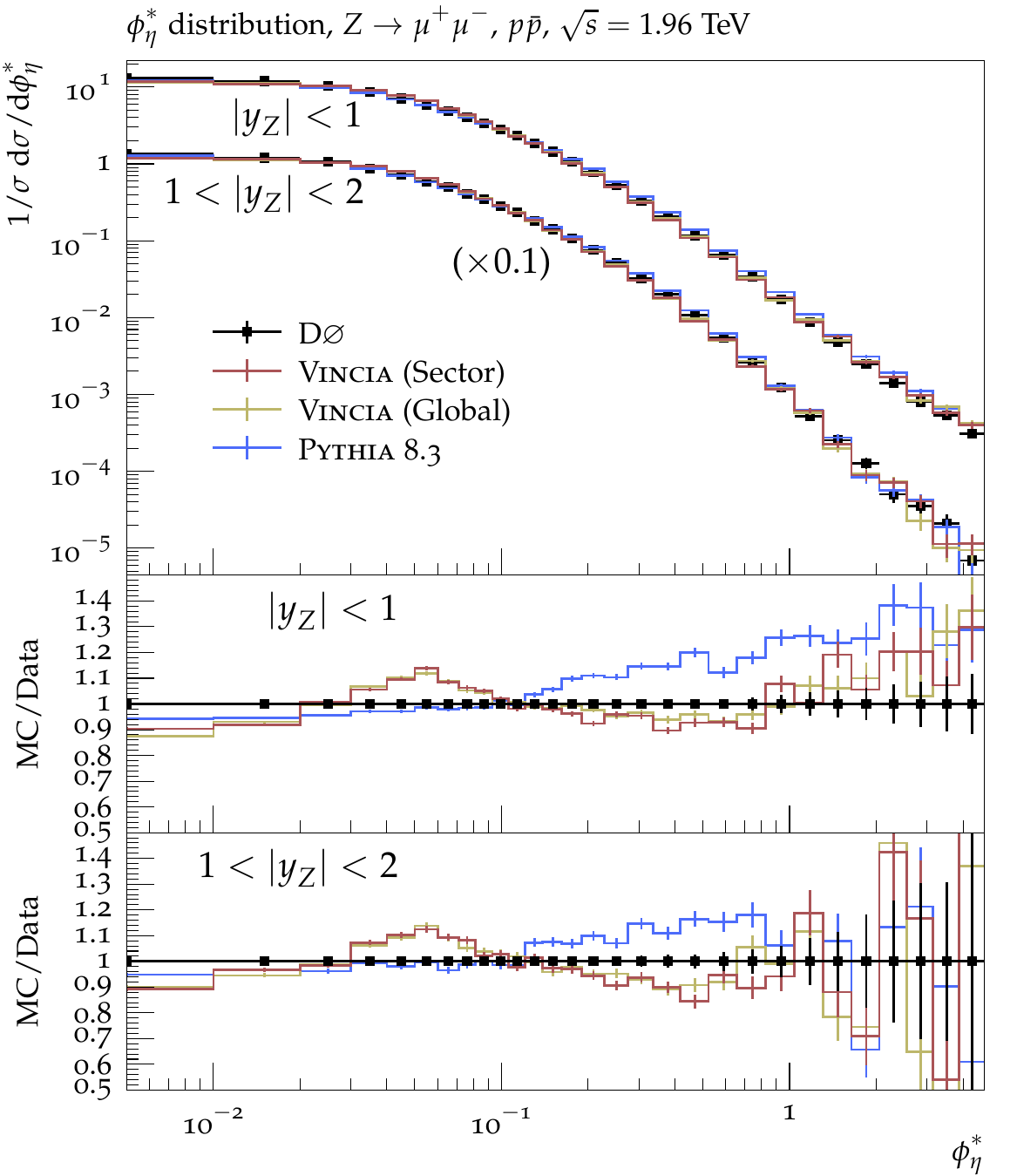}
    \caption{Distribution of the $\phi_\eta^*$ angle in the electron channel (\textit{left}) and muon channel (\textit{right}) for different rapidity bins in $p\bar p$ collisions at a 
centre-of-mass energy of $\sqrt{s} = 1.96~\tera e\volt$. Comparison of the sector shower against the default (global) \Vincia\ shower, \textsc{Pythia} 8.3 and D$\varnothing$ data \cite{Abazov:2010mk}.}
    \label{fig:phiZTevatron}
\end{figure}

As a first test of the sector shower implementation and specifically as a test of its FSR component, in \cref{fig:LEPEventShapes1} we show event-shape distributions  for hadronic $Z$ decays (see \cite{Ellis:1980wv,Donoghue:1979vi,Parisi:1978eg,Barber:1979yr,Farhi:1977sg,Brandt:1964sa} for definitions) compared to L3 measurements at $\sqrt{s} = 91.2$ GeV \cite{Achard:2004sv} for light- and heavy-flavour tagged samples separately. 

Although we are considering only pure shower predictions here, i.e., without matching or merging to fixed-order calculations, we find that both the sector and global  showers in \textsc{Vincia} give very good agreement with data.
All three showers considered here describe the thrust and $C$ parameter distributions equally well; the sector-shower predictions in decays to light flavours are in slightly closer agreement with data than the global shower.
The default \Pythia\ 8.3 shower is slightly closer to the experimentally measured distributions for the $D$ parameter in $Z$ decays to light flavours. We note, however, that its first emission is corrected to LO matrix elements and the second includes an approximate treatment of gluon-polarisation effects, whereas no such corrections are imposed on the 
\Vincia\ showers.
Given that the latter two agree with each other, we suspect this difference to mainly stem from the additional corrections that are implemented in \Pythia. This argument is supported by a study of the distribution obtained with the unmatched \Pythia\ 8.3 shower and the fact that the LO matched \Vincia\ 1 shower agreed better with data as well as \Pythia, cf.\ \cite{Giele:2011cb}. We plan to return to (iterative) matrix-element corrections and (multi-leg) matching/merging for \Vincia\ in a separate follow-up study. 

\begin{figure}[ht]
	\centering
	\includegraphics[width=0.49\textwidth]{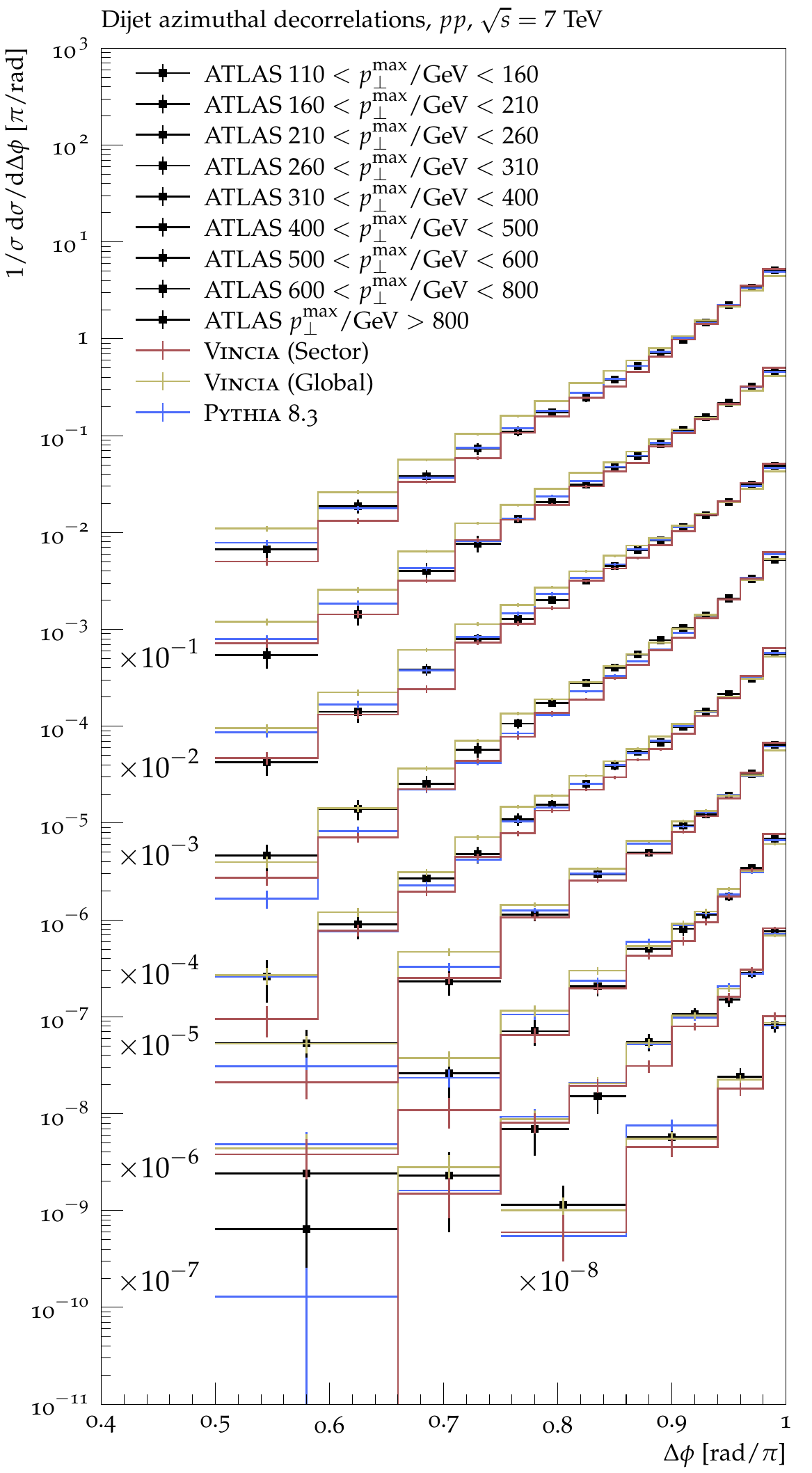}
	\includegraphics[width=0.49\textwidth]{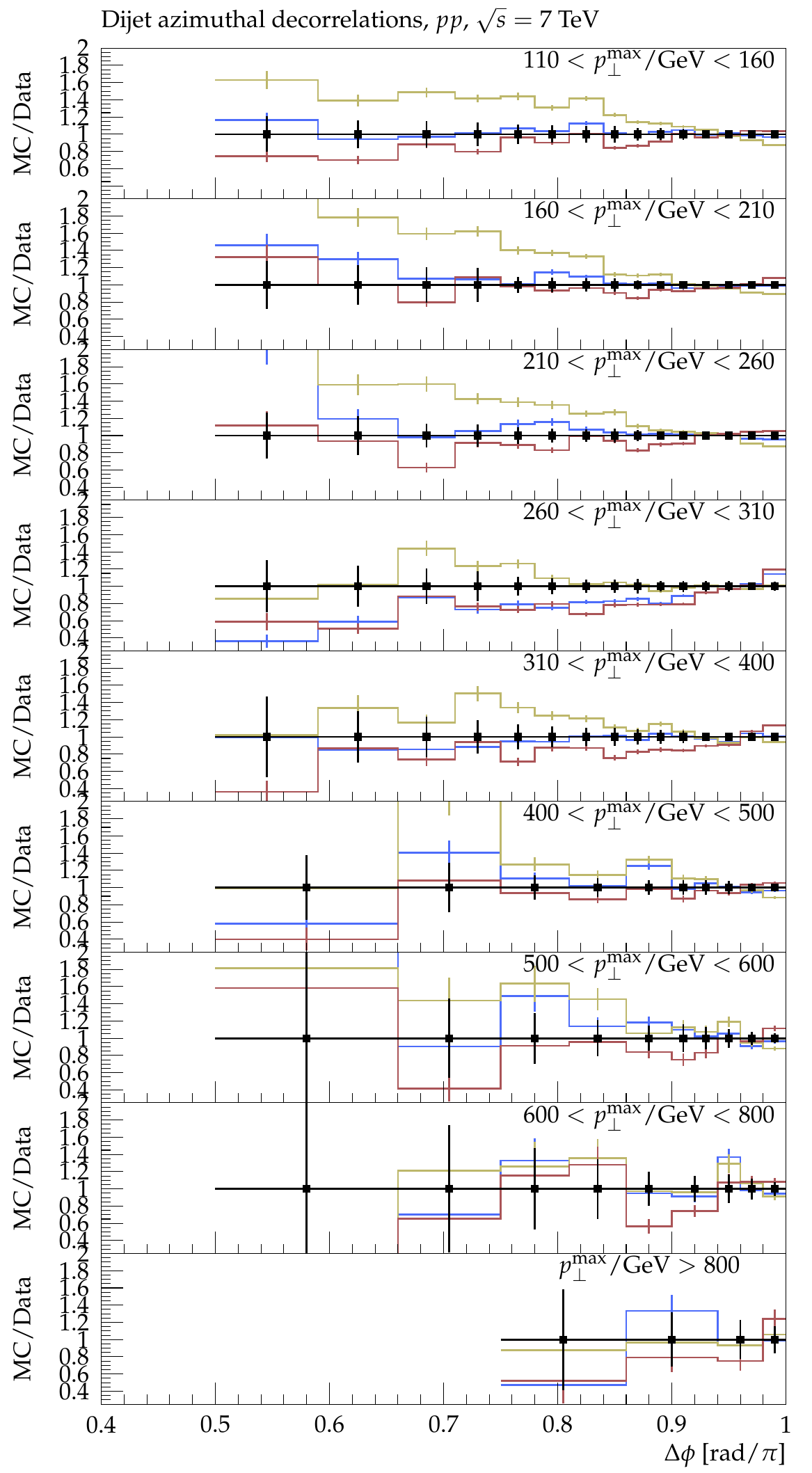}
	\caption{Dijet azimuthal decorrelations in $p p$ collisions at a centre-of-mass energy of $\sqrt{s} = 7~\tera e\volt$. Comparison of the sector shower against the default (global) \Vincia\ shower, \Pythia\ 8.3, and ATLAS data \cite{daCosta:2011ni}.}
	\label{fig:decorrelationAtlas}
\end{figure}

\paragraph{Drell-Yan}

To test the ISR component of the sector shower with minimal interplay with the FSR shower, we consider Drell-Yan processes as measured by 
the D$\varnothing$ experiment \cite{Abazov:2010mk} at $\sqrt{s} = 1.96~\tera e \volt$ in \cref{fig:phiZTevatron}.
We study the angle $\phi_\eta^*$, defined in \cite{Banfi:2010cf}, which relates to the opening angle of the Drell-Yan lepton pair and is sensitive to the $Z$ transverse momentum.

It is known that a pure unmatched parton shower with the evolution starting at the factorisation scale $\sim m_Z$ is incapable of describing these observables well, as they are sensitive to phase-space regions with $p_{\perp,\mathrm{jet}} > m_Z$. 
We here follow the default choice in both \Pythia\ and \Vincia\ and start showers off configurations containing no final-state (QCD) partons at the phase-space maximum.
As alluded to before, cf.\ \cref{sec:MERatios}, we leave a dedicated study of the impact of so-called \textquotedblleft power showers\textquotedblright\ in sector showers as well as matching and merging in the sector-shower framework to future work. 

\begin{figure}[ht]
	\centering
	\includegraphics[width=0.6\textwidth]{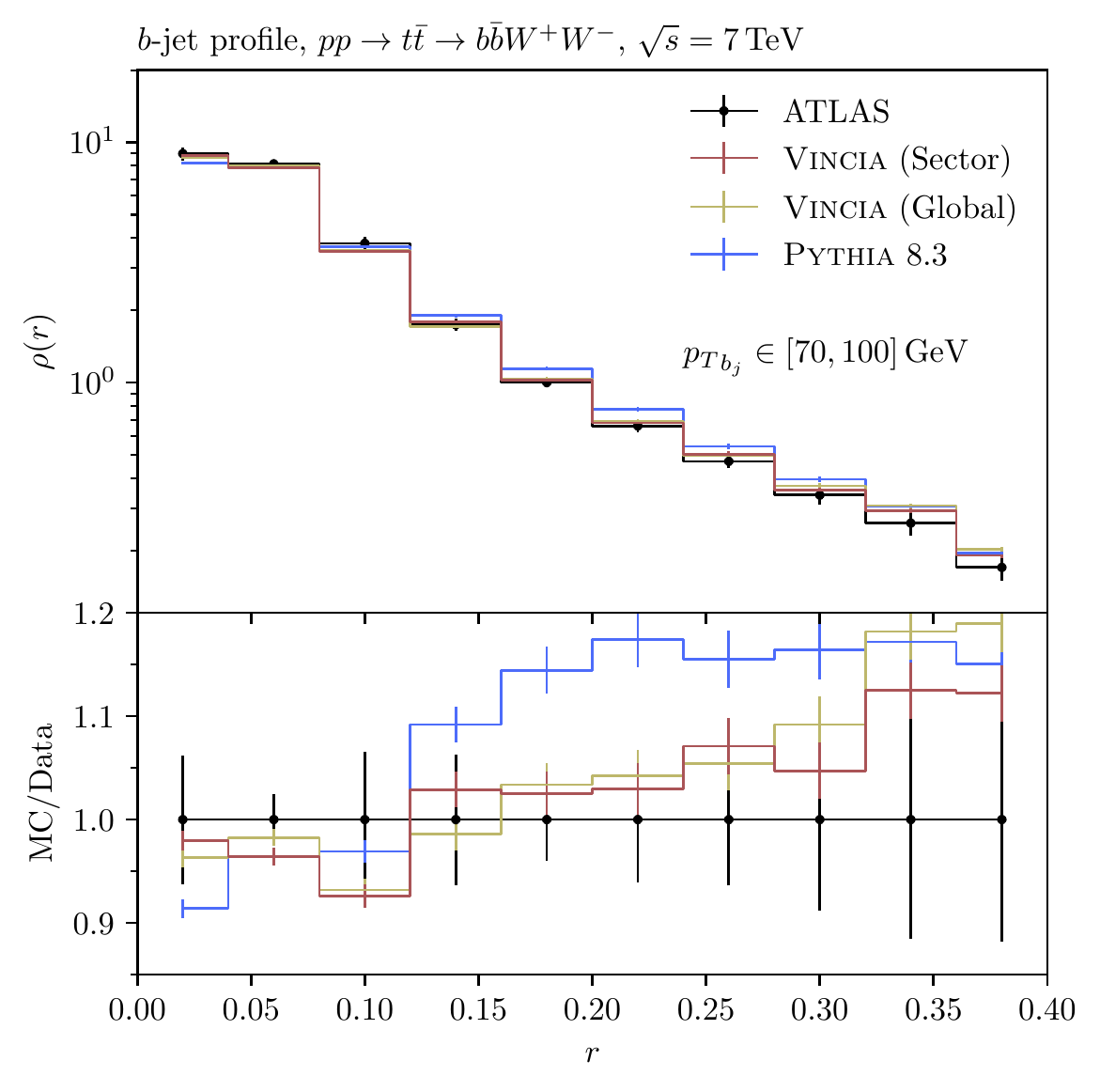}
	\caption{Comparison of the sector shower against the default (global) \Vincia\ shower, and \textsc{Pythia} 8.3 for the
		$b$-jet profile as measured by ATLAS in $pp \rightarrow t\bar{t}$ at $\sqrt{s}=7$ TeV \cite{Aad:2011kq}.}
	\label{fig:bjetprofileAtlas}
\end{figure}

With this choice, the data is very well described by the sector shower, although there is some larger deviation for large $\phi^*$. This is, however, the region a \textquotedblleft wimpy\textquotedblright\ shower starting at the factorisation scale would not describe well at all and fixed-order corrections have a great impact, cf.\ e.g.\ \cite{Fischer:2016Vincia,Hoche:2015sya}.
A peculiarity of Drell-Yan processes is that the first emission is described in the same way in the sector and global shower framework in \Vincia, as initial-state legs are sectorised in the first place. Only starting from the second branching do different initial-final sectors compete, and final-final ones do not show up until the third branching. This is observed in \cref{fig:phiZTevatron}, as there is virtually no difference between the predictions of the sector and global shower.

\paragraph{QCD Jets}

As a test of the complete sector shower with full interplay between its ISR and FSR components, we compare sector-shower predictions for dijet azimuthal decorrelations in proton-proton collisions at $\sqrt{s} = 7~\tera e \volt$ to measurements by the ATLAS experiment \cite{daCosta:2011ni} in \cref{fig:decorrelationAtlas}.

Of all analyses considered here, the different shower predictions differ the most for this observable.
Systematically, the sector shower is shifted towards smaller values compared to the global shower.
However, already for the first emission, many different sectors compete for the branching in the sector shower, cf.~\cref{sec:MERatios},
which may explain this bigger difference to the global shower than seen in the pure FSR shower and showers off Drell-Yan processes.
Moreover, dijet azimuthal decorrelations are sensitive to higher-order contributions and typically, NLO theory predictions are required to describe the data well, cf.~\cite{Wobisch:2015jea};
here however only shower predictions matched to at most LO for the first emission are considered.
The good agreement that we nevertheless see may be explained by noting that the distributions are normalised to the cross section.

\begin{figure}[ht]
 \centering
\includegraphics[width=0.6\textwidth]{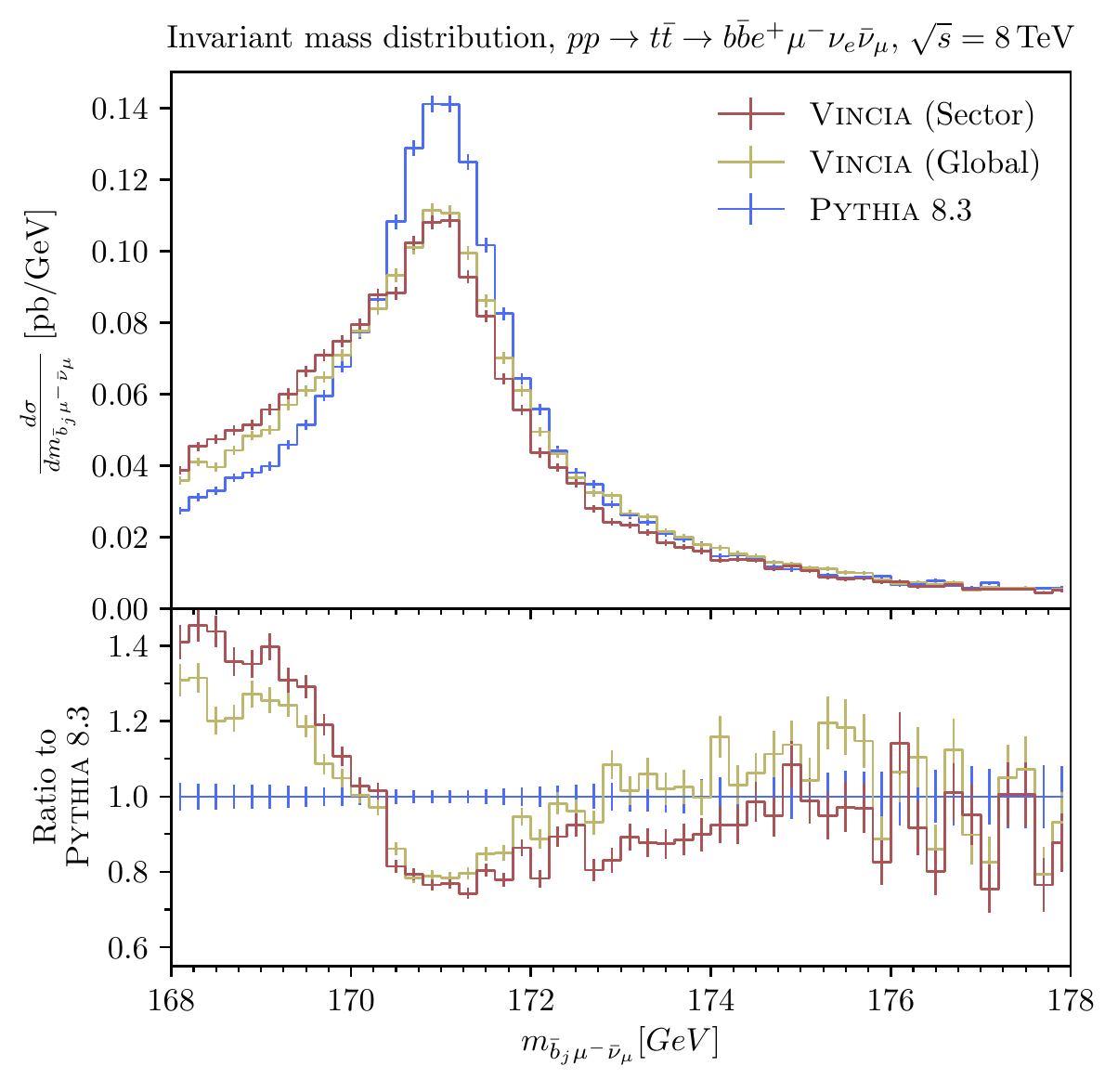}
\caption{Comparison of the sector shower against the default (global) \Vincia\ shower, and \Pythia\ 8.3 for invariant mass distribution of
the $\bar{t}$ decay system $\bar{b}_j \mu^- \bar{\nu}_\mu$ in dileptonic $pp \rightarrow t\bar{t}$ at $\sqrt{s}=8$ TeV .}
\label{fig:tdecmass}
\end{figure}

\paragraph{Coloured Resonance Decays}

As a final validation of the sector shower in resonance decays, we consider two observables already studied in \cite{Brooks:2019xso}, known to be sensitive to RF branchings. 
The first observable is the $b$-jet profile $\rho(r)$ as measured by ATLAS for $pp \rightarrow t\bar{t}$ at $\sqrt{s}=7$ TeV \cite{Aad:2011kq}.
As shown in \cref{fig:bjetprofileAtlas} we find that the sector shower reproduces the global shower results, namely that \Vincia gives rise to systematically narrower $b$-jets than \Pythia, 
and moreover that this is in closer agreement with the data.

The second observable we consider is the invariant mass of the $\bar{t}$ decay system $\bar{b}_j \ell^- \bar{\nu}_\ell$ in dileptonic $pp \rightarrow t\bar{t}$ at $\sqrt{s}=8$ TeV.
As already noted in \cite{Brooks:2019xso}, this observable is sensitive to numerous effects; therefore to isolate (and thereby highlight) any differences originating from the parton shower alone, we do the following. The analysis is performed at ``Monte Carlo truth'' level, namely the decay system is reconstructed with the correct pairing of leptons and $b$-jets, and the 
four-momentum of the neutrino is presumed to be known. In addition, we switch off 
the simulation of MPI, beam-remnant modelling, and hadronisation\footnote{In addition, we set \texttt{SigmaProcess:alphaSorder} and \texttt{SigmaProcess:alphaSvalue} to their default values in \Pythia~(such that the starting cross section is identical).}.
These actions strengthen our conclusion that the sector shower is consistent with the global shower, as demonstrated in \cref{fig:tdecmass}. Although there is again a small 
systematic shift of the sector shower towards smaller invariant masses relative to the global shower, this is well within an acceptable range of variation for showers of leading-logarithmic accuracy.
Indeed, the difference between the two \Vincia\ showers is significantly smaller than their mutual difference with respect to the predictions of \Pythia\ 8.3.
The variation between \Vincia\ and \Pythia\ is interesting. It was discussed in some detail in~\cite{Brooks:2019xso}; here we produce a brief summary.
In the \Pythia\ shower model, the recoil from each FSR emission must be taken by a single final-state parton. For the first gluon emission in $t\to bW$, therefore, the  $W$ is formally assigned to be the recoiler for emissions off the $b$ quark. At this stage, this choice is fully equivalent to \Vincia's RF kinematics (since $p_t - p_b = p_W$). After the first branching, the colour index of the top quark is now carried by the emitted gluon. In \Vincia, the ``recoiler'' for further emissions off that colour index effectively remains the crossed top minus the radiating parton,  $p_t-p_g = p_W+p_b$. In \Pythia, instead, one must choose it to be either the $W$ or the $b$. This choice is controlled by the flag \texttt{RecoilToColoured} whose default value \texttt{true} means that in ambiguous cases like this one the coloured parton will be picked over the uncoloured one. (This is to avoid imparting undesirably large recoil effects from QCD branchings onto non-QCD partons.) 
Due to the collinear enhancement for the preceding $b\to bg$ emission, however, the invariant mass of the $bg$ dipole (and hence the phase space for further radiation) tends to be small. This severely restricts further radiation and results in less ``out-of-cone'' radiation loss in \Pythia\ and hence a sharper peak structure in \cref{fig:tdecmass}. 
It should be clear from our discussion that we  consider this effect in \Pythia\ to be an artefact of having to choose between two undesirable options; either allowing an uncoloured parton to continue to receive the full recoil of further QCD branchings in the RF dipole (\texttt{RecoilToColoured = false}) or choosing a coloured one with a tiny phase space (\texttt{RecoilToColoured = true}) to receive it. We believe that \Vincia's coherent treatment of RF recoils is more physical. We also note that, since this difference arises in the treatment of the second emission and beyond, it is of NNLO origin in a fixed-order expansion and will persist even when the decay process is matched to NLO calculations, see~\cite{FerrarioRavasio:2019vmq,Brooks:2019xso}.

\section{Conclusion and Outlook}\label{sec:Conclusion}
We here presented the first implementation of an ISR and FSR parton shower that possesses only a single shower history. 
We extensively reviewed the shower algorithm which is ordered in a generalised \Ariadne\ transverse momentum and presented the construction of all required sector antenna functions 
from \Vincia's global final-final antenna functions; we also defined a sector resolution variable, which enables us to uniquely invert the shower algorithm.

The sector-shower implementation was validated against LO matrix elements and first predictions for physical observables were made and compared to the global \Vincia\ antenna shower, and to \Pythia\ 8.3. Where relevant, we emphasised the expected impact of matrix-element corrections, which are available in \Pythia\ e.g.\ for resonance decays and Drell-Yan processes, but are not yet available in the current version of \Vincia.

We found very good agreement of sector-shower predictions with experimental data from LEP, Tevatron, and LHC experiments, taking into account that only \textquotedblleft pure shower\textquotedblright\ predictions, formally at LL accuracy, were considered.
Moreover, we showed that, at least for the observables studied here, there is no disadvantage of the sector shower in describing data in comparison to the global \Vincia\ shower.

While this may appear to be ``just another LL shower'', it is worth pointing out that the sector-shower approach provides the means to significantly reduce the computational overhead connected to matching and merging shower predictions to fixed-order calculations and, moreover,
to significantly reduce the complexity of shower algorithms.
The presentation here may therefore be understood as a first step towards dedicated and highly-efficient matrix-element-correction and merging schemes, which we discuss in separate forthcoming publications.
We also plan to explore whether the sector approach can facilitate the inclusion of direct higher-order splittings and one-loop antenna functions in the shower, e.g.\ following a similar approach as those of~\cite{Hartgring:2013jma,Li:2016yez}, with the ultimate goal of reaching higher formal precision in parton showers, cf.~\cite{Hoche:2017iem,Hoche:2017hno,Nagy:2017ggp,Dulat:2018vuy,Dasgupta:2018nvj,beyondLL}.

The sector antenna shower presented here will be made publicly available as part of the \Vincia\ antenna shower in a forthcoming update to \Pythia\ 8.3.

\section*{Acknowledgements}
We acknowledge support from the Monash eResearch Centre and
eSolutions-Research Support Services through the MonARCH HPC
Cluster. 
CTP is supported by the Monash Graduate Scholarship, the Monash International Postgraduate Research Scholarship, and the J.~L.~William Scholarship.
HB is funded by the Australian Research Council via Discovery
Project DP170100708 -- \textquotedblleft Emergent Phenomena in Quantum
Chromodynamics\textquotedblright. 
This work was also supported in part
by the European Union's Horizon 2020 research and innovation programme
under the Marie Sklodowska-Curie grant agreement No 722105 --
MCnetITN3.

\clearpage

\appendix

\section{Helicity-Dependent Antenna Functions}
\label{sec:AppendixAntennaFcts}
In this section, we collect all helicity- and mass-dependent global antenna functions. The corresponding sector antenna functions can be build by the considerations presented in section \cref{sec:SectorAntennae}.

\subsection{Final-Final Antennae} 

\paragraph{QQEmitFF} The helicity-averaged antenna function for the process $q_Iq_K \mapsto q_i g_j q_k$ is is
\begin{align}
    \bar a_{g/q\bar q}^\text{FF,gl}(\yij, \yjk) = \frac{1}{\sIK}\left[ \frac{(1-\yij)^2 + (1-\yjk)^2}{\yij \yjk} - \frac{2 \mu_I^2}{\yij^2} - \frac{2 \mu_{K}^2}{\yjk^2} + 1 \right]
\end{align}
with the individual helicity contributions
\begin{align}
    \bar a_{q_+\bar q_+\mapsto q_+g_+\bar q_+}^\text{FF,gl} =\,& \frac{1}{\sIK} \left[ \frac{1}{\yij \yjk} - \frac{\mu_i^2}{\yij^2(1-\yjk)} - \frac{\mu_k^2}{\yjk^2(1-\yij)} \right], \\
    \bar a_{q_+\bar q_+\mapsto q_+g_-\bar q_+}^\text{FF,gl} = \,& \frac{1}{\sIK} \left[ \frac{(1-\yij)^2 + (1-\yjk)^2 -1}{\yij \yjk} + 2 \right. \\ 
    &\left. \qquad - \frac{\mu_i^2(1-\yjk)}{\yij^2} - \frac{\mu_k^2(1-\yij)}{\yjk^2} \right], \nonumber \\
    \bar a_{q_+\bar q_+\mapsto q_-g_+\bar q_+}^\text{FF,gl} = \,& \frac{1}{\sIK} \frac{\mu_i^2 \yjk^2}{\yij^2}\frac{1}{1-\yjk}, \\
    \bar a_{q_+\bar q_+\mapsto q_+g_+\bar q_-}^\text{FF,gl} = \,& \frac{1}{\sIK} \frac{\mu_k^2 \yij^2}{\yjk^2} \frac{1}{1-\yij}, \\
    \bar a_{q_+\bar q_-\mapsto q_+g_+\bar q_-}^\text{FF,gl} = \,& \frac{1}{\sIK} \left[ \frac{(1-\yij)^2}{\yij\yjk} - \frac{\mu_i^2}{\yij^2(1 - \yjk)} - \frac{\mu_k^2(1 - \yij)}{\yjk^2}\right], \\
    \bar a_{q_+\bar q_-\mapsto q_+g_-\bar q_-}^\text{FF,gl} = \,& \frac{1}{\sIK} \left[ \frac{(1-\yjk)^2}{\yij \yjk} - \frac{\mu_i^2(1 - \yjk)}{\yij^2} - \frac{\mu_k^2}{\yjk^2(1 - \yij)}\right], \\
    \bar a_{q_+\bar q_-\mapsto q_-g_+\bar q_-}^\text{FF,gl} = \,& \frac{1}{\sIK} \frac{\mu_i^2 \yjk^2}{\yij^2(1 - \yjk)}, \\
    \bar a_{q_+\bar q_-\mapsto q_+g_-\bar q_+}^\text{FF,gl} = \,& \frac{1}{\sIK} \frac{\mu_k^2 \yij^2}{\yjk^2(1 - \yij)}.
\end{align}

\paragraph{QGEmitFF} The helicity-averaged antenna function for the process $q_Ig_K \mapsto q_i g_j g_k$ is
\begin{align}
    \bar a_{g/qg}^\text{FF,gl}(\yij, \yjk) =\,& \frac{1}{\sIK}\left[ \frac{(1-\yij)^3 + (1-\yjk)^2}{\yij \yjk} - \frac{2 \mu_I^2}{\yij^2} \right. \\ 
    & \qquad + (1-\alpha)\frac{(1-\yjk)(\yik-\yij)}{\yjk} + 2 - \yij - \frac{\yjk}{2}\Bigg] \nonumber
\end{align}
with the individual helicity contributions
\begin{align}
  \bar a_{q_+g_+\mapsto q_+g_+g_+}^\text{FF,gl} = \,& \frac{1}{\sIK} \Bigg[\frac{1}{\yij \yjk} + (1-\alpha)(1-\yjk)\left(\frac{1-2\yij-\yjk}{\yjk} \right) \\ 
  &\left. \qquad  - \frac{\mu_i^2}{\yij^2(1 - \yjk)} \right], \nonumber \\
  \bar a_{q_+g_+\mapsto q_+g_-g_+}^\text{FF,gl} = \,& \frac{1}{\sIK} \left[ \frac{(1-\yij)\yik^2}{\yij \yjk} - \frac{\mu_i^2 (1 - \yjk)}{\yij^2} \right], \\
  \bar a_{q_+g_+\mapsto q_-g_+g_+}^\text{FF,gl} = \,& \frac{1}{\sIK} \frac{\mu_i^2 \yjk^2}{\yij^2(1-\yjk)},\\
  \bar a_{q_+g_-\mapsto q_+g_+g_-}^\text{FF,gl} = \,& \frac{1}{\sIK} \left[ \frac{(1-\yij)^3}{\yij \yjk} - \frac{\mu_i^2}{\yij^2(1-\yjk)} \right],\\
  \bar a_{q_+g_-\mapsto q_+g_-g_-}^\text{FF,gl} = \,& \frac{1}{\sIK} \Bigg[ \frac{(1-\yjk)^2}{\yij \yjk} + (1-\alpha)(1-\yjk)\left(\frac{1-2\yij-\yjk}{\yjk} \right) \\ 
  &\left.\qquad  - \frac{\mu_i^2 (1-\yjk)}{\yij^2} \right], \nonumber \\
  \bar a_{q_+g_-\mapsto q_-g_+g_-}^\text{FF,gl} = \,& \frac{1}{\sIK} \frac{\mu_i^2 \yjk^2}{\yij^2 (1-\yjk)}.
\end{align}

\paragraph{GGEmitFF} The helicity-averaged antenna function for the process $g_Ig_K \mapsto g_i g_j g_k$ is
\begin{align}
    \bar a_{g/gg}^\text{FF,gl}(\yij, \yjk) = \,& \frac{1}{\sIK}\left[ \frac{(1-\yij)^3 + (1-\yjk)^3}{\yij \yjk} + (1-\alpha)\frac{(1-\yij)(2\yik-2\yjk+\yij)}{2\yij} \right. \nonumber \\ 
    & \qquad  + (1-\alpha)\frac{(1-\yjk)(2\yik-2\yij+\yjk)}{2\yjk} \\ 
    & \left. \qquad + 3 - \frac{3}{2}\yij - \frac{3}{2} \yjk \right] \nonumber
\end{align}
with the individual helicity contributions
\begin{align}
    \bar a_{g_+g_+\mapsto g_+g_+g_+}^\text{FF,gl} = \,& \frac{1}{\sIK} \left[ \frac{1}{\yij\yjk}\right.\\
    & \left.\qquad + (1-\alpha)\left((1-\yij)\frac{1-2\yjk-\yij}{\yij} + (1-\yjk)\frac{1-2\yij-\yjk}{\yjk}\right)\right]\nonumber \\
    \bar a_{g_+g_+\mapsto g_+g_-g_+}^\text{FF,gl} = \,& \frac{1}{\sIK} \frac{\yik^3}{\yij \yjk} \\
    \bar a_{g_+g_-\mapsto g_+g_+g_-}^\text{FF,gl} = \,& \frac{1}{\sIK} \left[ \frac{ (1-\yij)^3}{\yij \yjk} +(1-\alpha)(1-\yij)\frac{1-2\yjk}{\yij} \right],\\
    \bar a_{g_+g_-\mapsto g_+g_-g_-}^\text{FF,gl} = \,& \frac{1}{\sIK} \left[ \frac{ (1-\yjk)^3}{\yij \yjk} + (1-\alpha)(1-\yjk)\frac{1-2\yij}{\yjk} \right].
\end{align}

\paragraph{GXSplitFF} The helicity-averaged antenna function for the process $g_IX_K \mapsto q_i \bar q_j X_k$ with an arbitrary coloured spectator $X$ is
\begin{align}
  \bar a_{\bar q/gX}^\text{FF,gl}(\yij, \yjk) & = \frac{1}{\sIK} \frac{1}{2}\frac{1}{\yij + 2\mu_q^2}\left[ \yik^2 + \yjk^2 +\frac{2 \mu_q^2}{\yij + 2\mu_q^2}\right]
\end{align}
with the individual helicity contributions
\begin{align}
  \bar a_{g_+X\mapsto q_+\bar q_-X}^\text{FF,gl} & = \frac{1}{\sIK}\frac{1}{2(\yij + 2\mu_q^2)}\left[ \yik^2 - \frac{\mu_q^2}{\yij + 2\mu_q^2}\frac{\yik}{1-\yik} \right],\\
  \bar a_{g_+X\mapsto q_-\bar q_+X}^\text{FF,gl} & = \frac{1}{\sIK}\frac{1}{2(\yij + 2\mu_q^2)}\left[\yjk^2 -\frac{\mu_q^2}{\yij + 2\mu_q^2}\frac{\yjk}{1-\yjk} \right],\\
  \bar a_{g_+X\mapsto q_+\bar q_+X}^\text{FF,gl} & = \frac{1}{\sIK}\frac{\mu_q^2}{2(\yij + 2\mu_q^2)^2}\left[\frac{\yik}{1-\yik} + \frac{\yjk}{1-\yjk} + 2 \right].
\end{align}

\subsection{Initial-Final Antennae} 

\paragraph{QQEmitIF} The helicity-averaged antenna function for the process $q_A q_K \mapsto q_a g_j q_k$ is 
\begin{align} 
    \bar a_{g/qq}^\text{IF,gl}(\yaj, \yjk) =\,& \frac{1}{\sAK}\left[\frac{(1-\yaj)^2 + (1-\yjk)^2}{\yaj \yjk} +\frac{1}{2}(2-\yaj)(2-\yjk) \right. \\
    & \left. \qquad-\frac{2\mu^2_a}{\yaj^2}\left((1-\yjk)\left(1- \frac{\yaj}{2}\right) -\frac{\yaj}{2}(1-\yaj)  \right) \right. \nonumber \\
    &\left. \qquad - \frac{2\mu_k^2}{\yjk^2}\left(1 - \frac{\yjk}{4}(2-\yjk)\left(2+\frac{\yaj^2}{1-\yaj}\right)\right) \right] \nonumber
\end{align}
with the individual helicity contributions
\begin{align}
    \bar a_{q_+q_+\mapsto q_+g_+q_+}^\text{IF,gl} =\,& \frac{1}{\sAK} \left[ \frac{1}{\yaj \yjk}-\frac{\mu^2_a}{\yaj^2} - \frac{\mu^2_k}{(1-\yaj)\yjk^2} \right], \\
    \bar a_{q_+q_+\mapsto q_+g_-q_+}^\text{IF,gl} =\,& \frac{1}{\sAK} \left[ \frac{(1-\yaj)^2 + [(1-\yjk)^2 - 1] (1-\yaj)^2}{\yaj\yjk} \right. \\    
    & \left. \qquad -\frac{\mu^2_a (1-\yjk-\yaj)^2}{\yaj^2} - \frac{\mu^2_k (1-\yaj)(1-\yjk)^2}{\yjk^2} \right],\\
    \bar a_{q_+q_+\mapsto q_-g_-q_+}^\text{IF,gl} =\,& \frac{1}{\sAK} \frac{\mu_a^2 \yjk^2}{\yaj^2}, \\
    \bar a_{q_+q_+\mapsto q_+g_+q_-}^\text{IF,gl} =\,& \frac{1}{\sAK} \frac{\mu_k^2 \yaj^2}{(1-\yaj)\yjk^2},\\
    \bar a_{q_+q_-\mapsto q_+g_+q_-}^\text{IF,gl} =\,& \frac{1}{\sAK} \left[ \frac{(1-\yaj)^2}{\yaj \yjk} - \frac{\mu_a^2(1-\yaj)}{\yaj^2} - \frac{\mu_k^2(1-\yaj)}{\yjk^2} \right], \\
    \bar a_{q_+q_-\mapsto q_+g_-q_-}^\text{IF,gl} =\,& \frac{1}{\sAK} \left[ \frac{(1-\yjk)^2}{\yaj \yjk} - \frac{\mu_a^2(1-\yjk)^2}{\yaj^2} - \frac{\mu_k^2(1-\yjk)^2}{\yjk^2(1-\yaj)} \right], \\
    \bar a_{q_+q_-\mapsto q_-g_-q_-}^\text{IF,gl} =\,& \frac{1}{\sAK} \frac{\mu^2_a \yjk^2}{\yaj^2}, \\
    \bar a_{q_+q_-\mapsto q_+g_-q_+}^\text{IF,gl} =\,& \frac{1}{\sAK} \frac{\mu^2_k \yaj^2}{\yjk^2 (1-\yaj)}.
\end{align}

\paragraph{QGEmitIF} The helicity-averaged antenna function for the process $q_A g_K \mapsto q_a g_j g_k$ is 
\begin{align}
    \bar a_{g/qg}^\text{IF,gl}(\yaj, \yjk) =\,& \frac{1}{\sAK} \left[\frac{(1-\yaj)^3 + (1-\yjk)^2}{\yaj \yjk} + (1-\alpha) \frac{1-2\yaj}{\yjk} \right. \\
    & \qquad - \frac{2\mu^2_a}{y_{aj}^2} \left( (1-\yjk) - \frac{\yaj}{2}\left[ (1-\yaj) - (2-\yjk)^2\right] \right) \nonumber \\
     & \left. \qquad  + \frac{3}{2} + \yaj - \frac{\yjk}{2} - \frac{\yaj^2}{2} \right] \nonumber
\end{align} 
with the individual helicity contributions
\begin{align}
    \bar a_{q_+g_+\mapsto q_+g_+g_+}^\text{IF,gl} =\,& \frac{1}{\sAK} \left[ \frac{1}{\yaj \yjk} + (1-\alpha) \frac{1-2\yaj}{\yjk} - \frac{\mu^2_a}{\yaj^2}\right], \\
    \bar a_{q_+g_+\mapsto q_+g_-g_+}^\text{IF,gl} =\,& \frac{1}{\sAK} \left[ \frac{(1-\yaj)^3 + (1-\yjk)^2 - 1}{\yaj \yjk} \right. \\
    &\left.\qquad - \frac{\mu_a^2(1-\yjk-\yaj)^2(1-\yaj)}{\yaj^2} + 3 - \yaj^2\right], \nonumber \\
    \bar a_{q_+g_+\mapsto q_-g_-g_+}^\text{IF,gl} =\,& \frac{1}{\sAK} \frac{\mu^2_a \yjk^2}{\yaj^2}, \\
    \bar a_{q_+g_-\mapsto q_+g_+g_-}^\text{IF,gl} =\,& \frac{1}{\sAK} \left[ \frac{(1-\yaj)^3}{\yaj\yjk} - \frac{\mu_a^2(1-\yaj)^2}{\yaj^2} \right], \\
    \bar a_{q_+g_-\mapsto q_+g_-g_-}^\text{IF,gl} =\,& \frac{1}{\sAK} \left[\frac{(1-\yjk)^2}{\yaj \yjk} + (1-\alpha) \frac{1-2\yaj}{\yjk} \right.\\
    & \left. \qquad - \frac{\mu_a^2(1-\yjk)^2}{\yaj^2} + 2\yaj - \yjk \right],\\ 
    \bar a_{q_+g_-\mapsto q_-g_-g_-}^\text{IF,gl} =\,& \frac{1}{\sAK} \frac{\mu^2_a \yjk^2}{\yaj^2}.
\end{align}

\paragraph{GQEmitIF} The helicity-averaged antenna function for the process $g_A q_K \mapsto g_a g_j q_k$ is
\begin{align}
  \bar a_{g/gq}^\text{IF,gl}(\yaj, \yjk) =\,& \frac{1}{\sAK}\left[\frac{(1-\yjk)^3 + (1-\yaj)^2}{\yaj\yjk} + \frac{1+\yjk^3}{\yaj(1-\yjk)}\right. \\
  &\left. \qquad - \frac{2\mu_k^2}{\yjk^2}\left(1 - \frac{\yjk}{4}(3-3\yjk^2+\yjk^3)\left(2+\frac{\yaj^2}{1-\yaj}\right)\right) \right. \nonumber \\
  &\left. \qquad +\frac{1}{2}(2-\yaj)(3-\yjk+\yjk^2) \right] \nonumber
\end{align} 
with the individual helicity contributions
\begin{align}
    \bar a_{g_+q_+\mapsto g_+g_+q_+}^\text{IF,gl} =\,& \frac{1}{\sAK} \left[ \frac{1}{\yaj \yjk} + \frac{1}{\yaj (1 - \yjk)} - \frac{\mu_k^2}{\yjk^2(1-\yaj)} \right], \\
    \bar a_{g_+q_+\mapsto g_+g_-q_+}^\text{IF,gl} =\,& \frac{1}{\sAK} \left[\frac{(1-\yaj)^2 + [(1-\yjk)^3 - 1](1-\yaj)^2}{\yaj\yjk} \right. \\
    & \left. \qquad - \frac{\mu_k^2(1-\yaj)(1-\yjk)^3}{\yjk^2} \right],\\
    \bar a_{g_+q_+\mapsto g_-g_-q_+}^\text{IF,gl} =\,& \frac{1}{\sAK} \frac{\yjk^3}{\yaj (1-\yjk)}, \\
    \bar a_{g_+q_+\mapsto g_+g_+q_-}^\text{IF,gl} =\,& \frac{1}{\sAK} \frac{\mu_k^2 \yaj^2}{\yjk^2(1-\yaj)},\\
    \bar a_{g_+q_-\mapsto g_+g_+q_-}^\text{IF,gl} =\,& \frac{1}{\sAK} \left[ \frac{(1-\yaj)^2}{\yaj \yjk} + \frac{1}{\yaj (1-\yjk)}- \frac{\mu_k^2(1-\yaj)}{\yjk^2} \right],\\
    \bar a_{g_+q_-\mapsto g_+g_-q_-}^\text{IF,gl} =\,& \frac{1}{\sAK}\left[\frac{(1-\yjk)^3}{\yaj
    \yjk}  -\frac{\mu_k^2(1-\yjk)^3}{\yjk^2(1-\yaj)}\right].
\end{align}

\paragraph{GGEmitIF} The helicity-averaged antenna function for the process $g_A g_K \mapsto g_a g_j g_k$ is
\begin{align}
  \bar a_{g/gg}^\text{IF,gl}(\yaj, \yjk) =\,& \frac{1}{\sAK} \left[ \frac{(1-\yaj)^3 + (1-\yjk)^3}{\yaj\yjk} + \frac{1 + \yjk^3}{\yaj (1-\yjk)} \right.\\ 
  & \left. \qquad + (1-\alpha)\frac{1-2 \yaj}{\yjk} + 3 - 2 \yjk \right] \nonumber
\end{align}
with the helicity contributions
\begin{align}
    \bar a_{g_+g_+\mapsto g_+g_+g_+}^\text{IF,gl} & = \frac{1}{\sAK} \left[ \frac{1}{\yaj \yjk} + (1-\alpha) \frac{1-2\yaj}{\yjk} + \frac{1}{\yaj (1-\yjk)} \right], \\
    \bar a_{g_+g_+\mapsto g_+g_-g_+}^\text{IF,gl} & = \frac{1}{\sAK} \left[ \frac{(1-\yaj)^3 + (1-\yjk)^3-1}{\yaj \yjk} + 6 - 3\yaj - 3\yjk + \yaj\yjk \right], \\
    \bar a_{g_+g_-\mapsto g_+g_+g_-}^\text{IF,gl} & = \frac{1}{\sAK} \left[ \frac{(1-\yaj)^3}{\yaj \yjk} +  \frac{1}{\yaj (1-\yjk)} \right],\\
    \bar a_{g_+g_-\mapsto g_+g_-g_-}^\text{IF,gl} & =  \frac{1}{\sAK} \left[ \frac{(1-\yjk)^3}{\yaj \yjk} + (1-\alpha)\frac{1-2\yaj}{\yjk} + 3\yaj - \yjk - \yaj \yjk \right].
\end{align}

\paragraph{XGSplitIF} The helicity-averaged antenna function for the process $X_A g_K \mapsto X_a q_j \bar q_k$, where $X$ is an arbitrary coloured parton in the initial state, is
\begin{equation}
    \bar a_{\bar{q}/Xg}^\text{IF,gl}(\yaj, \yjk) = \frac{1}{\sAK} \frac{\yAK}{2(\yjk + 2\mu_j^2)} \left[ \yak^2 + \yaj^2 + \frac{2\mu_j^2}{\yjk+2\mu_j^2}\right]~.
\end{equation}
with the individual helicity contributions
\begin{align}
    \bar a_{Xg_+\mapsto X\bar q_-q_+}^\text{IF,gl} & = \frac{1}{\sAK} \frac{1}{2}\frac{\yAK}{\yjk+2\mu_j^2}\left[ \yak^2 -\frac{\mu_j^2 \yak}{\yjk+2\mu_j^2(1-\yak)} \right], \\
    \bar a_{Xg_+\mapsto X\bar q_+q_-}^\text{IF,gl} & = \frac{1}{\sAK} \frac{1}{2} \frac{\yAK}{\yjk+2\mu_j^2}\left[ \yaj^2 -\frac{\mu_j^2 \yaj}{\yjk+2\mu_j^2(1-\yaj)} \right],\\
    \bar a_{Xg_+\mapsto X\bar q_+q_+}^\text{IF,gl} & = \frac{1}{\sAK} \frac{1}{2} \frac{\mu_j^2 \yAK}{(\yjk + 2\mu_j^2)^2}\left[\frac{\yaj}{(1-\yaj)} + \frac{\yak}{(1-\yak)} + 2 \right].
\end{align}

\paragraph{QXSplitIF} The helicity-averaged antenna function for the process $q_A X_K \mapsto g_a \bar q_j X_k$, where $X$ is an arbitrary coloured parton in the final state, is
\begin{align}
    \bar a_{\bar q/qX}^\text{IF,gl}(\yaj, \yjk) & = \frac{1}{\sAK} \left[ \frac{\yAK^2 + (1-\yAK)^2}{\yaj} + \frac{2 \mu_j^2 \yAK}{\yaj^2}\right]
\end{align}
with the individual helicity contributions
\begin{align}
    \bar a_{q_+X \mapsto g_+\bar q_-X}^\text{IF,gl} & = \frac{1}{\sAK}\left[ \frac{\yAK^2}{\yaj} - \frac{\mu_j^2 \yAK^2}{\yaj^2(1-\yAK)}\right],\\
    \bar a_{q_+X \mapsto g_-\bar q_-X}^\text{IF,gl} & = \frac{1}{\sAK}\left[ \frac{(1-\yAK)^2}{\yaj} - \frac{\mu_j^2(1-\yAK)}{\yaj^2}\right],\\
    \bar a_{q_+X \mapsto g_+\bar q_+X}^\text{IF,gl} & = \frac{1}{\sAK} \frac{\mu_j^2}{\yaj^2(1-\yAK)}.
\end{align}

\paragraph{GXConvIF} The helicity-averaged antenna function for the process $g_A X_K \mapsto q_a q_j X_k$, where $X$ is an arbitrary coloured parton in the final state, is
\begin{align}
    \bar a_{q/gX}^\text{IF,gl}(\yaj, \yjk) & = \frac{1}{\sAK}\frac{1}{2}\left[ \frac{1 + (1-\yAK)^2}{\yAK(\yaj - 2\mu_j^2)} - \frac{2\mu_j^2 \yAK}{(\yaj-2\mu_j^2)^2} \right].
\end{align}
with the individual helicity contributions
\begin{align}
    \bar a_{g_+X \mapsto q_+q_+X}^\text{IF,gl} & = \frac{1}{\sAK}\frac{1}{2}\left[ \frac{1}{\yAK(\yaj-2\mu_j^2)} - \frac{\mu_j^2 \yAK}{(\yaj-2\mu_j^2)^2(1-\yAK)}\right],\\
    \bar a_{g_+X \mapsto q_-q_-X}^\text{IF,gl} & = \frac{1}{\sAK}\frac{1}{2}\left[ \frac{(1-\yAK)^2}{\yAK(\yaj-2\mu_j^2)}
    -\frac{\mu_j^2 \yAK(1-\yAK)}{(\yaj-2\mu_j)^2}\right],\\
    \bar a_{g_+X \mapsto q_+q_-X}^\text{IF,gl} & = \frac{1}{\sAK}\frac{1}{2} \frac{\mu^2_j}{(\yaj - 2\mu_j^2)^2} \frac{\yAK^3}{1-\yAK}.
\end{align}

\subsection{Initial-Initial Antennae} 

\paragraph{QQEmitII} The helicity-averaged antenna function for the process $\bar q_A q_B \mapsto q_a g_j q_b$ is 
\begin{align}
    \bar a_{g/\bar q q}^\text{II,gl}(\yaj, \yjb) =\,& \frac{1}{\sAB}\left[ \frac{(1-\yaj)^2 + (1-\yjb)^2}{\yaj \yjb} + 1 -\frac{2 \mu^2_a (1-\yjb)}{\yaj^2} \right. \\
    & \left. \qquad  - \frac{2 \mu^2_b (1-\yaj)}{\yjb^2} \right] \nonumber
\end{align} 
with the individual helicity contributions
\begin{align}
    \bar a_{q_+q_+\mapsto q_+g_+q_+}^\text{II,gl} & = \frac{1}{\sAB} \left[ \frac{1}{\yaj \yjb} - \frac{\mu_a^2}{\yaj^2} - \frac{\mu_b^2}{\yjb^2}\right], \\
    \bar a_{q_+q_+\mapsto q_+g_-q_+}^\text{II,gl} & = \frac{1}{\sAB} \left[ \frac{\yAB^2}{\yaj\yjb} - \frac{\mu_a^2(1-\yjb)^2}{\yaj^2} - \frac{\mu_b^2(1-\yaj)^2}{\yjb^2}\right], \\
    \bar a_{q_+q_+\mapsto q_-g_-q_+}^\text{II,gl} & = \frac{1}{\sAB} \frac{\mu_a^2 \yjb^2}{\yaj^2}, \\
    \bar a_{q_+q_+\mapsto q_+g_-q_-}^\text{II,gl} & = \frac{1}{\sAB} \frac{\mu_b^2 \yaj^2}{\yjb^2}, \\
    \bar a_{q_+q_-\mapsto q_+g_+q_-}^\text{II,gl} & = \frac{1}{\sAB} \left[\frac{(1-\yaj)^2}{\yaj\yjb} - \frac{\mu_a^2}{\yaj^2} - \frac{\mu_b^2(1-\yaj)^2}{\yjb^2} \right], \\
    \bar a_{q_+q_-\mapsto q_+g_-q_-}^\text{II,gl} & = \frac{1}{\sAB} \left[ \frac{(1-\yjb)^2}{\yaj \yjb} - \frac{\mu_a^2(1-\yjb)^2}{\yaj^2} - \frac{\mu_b^2}{\yjb^2}\right],\\
    \bar a_{q_+q_-\mapsto q_-g_-q_-}^\text{II,gl} & = \frac{1}{\sAB} \frac{\mu_a^2 \yjb^2}{\yaj^2},\\
    \bar a_{q_+q_-\mapsto q_-g_+q_+}^\text{II,gl} & = \frac{1}{\sAB} \frac{\mu_b^2 \yaj^2}{\yjb^2}.
\end{align}

\paragraph{QGEmitII} The helicity-averaged antenna function for the process $\bar q_A g_B \mapsto q_a g_j g_b$ is 
\begin{align}
    \bar a_{g/qg}^\text{II,gl}(\yaj, \yjb) =\,& \frac{1}{\sAB}\left[ \frac{(1-\yaj)^3 + (1-\yjb)^2}{\yaj\yjb} + \frac{1+\yaj^3}{\yjb(1-\yaj)}\right. \\ 
    & \left. \qquad - \frac{2 \mu^2_a (1-\yjb)}{\yaj^2} + 2 - \yaj - \frac{\yjb}{2}\right] \nonumber
\end{align}
with the individual helicity contributions
\begin{align}
    \bar a_{q_+g_+\mapsto q_+g_+g_+}^\text{II,gl} & = \frac{1}{\sAB} \left[ \frac{1}{\yaj\yjb} + \frac{1}{\yjb(1-\yaj)} - \frac{\mu_a^2}{\yaj^2} \right].\\
    \bar a_{q_+g_+\mapsto q_+g_-g_+}^\text{II,gl} & = \frac{1}{\sAB} \left[ \frac{(1-\yaj) \yAB^2}{\yaj\yjb} - \frac{\mu_a^2(1-\yjb)^2}{\yaj^2}\right],\\
    \bar a_{q_+g_+\mapsto q_-g_-g_+}^\text{II,gl} & = \frac{1}{\sAB}\left[ \frac{\mu_a^2 \yjb^2}{\yaj^2} \right],\\
    \bar a_{q_+g_+\mapsto q_+g_-g_-}^\text{II,gl} & = \frac{1}{\sAB} \frac{\yaj^3}{\yjb(1-\yaj)}, \\
    \bar a_{q_+g_-\mapsto q_+g_+g_-}^\text{II,gl} & = \frac{1}{\sAB} \left[ \frac{(1-\yaj)^3}{\yaj\yjb} - \frac{\mu_a^2}{\yaj^2}\right],\\
    \bar a_{q_+g_-\mapsto q_+g_-g_-}^\text{II,gl} & = \frac{1}{\sAB} \left[ \frac{(1-\yjb)^2}{\yaj\yjb} + \frac{1}{\yjb(1-\yaj)} - \frac{\mu_a^2(1-\yjb)^2}{\yaj^2} \right].
\end{align}

\paragraph{GGEmitII} The helicity-averaged antenna function for the process $g_A g_B \mapsto g_a g_j g_b$ is 
\begin{align}
    \bar a_{g/gg}^\text{II,gl}(\yaj, \yjb) =\,& \frac{1}{\sAB} \left[ \frac{(1-\yaj)^3 + (1-\yjb)^3}{\yaj\yjb} + \frac{1+\yaj^3}{\yjb(1-\yaj)} \right. \\
    & \qquad  \left. + \frac{1+\yjb^3}{\yaj(1-\yjb)}  + 3 - \frac{3\yaj}{2} - \frac{3\yjb}{2} \right] \nonumber
\end{align}
with the individual helicity contributions
\begin{align}
    \bar a_{g_+g_+\mapsto g_+g_+g_+}^\text{II,gl} & = \frac{1}{\sAB} \left[ \frac{1}{\yaj \yjb} + \frac{1}{\yjb(1-\yaj)} +  \frac{1}{\yaj(1-\yjb)} \right],\\     
    \bar a_{g_+g_+\mapsto g_+g_-g_+}^\text{II,gl} & = \frac{1}{\sAB} \frac{\yAB^3}{\yaj \yjb}, \\
    \bar a_{g_+g_-\mapsto g_+g_+g_-}^\text{II,gl} & = \frac{1}{\sAB} \left[ \frac{(1-\yaj)^3}{\yaj \yjb} + \frac{1}{\yaj(1-\yjb)}\right],\\
    \bar a_{g_+g_-\mapsto g_+g_-g_-}^\text{II,gl} & = \frac{1}{\sAB} \left[ \frac{(1-\yjb)^3}{\yaj \yjb} + \frac{1}{\yjb(1-\yaj)} \right], \\
    \bar a_{g_+g_+\mapsto g_+g_-g_-}^\text{II,gl} & = \frac{1}{\sAB} \frac{\yaj^3}{\yjb(1-\yaj)},\\
    \bar a_{g_+g_+\mapsto g_-g_-g_+}^\text{II,gl} & = \frac{1}{\sAB} \frac{\yjb^3}{\yaj(1-\yjb)}.
\end{align}

\paragraph{QXSplitII} The helicity-averaged antenna function for the process $q_A X_B \mapsto g_a \bar q_j X_b$, where $X$ is an arbitrary coloured parton, is 
\begin{align}
    \bar a_{\bar q/qX}^\text{II,gl}(\yaj, \yjb) & = \frac{1}{\sAB}\left[ \frac{\yAB^2 + (1-\yAB)^2}{\yaj} + \frac{2\mu_j^2 \yAB}{\yaj^2}\right]
\end{align}
with the individual helicity contributions
\begin{align}
    \bar a_{q_+X \mapsto g_+\bar q_-X}^\text{II,gl} & = \frac{1}{\sAB}\left[ \frac{\yAB^2}{\yaj} - \frac{\mu^2_j \yAB^2}{\yaj^2(1-\yAB)} \right],\\
    \bar a_{q_+X \mapsto g_-\bar q_-X}^\text{II,gl} & = \frac{1}{\sAB}\left[ \frac{(1-\yAB)^2}{\yaj} - \frac{\mu_j^2(1-\yAB)}{\yaj^2} \right],\\
    \bar a_{q_+X \mapsto g_+\bar q_+X}^\text{II,gl} & = \frac{1}{\sAB}\frac{\mu_j^2}{\yaj^2(1-\yAB)}.
\end{align}

\paragraph{GXConvII} The helicity-averaged antenna function for the process $g_A X_B \mapsto q_a q_j X_b$, where $X$ is an arbitrary coloured parton, is 
\begin{align}
    \bar a_{q/gX}^\text{II,gl}(\sAB, \yaj, \yjb) & = \frac{1}{\sAB} \left[ \frac{1 + (1-\yAB)^2}{(\yaj-2\mu_j^2) \yAB} - \frac{2\mu_j^2 \yAB}{(\yaj-2\mu_j^2)^2}\right].
\end{align}
with the individual helicity contributions
\begin{align}
    \bar a_{g_+X \mapsto q_+q_+X}^\text{II,gl} & = \frac{1}{\sAB} \frac{1}{2}\left[ \frac{1}{(\yaj-2\mu_j^2)\yAB} - \frac{\mu_j^2}{(\yaj-2\mu_j^2)^2}\frac{\yAB}{1-\yAB} \right],\\
    \bar a_{g_+X \mapsto q_-q_-X}^\text{II,gl} & = \frac{1}{\sAB}\left[ \frac{(1-\yAB)^2}{(\yaj-2\mu_j^2)\yAB} - \frac{\mu_j^2\yAB(1-\yAB)}{(\yaj-2\mu_j^2)^2} \right],\\
    \bar a_{g_+X \mapsto q_+q_-X}^\text{II,gl} & = \frac{1}{\sAB} \frac{\mu_j^2}{(\yaj - 2\mu_j^2)^2} \frac{\yAB^3}{1-\yAB}.
\end{align}

\section{Collinear Limits of Sector Antennae}
\label{sec:CollinearLimits}
To ensure that the sector (as well as the global) antenna functions reproduce the correct collinear limits, they are numerically tested against DGLAP splitting kernels. For two final-state partons (i.e.\ in final-final or initial-final antennae), we compare antenna functions against massive DGLAP kernels in the quasi-collinear limit, while for initial-final and initial-initial configurations, we compare them to massless DGLAP kernels in the collinear limit.

Below, we collect all helicity- and mass-dependent DGLAP kernels as implemented in \Vincia \cite{Kleiss:2020rcg}, denoting them by $P_{I_\pm \mapsto i_\pm j_\pm}(z)$, where $z \equiv z_i$ is the energy fraction taken by parton $i$.

The helicity-averaged splitting kernel for gluon emissions $g\mapsto gg$ is given by
\begin{equation}
    P_{g \mapsto gg}(z) = 2\frac{(1-z(1-z))^2}{z(1-z)}
\end{equation}
and the individual helicity contributions are
\begin{align}
    P_{g_+ \mapsto g_+ g_+}(z) &= \frac{1}{z(1-z)}, \\
    P_{g_+ \mapsto g_- g_+}(z) &= \frac{(1-z)^3}{z}, \\
    P_{g_+ \mapsto g_+ g_-}(z) &= \frac{z^3}{1-z}.
\end{align}
For gluon emissions from a quark, $q\mapsto qg$, the helicity-averaged splitting kernel is
\begin{equation}
    P_{q \mapsto qg}(z) = \frac{1+z^2}{1-z} - 2\mu_q^2
\end{equation}
and the individual helicity contributions are
\begin{align}
    P_{q_+ \mapsto q_+ g_+}(z) &= \frac{1}{1-z}-\mu_q^2\frac{1}{z}, \\
    P_{q_+ \mapsto q_- g_+}(z) &= \mu_q^2\frac{(1-z)^2}{z}, \\
    P_{q_+ \mapsto q_+ g_-}(z) &= \frac{z^2}{1-z}-\mu_q^2z.
\end{align}
Lastly, for gluon splittings $g\mapsto q\bar q$, the helicity-averaged splitting kernel is
\begin{equation}
    P_{g \mapsto q\bar q}(z) = z^2 +(1-z)^2 + 2\mu_q^2
\end{equation}
and the individual helicity contributions are given by
\begin{align}
    P_{g_+ \mapsto q_+ \bar q_-}(z) &= z^2 -\mu_q^2 \frac{z}{1-z}, \\
    P_{g_+ \mapsto q_- \bar q_+}(z) &= (1-z)^2 - \mu_q^2 \frac{1-z}{z},\\
    P_{g_+ \mapsto q_+ \bar q_+}(z) &= \mu_q^2\left(\frac{z}{1-z}+\frac{1-z}{z} +2 \right).
\end{align}

\section{Explicit Expressions for Trial Generators}\label{sec:ExplicitTrialGens}
Here, we collect all additional trial integrals, \cref{eq:TrialIntegralDef}, needed for the sector shower in terms of the phase space variables $x_\perp$ and $\zeta$. We denote trial functions used only in the sector shower by a superscript \textquotedblleft sct\textquotedblright\ and express trial integrals in terms of \cref{eq:EvolutionMasterInt}.

\subsection{Trial Integrals}

\paragraph{FF Branchings} For final-final configurations, the trial integral we are inverting analytically is given by
\begin{equation}
\label{eq:trialIntFF}
\mathcal A(x^\text{FF}_{\perp1}, x^\text{FF}_{\perp2}) = \Kallen \frac{\mathcal C}{4 \uppi} \int\limits_{x^\text{FF}_{\perp2}}^{x^\text{FF}_{\perp1}} \alphastrong(x_\perp^\text{FF}) \bar a_\text{trial}(x^\text{FF}_\perp, \zeta^\text{FF}) \sIK \vert J(x^\text{FF}_\perp, \zeta^\text{FF})\vert \D x^\text{FF}_\perp \D \zeta^\text{FF}.
\end{equation}

To ensure that the trial function overestimates the sector antenna function, we multiply the gluon splitting trial function,
\begin{equation}
\bar a_{\text{trial-split}}^\text{FF} = \frac{1}{\sIK} \frac{1}{2(\yij+2\mu_j^2)} \, , \label{eq:GSplitFFTrial}
\end{equation}
by two to take into account that the splitting gluon is part of two antennae and overestimate the additional collinear parts of gluon-emission sector antenna functions in \cref{eq:DefSectorAntennaFF} by 
\begin{align}
	\bar a_{\text{trial-coll-}I}^\text{FF,sct} &= \frac{1}{\sIK} \frac{2}{\yij(1-\yjk)} \, , \\
	\bar a_{\text{trial-coll-}K}^\text{FF,sct} &= \frac{1}{\sIK} \frac{2}{\yjk(1-\yij)} \, .
\end{align}

For these additional collinear trial functions, both in gluon emissions and gluon splittings, we choose the corresponding non-singular invariant,
\begin{equation}
\zeta^{\text{FF}} = \begin{cases}
\zeta_-^{\text{FF}} = \yjk+\mu_j^2 & I\text{-collinear}\\
\zeta_+^{\text{FF}} = \yij+\mu_j^2 & K\text{-collinear}
\end{cases}, \label{eq:DefZetaFF}
\end{equation}
where $\mu_j^2 = 0$ for gluon emissions.
The Jacobian to transform from $(\yij, \yjk)$ to $(x_{\perp}^{\text{FF}}, \zeta^{\text{FF}})$ is given by
\begin{equation}
\vert J \vert = \vert J_\pm\vert = \frac{1}{\zeta_\pm^{\text{FF}}}\, .
\end{equation}
In terms of \cref{eq:EvolutionMasterInt}, the additional trial integrals are given by
\begin{align}
	\mathcal A_{\text{coll}}\left(x^\text{FF}_{\perp1}, x^\text{FF}_{\perp2}\right) &= \Kallen \frac{\mathcal C}{2\uppi} \Atrial \left(x^\text{FF}_{\perp1}, x^\text{FF}_{\perp2},\frac{1}{1-\zeta^\text{FF}_\pm}\right) \, ,\\
	\mathcal A_{\text{split}}\left(x^\text{FF}_{\perp1}, x^\text{FF}_{\perp2}\right) &= \Kallen \frac{\mathcal C}{2\uppi}\Atrial\left(x^\text{FF}_{\perp1}, x^\text{FF}_{\perp2}, \frac{1}{2}\right) \, ,
\end{align}
which are the same for the $I$-collinear and $K$-collinear regions, and the choice of $\zeta^\text{FF}_\pm$ as the corresponding non-singular invariant is implicit.

\paragraph{RF Branchings} 
For resonance-final configurations, the trial integral is given by:
\begin{align}
	\label{eq:trialIntRF}
	\mathcal A(x^\text{RF}_{\perp1}, x^\text{RF}_{\perp2}) = \KallenRF \frac{\mathcal C}{4 \uppi}
	\int\limits_{x^\text{RF}_{\perp2}}^{x^\text{RF}_{\perp1}} \alphastrong(x_\perp^\text{RF})
	\bar a_\text{trial}(x^\text{RF}_\perp, \zeta^\text{RF}) 
	\frac{(\sAK + m_j ^2 +m_k^2 -m_K^2)^2}{(1-\yjk)^3} \nonumber \\
	\times \vert J(x^\text{RF}_\perp, \zeta^\text{RF})\vert \D x^\text{RF}_\perp \D \zeta^\text{RF}.
\end{align}

The trial antenna for ($K$-) gluon splittings is given by
\begin{equation}
\bar a^\text{RF,sct}_{\text{trial-split-}K} = \left[ \frac{1-\yjk}{\sAK+2m_j^2}\right]  \frac{2 \left(1-\yjk\right)^2 }{(\yjk+2\mu_j^2)} \, .
\end{equation}
The first factor in square brackets simply results from the dimensionful normalisation of the trial antenna. 

The additional trial antenna for gluon emission collinear with parton $K$ is given by:
\begin{equation}
	\bar a^\text{RF,sct}_{\text{trial-coll-}K} = \left[\frac{1}{\sAK}\right]\frac{2}{\yjk (1-\yaj)}\left[2(1-\yjk)\right]^3.
\end{equation}
The final term inside the brackets is simply $\yAK/{\yAK}_\mathrm{min} \ge 1$, and is included to cancel the factor appearing in \cref{eq:phaseSpaceFac}.
We also note that since $A$ is a resonance, and therefore cannot be a gluon, there is no additional $A$-collinear piece present in the antenna function and hence there is no need for the corresponding overestimate.

We make the following choice for the complementary phase space variable\footnote{We note that this choice is not completely identical to that made in \cite{Brooks:2019xso} for gluon splittings; however, this was a more natural choice for the normalisation chosen in \cref{eq:DimensionlessInvariants} and the only impact of this choice is in the efficiency of the exploration of phase space, expected to be minimal.}
\begin{equation}
	\zeta^{\text{RF}} = \zeta_+^{\text{RF}} = \yaj \, ,
\end{equation}
leading to the Jacobian
\begin{equation}
	\vert J\vert = \vert J_+ \vert = \frac{1}{\yaj-\mu_j^2} \, , \label{eq:DefJRF}
\end{equation}
where $\mu^2_j = 0$ for gluon emissions.

In terms of \cref{eq:EvolutionMasterInt}, the sector trial integrals are given by
\begin{align}
	\mathcal A_{\text{coll-}K}\left(x^\text{RF}_{\perp1}, x^\text{RF}_{\perp2}\right) &= 
	\KallenRF \frac{\mathcal C}{2\uppi}\Atrial\left(x^\text{FF}_{\perp1}, x^\text{FF}_{\perp2}, \frac{8}{1-\zeta^\text{RF}_+}\right) \\
	\mathcal A_{\text{split}}\left(x^\text{RF}_{\perp1}, x^\text{RF}_{\perp2}\right) &=
	\KallenRF \frac{\mathcal C}{2\uppi}\Atrial\left(x^\text{FF}_{\perp1}, x^\text{FF}_{\perp2}, 1\right).
\end{align}

\paragraph{IF Branchings} For initial-final configurations, the trial integral is
\begin{equation}
\mathcal A(x_{\perp1}, x_{\perp2}) = \frac{\mathcal C}{4 \uppi} \tilde R_f  \int\limits_{x_{\perp2}}^{x_{\perp1}} \alphastrong(x_\perp^\text{IF}) \bar a_\text{trial}(x^\text{IF}_\perp, \zeta^\text{IF})\yAK^\gamma \frac{\sAK}{1-\yjk} \vert J(x^\text{IF}_\perp, \zeta^\text{IF})\vert \D x^\text{IF}_\perp \D \zeta^\text{IF}
\end{equation}
as $x_A/x_a = \yAK$, cf.\ \cref{sec:Kinematics}.

Similarly to the case of final-final splittings, we overestimate gluon splittings in the final state by
\begin{equation}
\bar a^\text{IF,sct}_{\text{trial-split-}K} = 2 \cdot \frac{1}{\sAK} \frac{1}{2(\yjk+2\mu_j^2)} \left[ \frac{\sAK + m^2_j}{\sAK} \right] \, ,
\end{equation}
where the factor of 2 again takes into account that the gluon is part of two antennae and the additional factor in the brackets, which is $\geq 1$, cancels unwanted factors in the trial integral. Gluon splittings and gluon conversions in the initial state are already sectorised and do not need additional sector overestimates.
In the sector shower, we complement this by another overestimate of the $K$-collinear part which is added for sector antennae. Hence, the only additional gluon-emission trial functions is
\begin{equation}
	\bar a^\text{IF,sct}_{\text{trial-coll-}K} = \frac{1}{\sAK} \frac{2}{\yjk (1-\yaj)} \, .
\end{equation}
We emphasise that although the additional singularity $\yaj \to 1$ is kinematically allowed, it is regularised by the parton shower cutoff $Q^2_\text{cut}$, as it lies in the $a$-$k$ (soft-)collinear sector. 
Hence, the sector-shower phase space limit
\begin{equation}
	\yaj^\text{max} = \frac{\saj}{\saj + \sak^\text{min}} = \frac{1}{1+x_A\frac{Q^2_\text{cut}}{\sAK}}
\end{equation}
is imposed in the shower evolution.

We use the $K$-collinear phase space variable
\begin{equation}
	\zeta^\text{IF} = \zeta^\text{IF}_+ = \yaj-\mu^2_j \, ,
\end{equation}
with the Jacobian for the transformation $(\yaj,\yjk)\mapsto(x_\perp^\text{IF},\zeta^\text{IF})$
\begin{equation}
	\vert J \vert = \vert J_+ \vert = \frac{1}{\zeta^\text{IF}_+} \, .
\end{equation}
For final-state gluon splittings, we therefore find the sector trial integral
\begin{equation}
	\mathcal A_{\text{split-}K} (x_{\perp1}^\text{IF},x_{\perp2}^\text{IF}) = \frac{\mathcal C}{2\uppi} \tilde R_f\Atrial\left(x^\text{IF}_{\perp1}, x^\text{IF}_{\perp2}, \frac{1}{2}\right) \, , \label{eq:TrialKSplitIF}
\end{equation}
while for gluon emissions, we have the sector trial integral
\begin{equation}
	\mathcal A_{\text{coll-}K} (x_{\perp1}^\text{IF},x_{\perp2}^\text{IF}) = \frac{\mathcal C}{2\uppi}\tilde R_f\Atrial\left(x^\text{IF}_{\perp1}, x^\text{IF}_{\perp2}, \frac{1}{1-\zeta^\text{IF}_+}\right) \, , \label{eq:TrialKCollIF}
\end{equation}
again expressed in terms of \cref{eq:EvolutionMasterInt}.

It should be pointed out that the additional sector integrals \cref{eq:TrialKSplitIF,eq:TrialKCollIF} are given only for $\gamma = 1$. In practice, $\gamma$ is chosen to be either $0$ or $1$ for the others.

\paragraph{II Branchings}
No additional trial generators are needed for the initial-initial case, because global antenna functions are sectorised in the first place, as there is no emission into the initial state.
The same trial integrals are used for the global and the sector shower.

There is, however, the possibility to express these trial integrals in terms of the dimensionless phase space variables $x_\perp^\text{II}$ and $\zeta^\text{II}$. 
The trial integral is then given by
\begin{equation}
\mathcal A(x^\text{II}_{\perp1}, x^\text{II}_{\perp2}) = \frac{\mathcal C}{4 \uppi} \tilde R_f \int\limits_{x^\text{II}_{\perp2}}^{x^\text{II}_{\perp1}} \alphastrong(x_\perp^\text{II}) \bar a_\text{trial}(x^\text{II}_\perp, \zeta^\text{II})\yAB^\gamma \frac{\sAB}{1-\yaj-\yjb} \vert J(x^\text{II}_\perp, \zeta^\text{II})\vert \D x^\text{II}_\perp \D \zeta^\text{II}
\end{equation}
as $x_Ax_B/x_ax_b = \yAB$, cf.\ \cref{sec:Kinematics}.

\subsection{Zeta Integrals}
Apart from the trivial integral
\begin{equation}
	 I_\text{lin} (\zeta) = \int 1 \D \zeta = \zeta \, 
\end{equation}
we make use of the integral
\begin{equation}
    I_\text{log} (\zeta) = \int \frac{1}{1-\zeta} \D \zeta = -\log(1-\zeta) \, .
\end{equation}
Trial values for $\zeta$ are then found by inverting \cref{eq:ZetaTrial} with the solution
\begin{equation}
	\zeta = I^{-1}_\zeta \left[ R_\zeta\left( I(\zeta_\text{max}) - I(\zeta_\text{min}) \right) + I(\zeta_\text{min}) \right] \, ,
\end{equation}
where $I^{-1}_\zeta$ denotes the inverse of $I_\zeta$, given by
\begin{align}
	I^{-1}_\text{lin} (r) &= r \, ,\\
	I^{-1}_\text{log} (r) &= 1-\expo{-r} \, .
\end{align}

\FloatBarrier
\clearpage
\section{Comparisons to High-Multiplicity Matrix Elements}
\label{sec:hiMultiMERatios}
Here, we present validations of the sector shower against high-multiplicity tree-level matrix elements. For the sake of clarity, we do not include separate sector histograms, as too many histories contribute to each final state, although for a given kinematic configuration only a single one contributes at a time.

\begin{figure}[ht]
	\centering
	\includegraphics[width=0.49\textwidth]{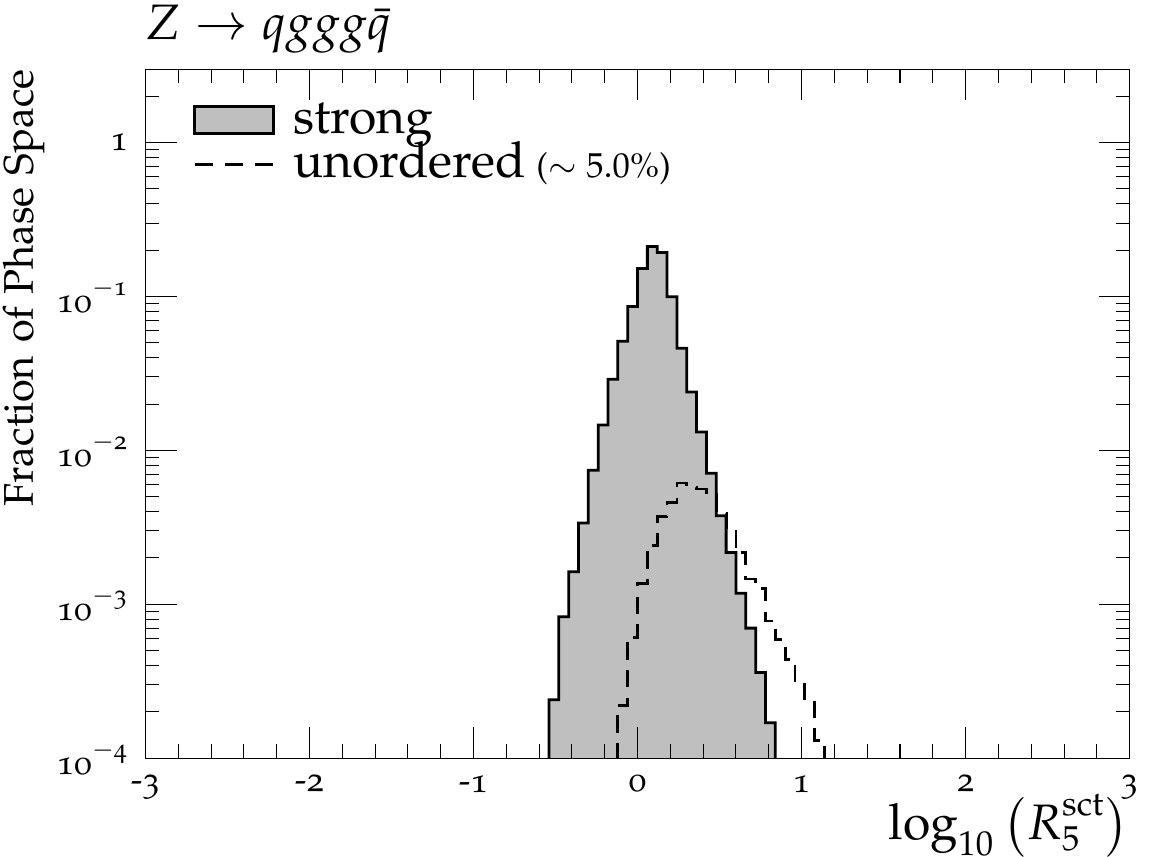}
	\includegraphics[width=0.49\textwidth]{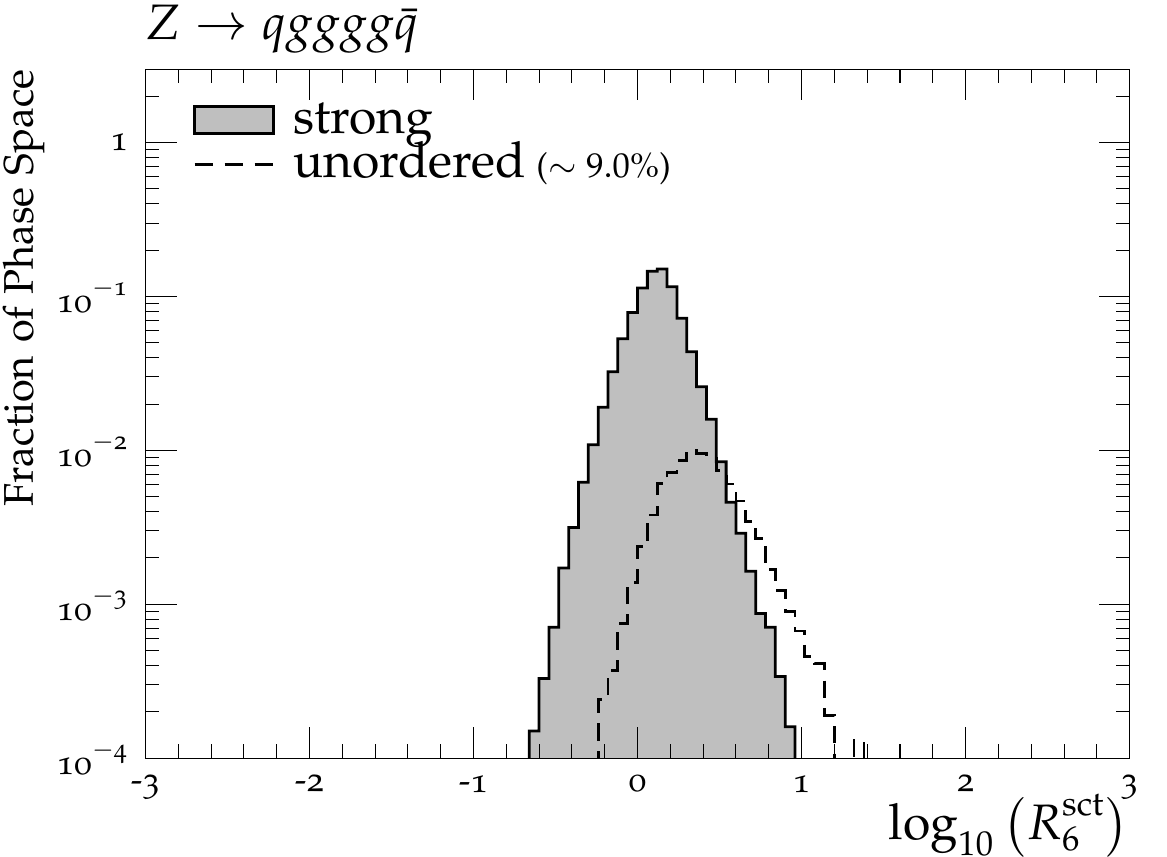}\\
	\includegraphics[width=0.49\textwidth]{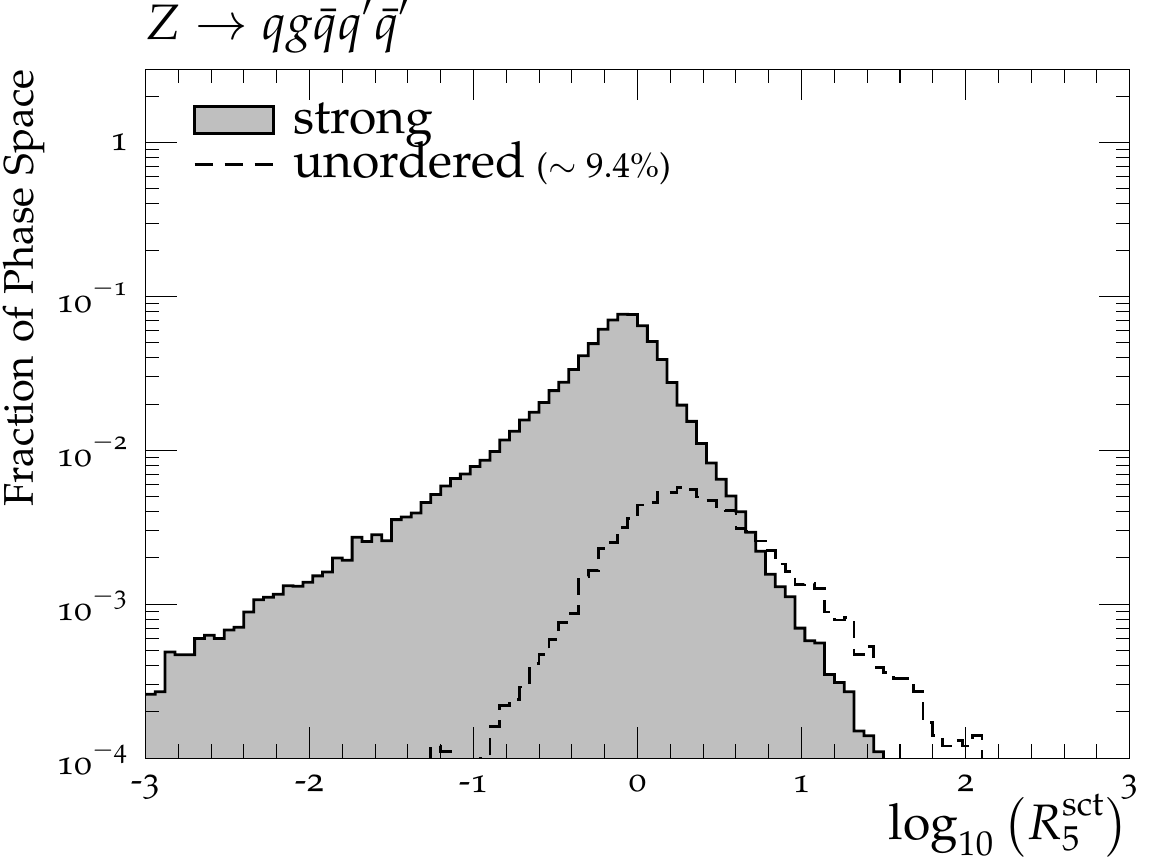}
	\includegraphics[width=0.49\textwidth]{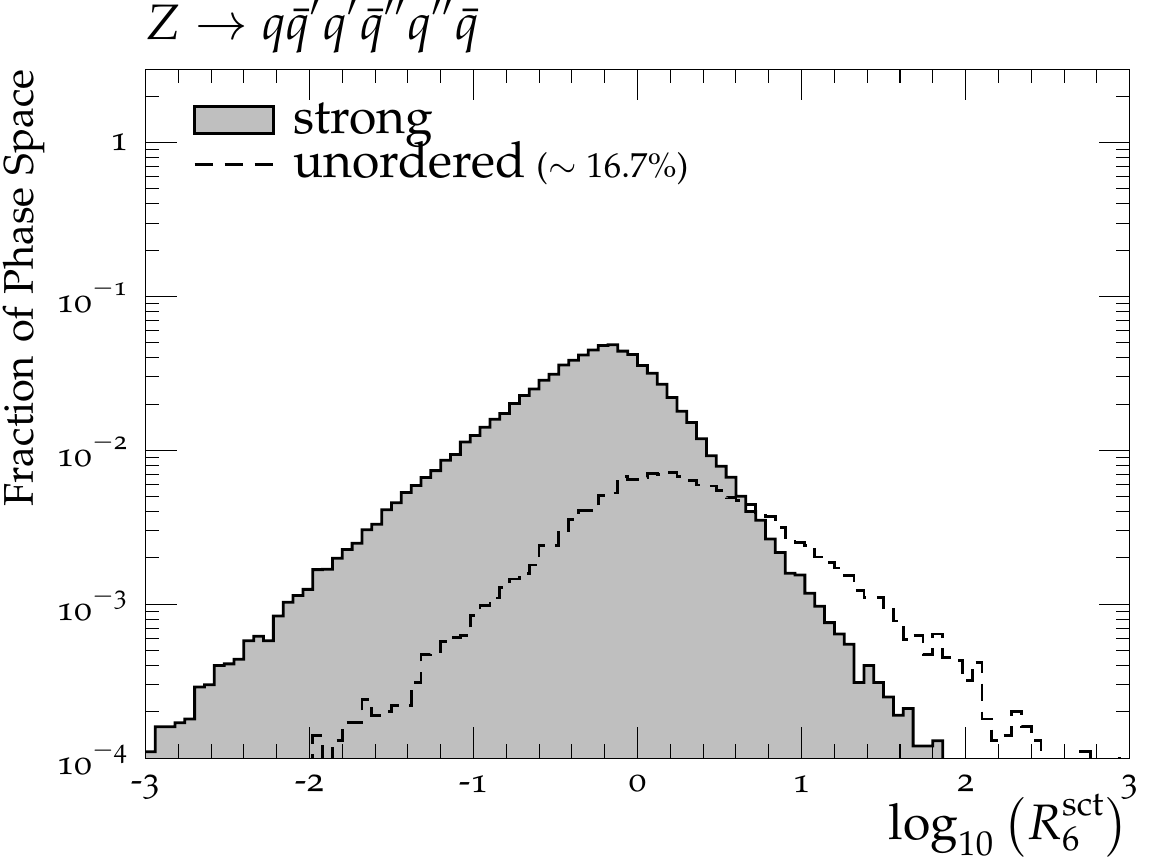}
	\caption{Ratios $R_5^\text{sct}$ of the sector shower approximation to LO matrix elements for $Z$ decays to 5- and 6- parton final states in a flat phase space scan.}
	\label{fig:MERatiosZhiMulti}
\end{figure}

\begin{figure}[ht]
	\centering
	\includegraphics[width=0.49\textwidth]{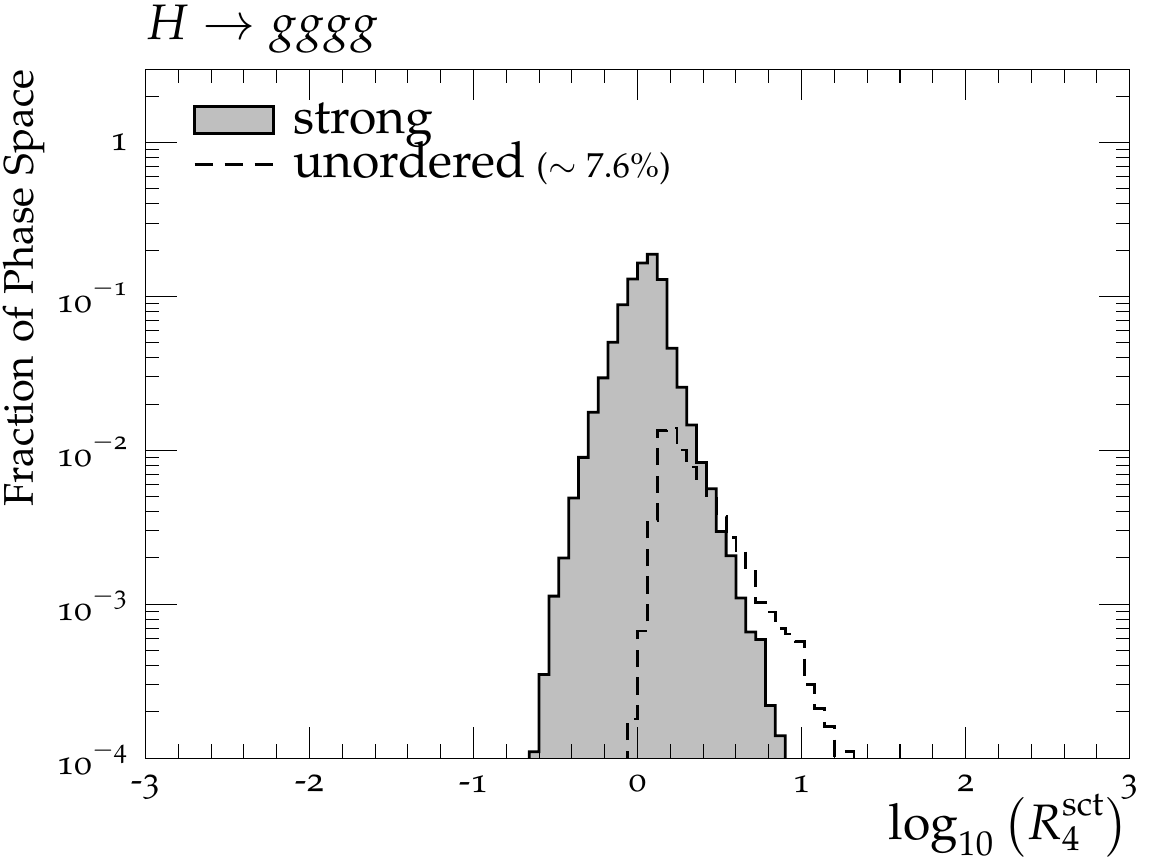}
 	\includegraphics[width=0.49\textwidth]{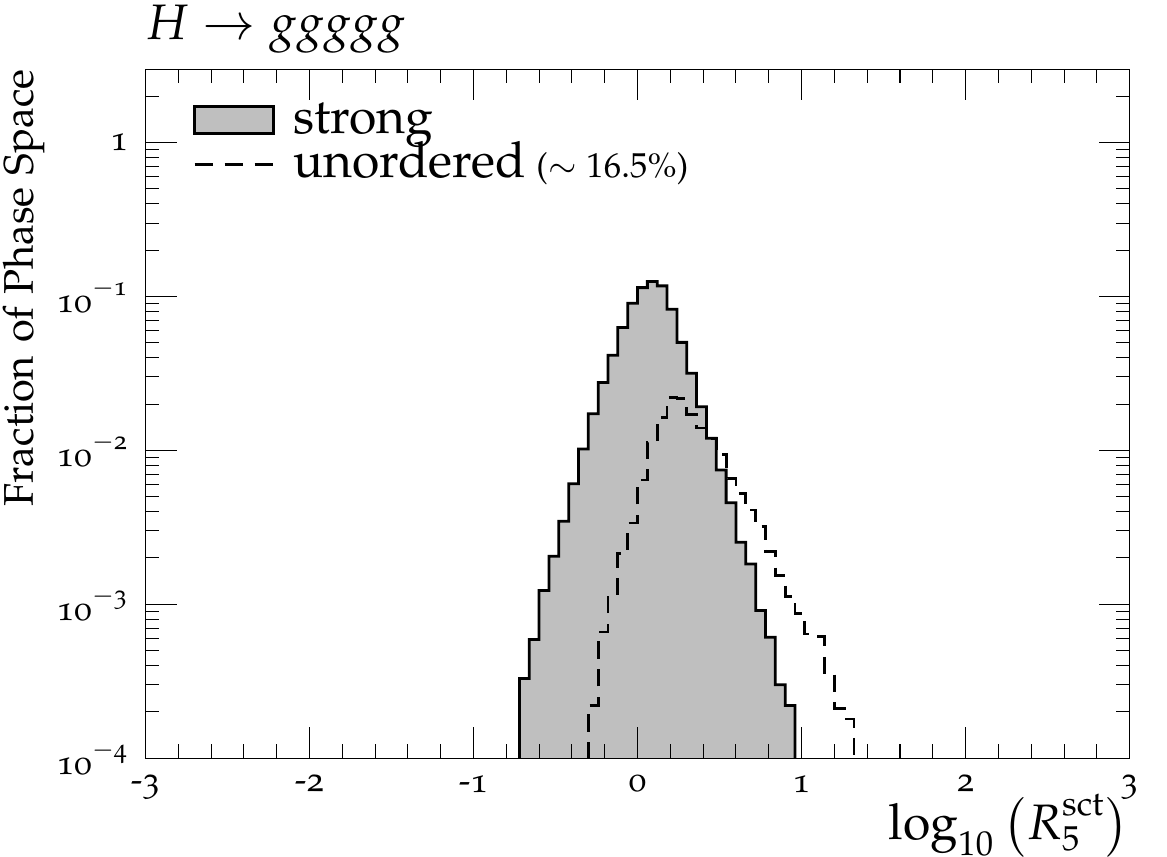}
	\caption{Ratios $R_n^\text{sct}$ of the sector-shower approximation to LO matrix elements for $H \to ng$ with $n=4,5$ in a flat phase space scan.}
	\label{fig:MERatiosHhiMulti}
\end{figure}

\begin{figure}[ht]
	\centering
	\includegraphics[width=0.49\textwidth]{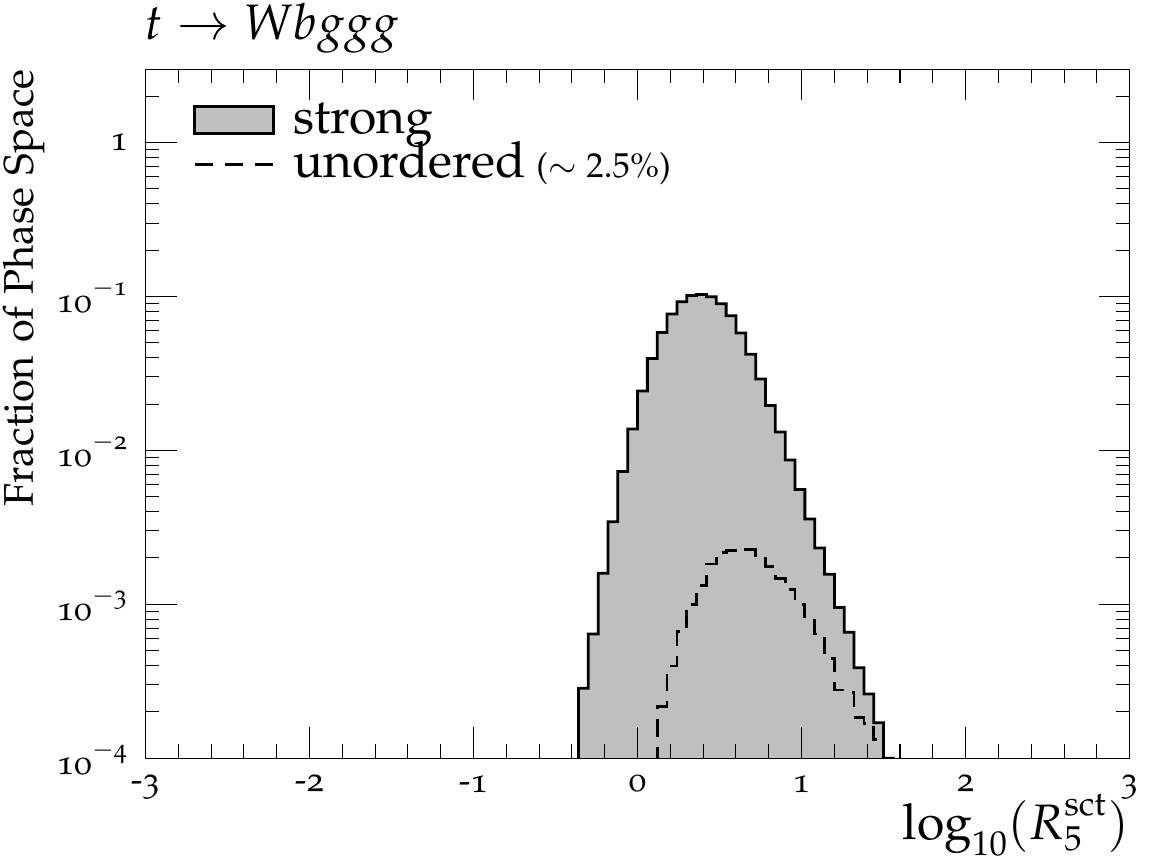}
	\includegraphics[width=0.49\textwidth]{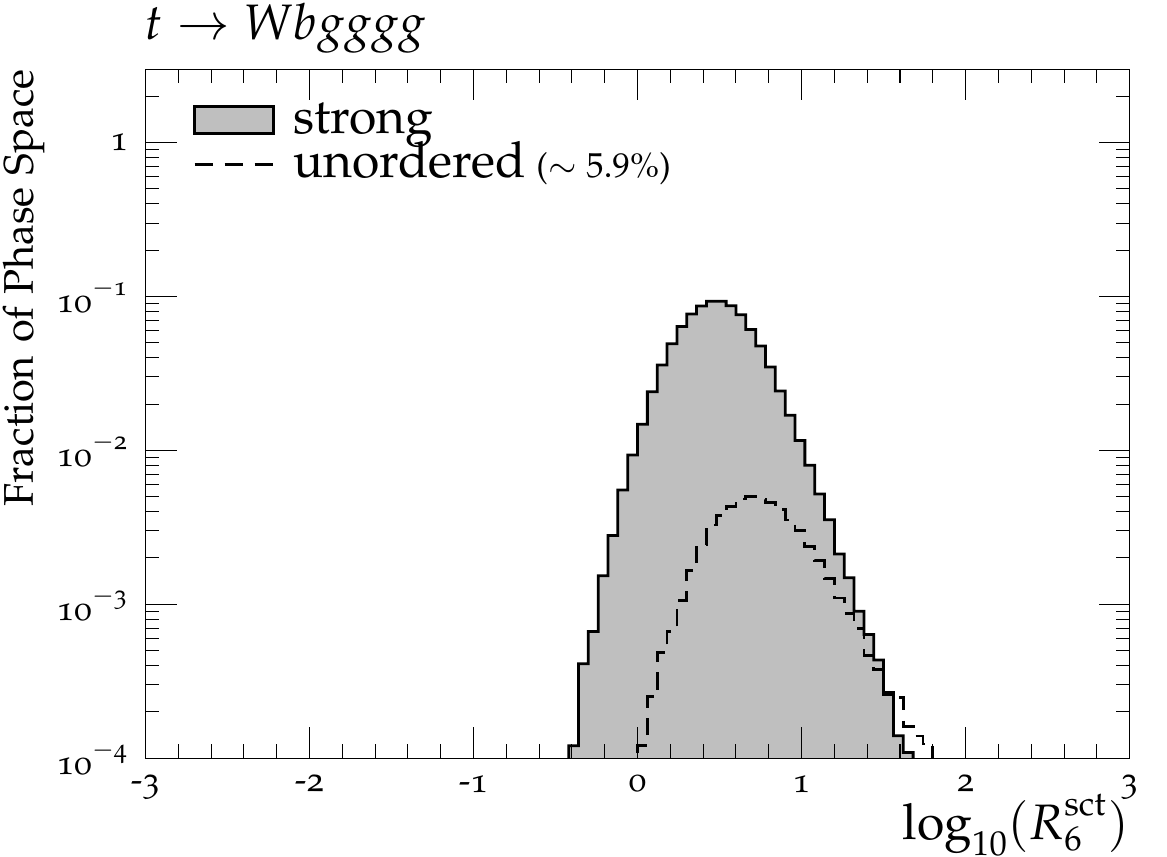}
	\caption{Ratios $R_n^\text{sct}$ of the sector-shower approximation to LO matrix elements for top-quark decay processes $t \to W^+ b$ with additional gluons in a flat phase space scan.}
	\label{fig:MERatiosResHiMulti}
\end{figure}

\begin{figure}[ht]
	\centering
	\includegraphics[width=0.49\textwidth]{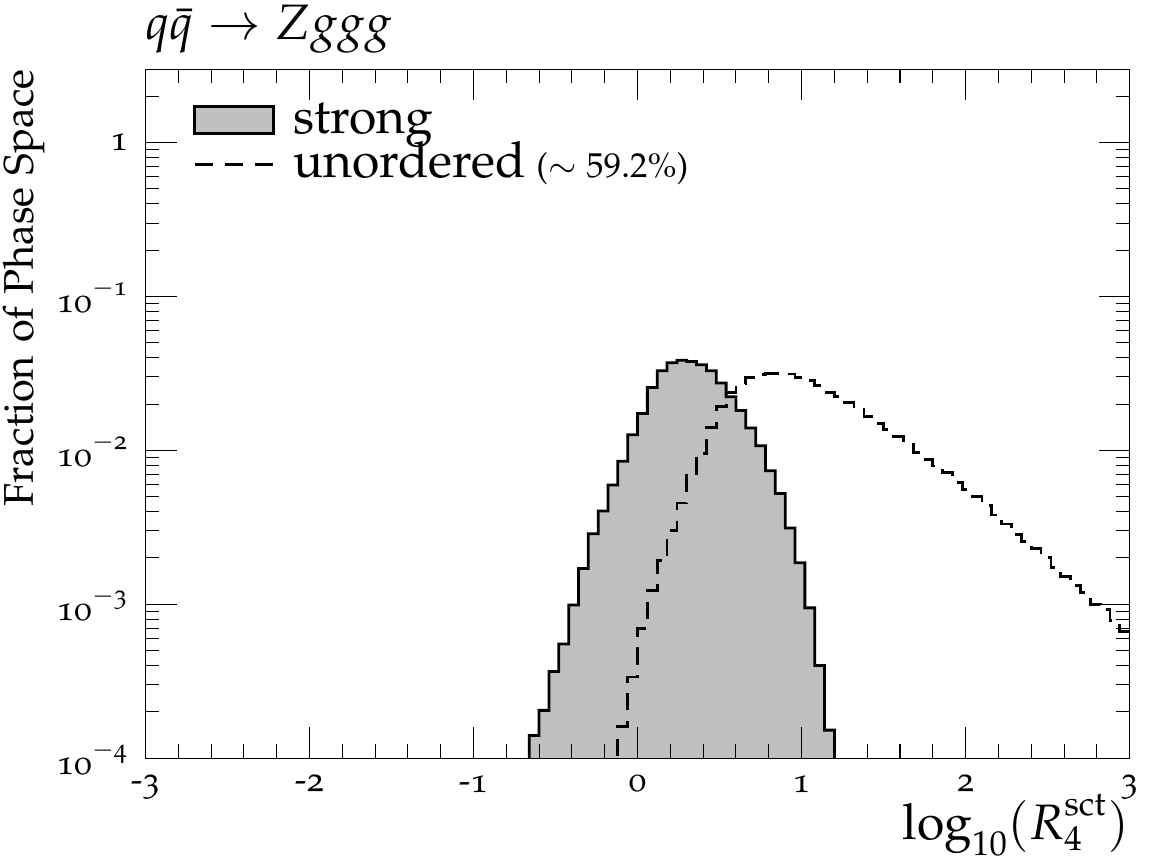}
	\includegraphics[width=0.49\textwidth]{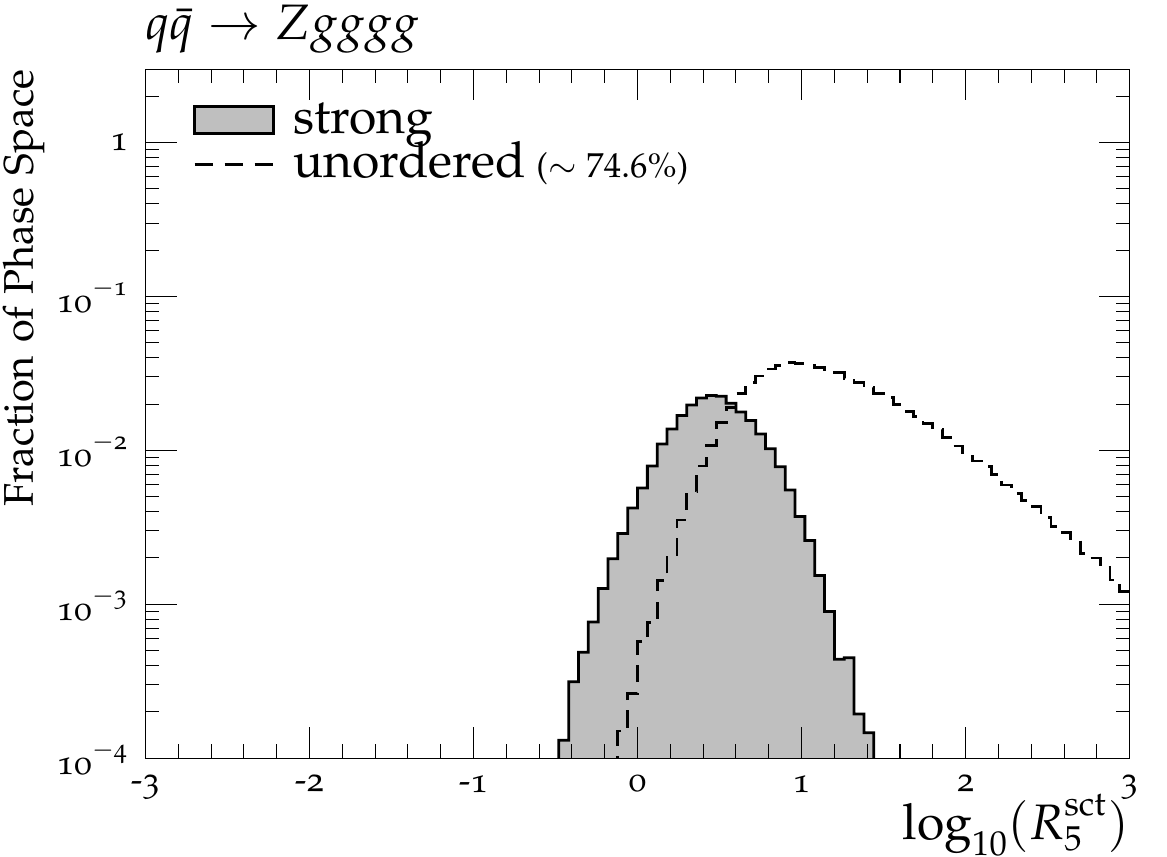}
	\caption{Ratios $R_n^\text{sct}$ of the sector-shower approximation to LO matrix elements for Drell-Yan processes $q\bar q \to Z$ with additional gluons in a flat phase space scan.}
	\label{fig:MERatiosDYhiMulti}
\end{figure}

\begin{figure}[ht]
	\centering
	\includegraphics[width=0.49\textwidth]{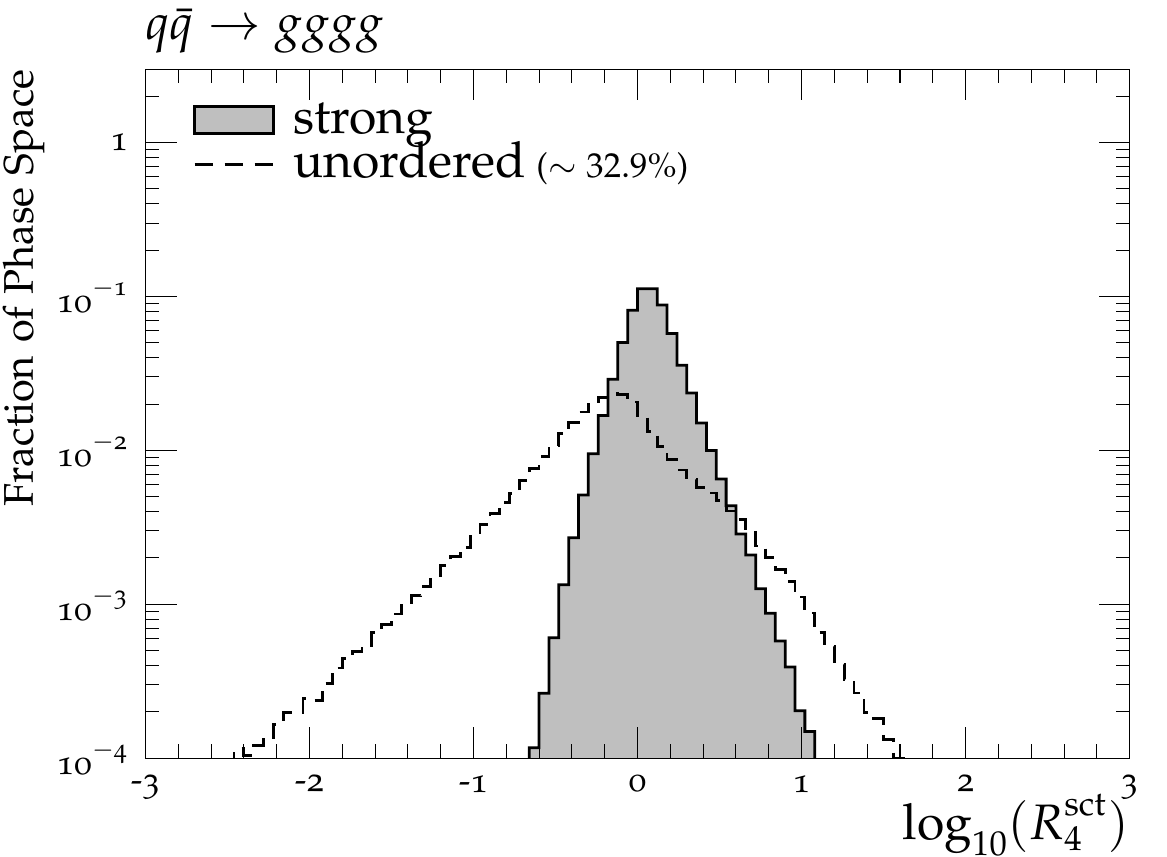}
	\includegraphics[width=0.49\textwidth]{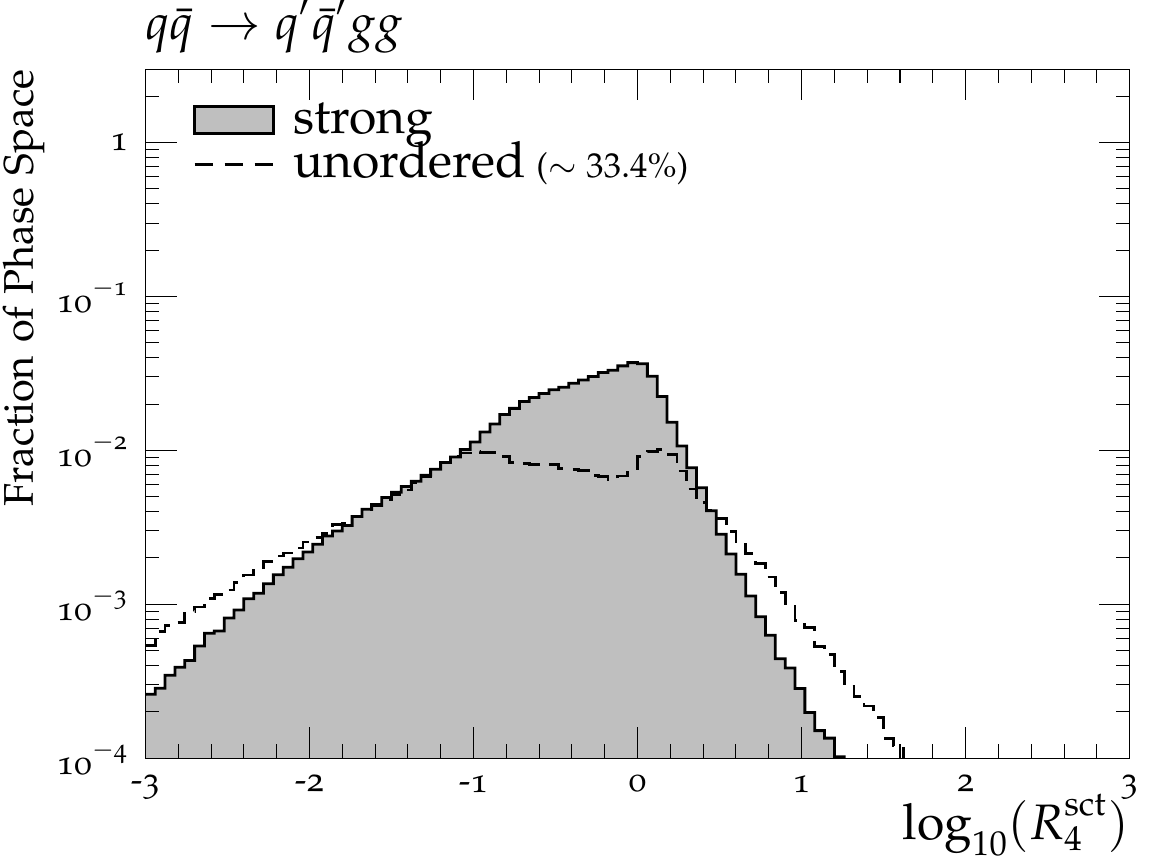}
	\caption{Ratios $R_n^\text{sct}$ of the sector-shower approximation to LO matrix elements for $q\bar q$-annihilation processes to 4 partons in a flat phase space scan.}
	\label{fig:MERatiosQQbHiMulti}
\end{figure}

\begin{figure}[ht]
	\centering
	\includegraphics[width=0.49\textwidth]{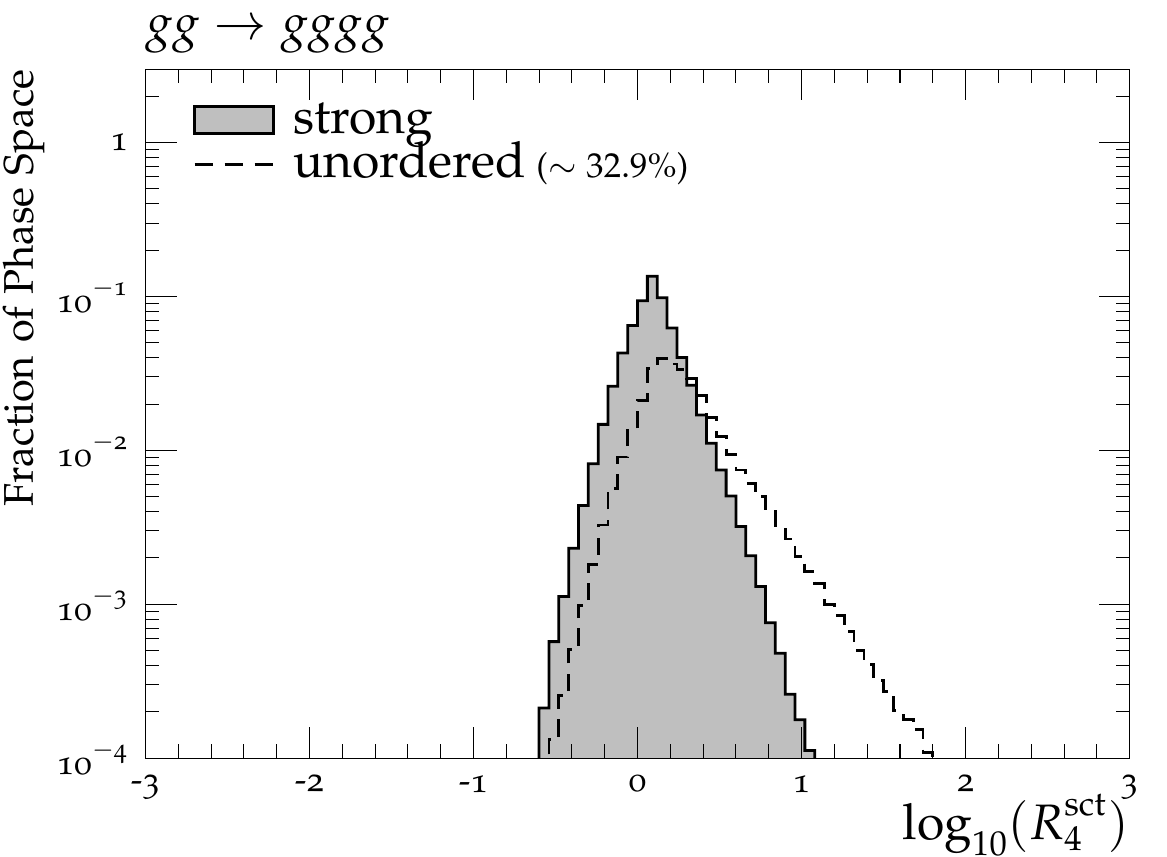}
	\includegraphics[width=0.49\textwidth]{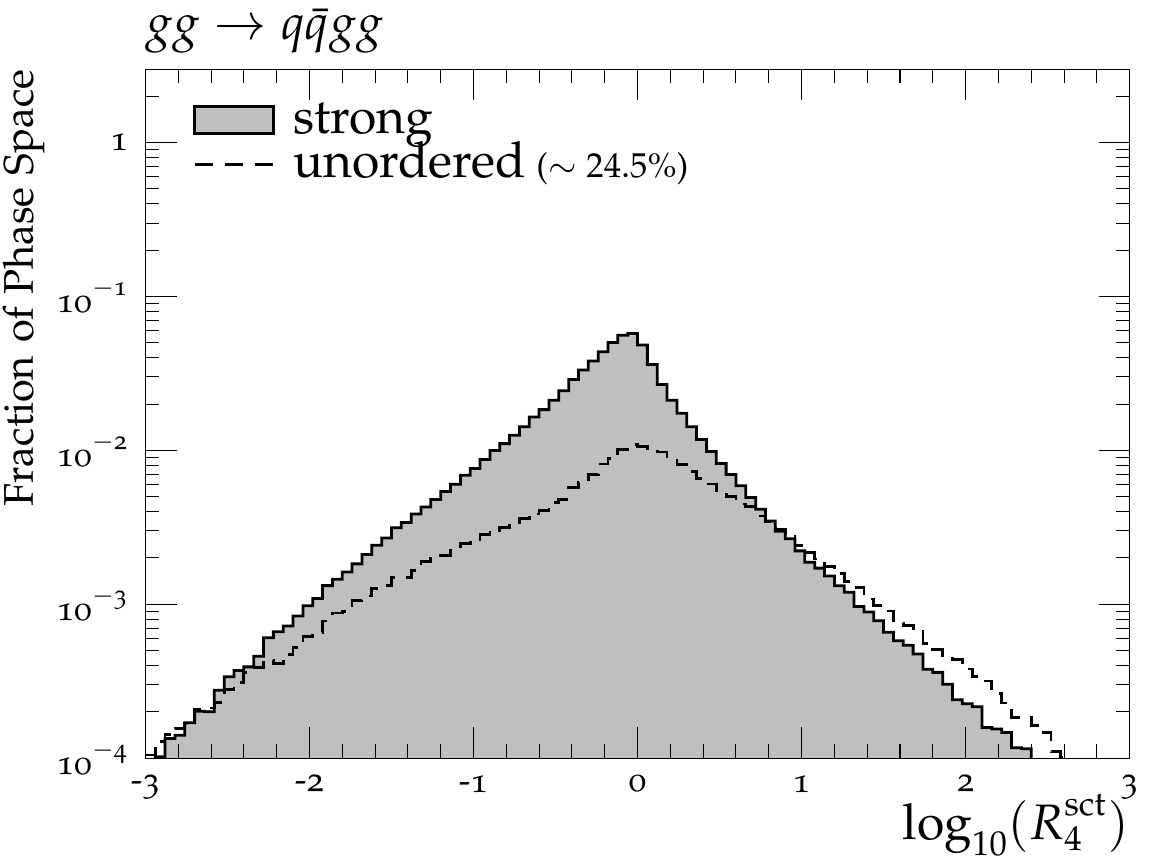}
	\caption{Ratios $R_n^\text{sct}$ of the sector-shower approximation to LO matrix elements for $gg$-fusion processes to 4 partons in a flat phase space scan.}
	\label{fig:MERatiosGGhiMulti}
\end{figure}

\FloatBarrier
\clearpage
\section{Tune parameters}
\label{sec:tune}
The full set of parameters changed relative to the \Pythia\ 8.3 default tune are shown in the following table:

\begin{table}[h]
\begin{tabular}{l c c }
\toprule
  Name &                      Now &      Default \\
\midrule
  \texttt{BeamRemnants:primordialKThard}                 &                  0.4 &      1.8 \\
  \texttt{BeamRemnants:primordialKTsoft}                 &                  0.25 &      0.90 \\
  \texttt{ColourReconnection:range}                      &                  1.75 &      1.80 \\
  \texttt{Diffraction:mMinPert}                          &               1000000.0 &     10.0 \\
  \texttt{MultipartonInteractions:alphaSorder}           &                        2 &            1 \\
  \texttt{MultipartonInteractions:alphaSvalue}           &                  0.119 &      0.130 \\
  \texttt{MultipartonInteractions:ecmPow}                &                  0.210 &      0.215 \\
  \texttt{MultipartonInteractions:expPow}                &                  1.75 &      1.85 \\ 
  \texttt{MultipartonInteractions:pT0Ref}                &                  2.24 &      2.28 \\
  \texttt{SigmaProcess:alphaSorder}                      &                        2 &            1 \\ 
  \texttt{SigmaProcess:alphaSvalue}                      &                  0.119 &      0.130 \\ 
  \texttt{StringFlav:etaPrimeSup}                        &                  0.10 &      0.12 \\ 
  \texttt{StringFlav:etaSup}                             &                  0.5 &      0.6 \\ 
  \texttt{StringFlav:mesonCvector}                       &                  1.30 &      0.88 \\
  \texttt{StringFlav:mesonSvector}                       &                  0.53 &      0.55 \\ 
  \texttt{StringFlav:mesonUDvector}                      &                  0.42 &      0.50 \\ 
  \texttt{StringFlav:popcornSmeson}                      &                  0.75 &      0.50 \\ 
  \texttt{StringFlav:popcornSpair}                       &                  0.75 &      0.90 \\ 
  \texttt{StringFlav:probQQ1toQQ0}                       &                0.025 &    0.0275 \\ 
  \texttt{StringFlav:probQQtoQ}                          &                0.077 &    0.081 \\ 
  \texttt{StringFlav:probSQtoQQ}                         &                  1.000 &      0.915 \\ 
  \texttt{StringFlav:probStoUD}                          &                  0.205 &      0.217 \\ 
  \texttt{StringPT:sigma}                                &                  0.305 &      0.335 \\ 
  \texttt{StringZ:aExtraDiquark}                         &                  0.90 &      0.97  \\
  \texttt{StringZ:aLund}                                 &                  0.55 &      0.68 \\ 
  \texttt{StringZ:bLund}                                 &                  0.78 &      0.98 \\ 
  \texttt{StringZ:rFactB}                                &                  0.850 &      0.855 \\ 
  \texttt{StringZ:rFactC}                                &                  1.15 &      1.32 \\ 
\bottomrule
\end{tabular}
\label{tab:tune}
\end{table}

\bibliographystyle{JHEP}
\bibliography{ms}

\end{document}